\documentclass[a4paper,11pt]{article}

\PassOptionsToPackage{dvipsnames,table}{xcolor}

\usepackage{jheppub}
\usepackage{lineno}

\usepackage{amssymb}
\usepackage{amsfonts}
\usepackage{mathrsfs}
\usepackage{subcaption}
\usepackage{bbm}
\usepackage{longtable}
\usepackage[nohyperlinks]{acronym}
\usepackage{pifont}
\usepackage{enumerate}
\usepackage{pdflscape}
\usepackage{hhline}
\usepackage{multirow}
\usepackage{booktabs}
\usepackage{adjustbox}
\usepackage{tabulary}
\usepackage{bbold}
\usepackage{centernot}
\usepackage{mathtools}
\usepackage{stmaryrd}
\usepackage{yfonts}
\usepackage{braket}
\usepackage{tikz}
\usetikzlibrary{decorations.pathmorphing}
\usepackage{comment}

\usepackage[many]{tcolorbox}
\tcbuselibrary{breakable, skins}
\usepackage{setspace}

\definecolor{main}{HTML}{5989cf}
\definecolor{sub}{HTML}{cde4ff}
\definecolor{lightgray}{gray}{0.95}
\definecolor{midgray}{gray}{0.9}

\newtcolorbox{boxD}{
    colback = sub,
    colframe = main,
    boxrule = 0pt,
    toprule = 3pt,
    bottomrule = 3pt
}

\makeatletter
\newcommand{\xMapsto}[2][]{\ext@arrow 0599{\Mapstofill@}{#1}{#2}}
\def\Mapstofill@{\arrowfill@{\Mapstochar\Relbar}\Relbar\Rightarrow}
\makeatother

\newcolumntype{R}[1]{>{\raggedleft\arraybackslash}p{#1}}

\allowdisplaybreaks

\title{\boldmath Axio-Dilaton Dark Energy:  A Dynamical Systems and Bayesian Inference  Analysis}

\author[a]{Mario Ramos-Hamud,}
\author[b]{Gabriela García-Arroyo,}
\author[a,c]{Fernando Quevedo,}
\author[b]{J.~Alberto Vázquez}

\affiliation[a]{DAMTP, University of Cambridge,
Wilberforce Road, Cambridge, Cambridgeshire, UK}
\affiliation[b]{Instituto de Ciencias Físicas, Universidad Nacional Autónoma de México,
Cuernavaca, Morelos, 62210, México}
\affiliation[c]{New York University Abu Dhabi,
Saadiyat Island, Abu Dhabi, PO Box 128199, UAE}

\emailAdd{mr895@cam.ac.uk}
\emailAdd{arroyo@icf.unam.mx}
\emailAdd{fq201@cam.ac.uk}
\emailAdd{javazquez@icf.unam.mx}

\abstract{We initiate a systematic  cosmological investigation of an axio-dilaton system governed by a  curved field-space metric as generically motivated by supergravity and string compactifications. The scalar sector consists of a light scalar dilaton field evolving along an asymptotically exponential potential and a pseudoscalar field spanning an essentially flat direction, kinetically coupled through an exponential field-space metric. Deviations from the asymptotic regime are incorporated by introducing a generalised Albrecht-Skordis potential containing a polynomial factor. Utilising a dynamical systems approach alongside numerical evolution and Bayesian inference, we constrain the model using Type Ia Supernovae (SNe Ia), Baryon Acoustic Oscillations (BAO) and Planck 2018 distance-prior data.  We identify a parameter degeneracy indicating that steeper potentials demand stronger field-space kinetic couplings to sustain late-time cosmic acceleration. Finally, a joint likelihood analysis reveals that while the model successfully accounts for dark energy dynamics, it is only slightly statistically favoured  over $\Lambda$CDM by current observations. }

\begin{document}
\maketitle
\flushbottom

\acrodef{FLRW}{Friedmann-Lemaître-Robertson-Walker}
\acrodef{EOS}{equation of state} 
\acrodef{EDE}{early dark energy} 
\acrodef{EXP}{Exponential}
\acrodef{CL}{Clemson and Liddle}
\acrodef{AS}{Albrecht-Skordis}
\acrodef{SNe Ia}{Type Ia Supernovae}
\acrodef{BAO}{Baryon Acoustic Oscillations}
\acrodef{BGS}{Bright Galaxy Sample}
\acrodef{LRG}{Luminous Red Galaxies}
\acrodef{ELG}{Emission Line Galaxies}
\acrodef{PLK}{Planck}
\acrodef{PP}{Pantheon+}
\acrodef{BP}{Binned Pantheon}
\acrodef{CMB}{Cosmic Microwave Background}


\section{Introduction}

Over the past quarter-century, high-precision observational data have substantially advanced our understanding of the cosmic evolution, especially through the unexpected discovery of the current acceleration of the universe \cite{SupernovaSearchTeam:1998fmf, SupernovaCosmologyProject:1998vns}. Despite its remarkable success in explaining a vast array of observational data, the standard $\Lambda$CDM paradigm is also being challenged. The persistent discrepancy in the measured values of the Hubble parameter at low and high redshifts \cite{CosmoVerseNetwork:2025alb}, together with the recent DESI data on baryon acoustic oscillations (BAO) \cite{DESI:2024mwx, DESI:2025zgx}, leave open the possibility that dark energy is not constant but evolves with cosmic time, potentially indicating a departure from the standard $\Lambda$CDM paradigm toward a time-dependent, dynamical dark energy component.

A thoroughly explored approach to modelling dynamical dark energy is to postulate the existence of a real Brans-Dicke scalar field, with an ad-hoc potential, that can be arbitrarily adjusted to match observations (for general reviews, see for instance \cite{Peebles:2002gy, Copeland:2006wr, Peebles:2022bya}). However, this conventional single-field approach suffers from several conceptual and phenomenological limitations:

\begin{itemize}
\item The scalar field itself and its potential are introduced by hand, which severely limits the explanatory and predictive power of the framework.
\item By choosing a specific potential it is implicitly assumed that the background vacuum energy vanishes, for which there is no explanation. Unlike $\Lambda$CDM, which can rely on the string landscape to explain the smallness of the cosmological constant, plain quintessence models fail to  address the core problem of dark energy: why its energy scale is so hierarchically small  compared with the Standard Model scales, such as the electron mass contribution to the vacuum energy (for reviews emphasising this point, see \cite{Weinberg:1988cp,Burgess:2013ara}).
\item A light scalar field is subject to exceptionally stringent observational constraints from the running of coupling constants and fifth force experiments. Since a single scalar field manifold is flat, its kinetic energy can always be canonically normalised, and therefore a  single scalar field does not have two-derivative interactions to compete with the two-derivative gravitational interactions that are typically considered to address different cosmological questions \cite{Brax:2023qyp, Smith:2025grk}.
\item Choosing a single light scalar field is done purely for mathematical simplicity; it is neither unique nor ubiquitous in ultraviolet-complete fundamental frameworks such as string theory (for recent reviews on dark energy from string theory, see \cite{Cicoli:2023opf, Andriot:2026lac}).
\item So far, many attempts to achieve asymptotic acceleration from exponential potentials in a flat (curvature-free) universe have failed in single-field models\footnote{For possible workarounds, see \cite{Adam:2025kve, SanchezLopez:2025uzw} for recent studies on data fitting of minimally and non-minimally coupled single-field models with various potentials, and \cite{Pourtsidou:2025sdd} for a Bayesian analysis of single-field quintessence with momentum coupling to dark matter.}. To alleviate the situation, different avenues have been explored in the literature such as including spatial curvature  \cite{Andriot:2024jsh, Bhattacharya:2024hep}, considering multi-field quintessence \cite{Cicoli:2020cfj, Brinkmann:2022oxy, Russo:2022pgo, Alestas:2025syk, Grimm:2025cpq}, or interacting two-scalar-field scenarios \cite{Garcia-Arroyo:2024tqq, Garcia-Arroyo:2026mdt}.. 
\end{itemize}

It is therefore important to adopt a top-down approach that goes beyond single-field models and explores whether a fundamental theory, such as string theory, contains the necessary ingredients to account for the observations, should the current hints of dynamical dark energy ultimately be confirmed. Within the richness of elements present in string compactifications, moduli fields stand out as one of the most generic low-energy remnants from string compactifications. Among these many moduli, the dilaton-like fields (which parametrise properties such as the overall volume of the extra dimensions or the inverse string coupling) never appear in isolation: they are universally accompanied by an axionic partner that completes a complex scalar field layout.  Therefore, studying the cosmological implications of a dilatonic field while neglecting its axion partner fails to capture the indispensable low-energy physics dictated by string theory. This provides a strong motivation to seriously study the cosmological implications of axio-dilaton systems.

A foundational approach toward addressing dark energy from string theory was to obtain de Sitter solutions in which all the moduli are stabilised, giving rise to the now well-known string landscape. The string landscape provides a first-principles derivation of the $\Lambda$CDM scenario with an added conceptual bonus: contrary to other proposals, the landscape provides a statistical mechanism to explain why the cosmological constant can be hierarchically smaller than the quantum corrections to the vacuum energy coming from the Standard Model. If dark energy is indeed dynamical, this landscape approach has to be adapted to explain the potential evolution of dark energy. The simplest modification is to study the dynamics of the ubiquitous axio-dilaton system.

Furthermore, a string inspired scenario known as Yoga dark energy \cite{Burgess:2021obw} has been put forward to potentially explain the smallness of the dark energy and precisely incorporates the dynamics of the axio-dilaton system. Even though a full-fledged realisation of this scenario is not yet available, comparisons of this class of models with current observations have recently commenced \cite{Smith:2025icl,Smith:2025uaq}. 

In this work, we focus explicitly on the cosmological implications of the axio-dilaton system with the fields $\phi_1$ and $\phi_2$ are coupled through the kinetic function $W(\phi_1)=e^{-\xi\phi_1/M_p}$. We further consider the corrected runaway potential
\begin{equation}
V(\phi_1)=V_0 \; f(\phi_1) e^{-\lambda\phi_1/M_P},
\label{Eq_Potential}
\end{equation}
where $f(\phi_1)=A+(\phi_1/M_P-B)^n$ is a polynomial of degree $n$. Throughout this work, we treat $\xi$, $\lambda$, $A$, and $B$ as free parameters. Therefore, we extend the multi-field framework of \cite{Cicoli:2020cfj} by considering the prefactor $f(\phi_1)$ in the potential and by performing a comprehensive Bayesian analysis\footnote{A two-field Bayesian analysis was also considered in \cite{Alestas:2025syk} for a flat field space.} of the system to study the viability of its cosmological solutions against experimental data. 

Crucially, we first map the multi-dimensional phase space of the system using a dynamical systems approach. We then translate these physical dynamics into a detailed Bayesian statistics and parameter inference framework. We explore a 6-parameter model consisting of the three standard baseline cosmological parameters ($\Omega_m, \Omega_b h^2, H_0$) together with three parameters that determine the specific couplings, field-space curvature, and polynomial slopes of our scalar system. Using the \texttt{SimpleMC} code, we subject this six-parameter model to a joint-likelihood analysis of Type Ia supernovae, Planck distance priors, and the latest DESI BAO data, thereby assessing the empirical viability of the axio-dilaton framework as a theoretically well-motivated alternative to $\Lambda\mathrm{CDM}$.

The remainder of this paper is organised as follows. In Section 2, we review the theoretical formulation of the axio-dilaton action and its field-space geometry, which serves as a way to define the system under investigation. Section 3 is devoted to the dynamical systems analysis. In Section 4, we present the results of a direct numerical calculation, while section 5 presents the full Bayesian analysis. In Appendix A, we conduct a stability study of all the fixed points, and then appendix B extends our results by considering arbitrary powers of the polynomial factor in the scalar potential.

\section{Axio-dilaton system}

The most general Lagrangian density for a two-field scalar system is given by
\begin{equation}
\frac{\mathcal{L}}{\sqrt{-g}} = \frac{1}{2} \mathcal{G}_{ij} \partial_\mu \phi^i \partial^\mu \phi^j - V(\phi^i),
\end{equation}
where $\mathcal{G}_{ij}$ denotes the dimensionless field-space metric that governs the kinetic couplings among the scalar fields. In this work, we consider a curved target-space metric of the form
\begin{equation}
d\mathfrak{s}^2 = d\phi_1^2 + W^2(\phi_1)  d\phi_2^2 \quad \text{with} \quad W(\phi_1) = e^{-\xi \phi_1/M_P},
\end{equation}
where $\xi$ is a positive constant and without loss of generality, we redefine the field such that $W(0) = 1$. Motivated by supergravity constructions, such as axion–saxion supermultiplets, we assume the existence of an approximate flat direction along $\phi_2$. As a consequence, the scalar potential depends solely on  $\phi_1$. In particular, inspired by string compactifications in the asymptotic region of moduli space -- where the effective theory is under perturbative control -- the potential is often assumed to take the form
\begin{equation}
V_{\text{asympt}}(\phi_1) = V_0 \; e^{-\lambda \phi_1/M_P},
\end{equation}
with $V_0 \sim \mathcal{O}(1)$ (in Planck units) and $\lambda$ a positive constant. A natural generalisation of this asymptotic potential consists of including sub-leading corrections through a prefactor function, leading to Eq.~(\ref{Eq_Potential}). There, $f(\phi_1)$ is taken to be a polynomial encoding deviations from the asymptotic regime. As emphasised in \cite{Burgess:2022nbx}, there is a natural source of the polynomial factor in string compactifications, corresponding to logarithmic corrections to the scalar potential which are induced by the renormalisation group running of physical parameters. For a modulus field $\tau$, the leading order term in the scalar potential in a $1/\tau$ expansion is $f(\log\tau)/\tau^m$ with $f(\log\tau)$ determined from a resummation of logarithms as usual in renormalisation group calculations. As long as the spectrum contains particles whose masses depend on different powers of $\tau$, the standard logarithmic dependence on mass ratios in the RG flow generically becomes a logarithmic dependence on $\tau$. Keeping the first few terms in the expansion of $f$ gives a generalized Albrecht-Skordis potential for the canonically normalized field $\phi_1\propto \log\tau$. See \cite{Burgess:2022nbx, Burgess:2021obw} for further details.  

In this work, we consider the general form
\begin{equation}
    f(\phi_1)= A+ \left(\frac{\phi_1}{M_P} -B\right)^n
,
\end{equation}
where $A$ and $B$ are dimensionless constants and $n>0$. We perform a systematic study of the cosmological implications of the exponent $n$ by combining a dynamical systems analysis with numerical evolution and Bayesian parameter inference using observational data. For concreteness, in the dynamical system analysis, we focus on three representative cases corresponding to $n=0,1,2$, and subsequently present numerical results extending our analysis to arbitrary values of $n$.

We assume a homogeneous and isotropic Universe described by the standard  \ac{FLRW} metric.
\begin{equation}
    ds^2= -dt^2+ a^2(t) \left(\frac{dr^2}{1-kr^2} + r^2 d\Omega^2_2  \right),
\end{equation}
where $a(t)>0$ is the scale factor and $k=0,\pm 1$ denotes the  spatial curvature. In the following, we restrict our analysis to the spatially flat case, $k=0$. In addition to the scalar fields, we include a barotropic perfect fluid characterised by the \ac{EOS}
\begin{equation}
    p_b= \omega_b \;\rho_b = (-1+\gamma) \rho_b \quad \text{such that} \quad \dot{\rho}_b= -3H (1+\omega_b)\rho_b = -3 H \gamma\; \rho_b,  
\end{equation}
with $0\leq\gamma<2$. The cases $\gamma=0,1, 4/3$ correspond to dark energy, pressureless matter, and radiation, respectively. Meanwhile, the effective density and pressure of the scalar sector in general are given by
\begin{equation}
    \rho_{\phi_i}= \frac{1}{2} \phi^2_i+ V[\phi_i] \quad \text{and} \quad p_{\phi_i}= \frac{1}{2}\phi^2_i-V[\phi_i], \quad \text{such that} \quad \omega_{\phi_i}= \frac{p_i}{\rho_i},
\label{Eq:EoS-fields}
\end{equation}
where $V[\phi_i]$ means the potential for the field $\phi_i$, but as mentioned above, we assume $V[\phi_2]=0$. 

The equations of motion for the scalar fields are
\begin{subequations}
    \begin{equation}
        \ddot{\phi}_1+ 3H \dot{\phi}_1- WW_{\phi_1 } \dot{\phi}^2_2+V_{\phi_1}=0,
    \end{equation}
    \begin{equation}
        \ddot{\phi}_2+ 3H \dot{\phi}_2+ 2\frac{W_{\phi_1 }}{W} \dot{\phi}_1 \dot{\phi}_2=0.
    \end{equation}
\end{subequations}
where $\cdot{} \equiv d/dt$ and $W_{\phi_1}=dW/d\phi_1$. Finally, the first Friedmann equation reads
\begin{equation}
    H^2= \frac{1}{3 M^2_p} \sum_i \rho_i =  \frac{1}{3 M^2_P}  \left(  \frac{\dot{\phi}^2_1}{2}+ \frac{W^2}{2} \dot{\phi}^2_2+V+ \rho_b \right),
\label{eq:1st-Friedmann}    
\end{equation}
where $H\equiv \dot{a}/a$ denotes the Hubble expansion rate.

\section{Dynamical systems}

The equations of motion for scalar fields with an exponential potential, together with the associated Friedmann equation, have been extensively analysed in the literature, see e.g.  \cite{Copeland:1997et, Cicoli:2020cfj}. Departures from a purely exponential potential lead to a more intricate dynamical structure and require the introduction of additional auxiliary quantities \cite{Zhou:2007xp, Fang:2008fw, Matos_2009, Urena-Lopez:2011gxx}:
\begin{equation}
\alpha \equiv- M_P \frac{V_{\phi_1}}{V} = \lambda - \frac{f_{\phi_1}}{f}\quad \text{and} \quad \Gamma\equiv \frac{V_{\phi_1 \phi_1} V}{V^2_{\phi_1}}.
\label{eq:alpha-lambda}
\end{equation}
Introducing the dimensionless variables $z_i\equiv \phi_i/M_P$, the potential given in Eq.~(\ref{Eq_Potential}) can be written as
\begin{equation}
    V(z_1)= V_0 \left[A+ (z_1 -B)^n\right]e^{-\lambda z_1}.
\label{eq:AS-General-Pot}
\end{equation}
For this class of potentials, the auxiliary functions $\alpha$ and $\Gamma$ take the explicit form
\begin{subequations}
    \begin{equation}
    \alpha(z_1)=\lambda -\frac{n \mathfrak{z}^{-1+n}}{A+\mathfrak{z}^n}
    \end{equation}
and
    \small{
    \begin{equation}
    \Gamma(z_1) = \frac{
    (A + \mathfrak{z}^n ) [ n^2 \mathfrak{z}^n + \mathfrak{z}^2 (A + \mathfrak{z}^n) \lambda^2 - n \mathfrak{z}^n \big(1 + 2\lambda \mathfrak{z}\big)]
    }{
    [ \mathfrak{z} (A + \mathfrak{z}^n) \lambda-n \mathfrak{z}^n]^2
    },
\end{equation}
}
\label{eq:General-Alpha&Gamma}
\end{subequations}
where we introduced $\mathfrak{z}= z_1-B $ for notational convenience. Since the departure from an exponential potential generically implies $\alpha\neq \lambda$ and introduces an explicit field dependence in $\alpha$, the standard system of equations must be enlarged accordingly. To obtain a closed dynamical system, we therefore promote $\alpha$ to a dynamical variable and introduce the set
\begin{equation}
      x_1\equiv \frac{\dot{z}_1}{\sqrt{6}H }, \quad x_2\equiv \frac{W \dot{z}_2}{\sqrt{6}H}, \quad y\equiv  \frac{\sqrt{V}}{\sqrt{3}H} \quad \text{and}\quad \alpha \equiv-  \frac{V_{z_1}}{V}.  
     \label{eq:variables-DS}
\end{equation}
In terms of these variables, the evolution equations can be cast into the dynamical system
\begin{tcolorbox}
\setlength{\abovedisplayskip}{0pt} 
\setlength{\belowdisplayskip}{0pt} 
\begin{subequations}
\label{eq:DS}
\begin{align}
   h_1\equiv x_1' &= -3x_1 + \frac{3}{2}\Pi\, x_1
            + \sqrt{\frac{3}{2}}\Bigl(
                \alpha y^2
            -2\xi x_2^2
              \Bigr), 
              \label{eq:DS-Eq1} \\
    h_2\equiv x_2' &= -3x_2 + \frac{3}{2}\Pi\, x_2
            + \sqrt{6} \xi x_1 x_2, \label{eq:DS-Eq2} \\
    h_3\equiv y'   &= \frac{3}{2}\Pi\, y
            - \sqrt{\frac{3}{2}} \alpha\,x_1 y, \label{eq:DS-Eq3}\\
  h_4\equiv \alpha'&= \sqrt{6} (1-\Gamma) \alpha^2 x_1,
            \label{eq:DS-Eq4}
\end{align}
\label{eq:General-DS}
\end{subequations}
\end{tcolorbox}
\noindent
where $'\equiv d/dN$ denotes differentiation with respect to the number of e-folds $N=\ln a$. The quantity $\Pi$ appearing above is defined as
\begin{equation}
\label{eq:Pi}
    \Pi \equiv 2x_1^2 + 2x_2^2 + \Omega_b\,\gamma,
\end{equation}
which is directly related to the deceleration parameter $q$ via
\begin{equation}
\label{eq:q}
    q + 1 = -\frac{\dot H}{H^2} = \frac{3}{2}\,\Pi.
\end{equation}
Introducing $\Omega_b= \rho_b/(3H^2 M_P^2)$, the Friedmann constraint in Eq.~(\ref{eq:1st-Friedmann}) becomes $\Omega_b= 1-\Omega_\phi$, with
\begin{equation}
\Omega_\phi = x_1^2 + x_2^2 + y^2 \leq 1.
\label{Eq:constraint}
\end{equation}
Consequently, the physically admissible phase space is restricted to the interior of the unit half-ball in three dimensions, which is the region of interest for the subsequent dynamical analysis. Additionally, the scalar-sector equation of state (\ac{EOS}) in Eq.~(\ref{Eq:EoS-fields}) parameter can be expressed in terms of the dynamical variables as: 
\begin{equation}
    \omega_\phi\equiv \frac{p_\phi}{\rho_\phi}= \frac{x^2_1+x^2_2-y^2}{x^2_1+x^2_2+y^2}.
\end{equation}
The system of equations~(\ref{eq:DS-Eq1})–(\ref{eq:DS-Eq4}) is not strictly autonomous since the evolution equation for $\alpha$ contains an explicit dependence on the field $z_1$ through the definition of $\Gamma$. In principle, a complete description of the dynamics would therefore require supplementing the system with the evolution equation for $z_1$,
\begin{equation}
    z_1'\equiv \sqrt{6} x_1,
\label{eq:extra-variable}
\end{equation}
thus promoting the system to five dimensions. However, when analysing the fixed points of the dynamics, Eq.~(\ref{eq:extra-variable}) implies $x_1 = 0$. This condition is already captured by Eq.~(\ref{eq:DS-Eq4}) within the original four-dimensional system. Consequently, for the purpose of determining and classifying the fixed points, it is sufficient\footnote{Even though in this case the evolution of $z_1$ plays no role in the identification and classification of fixed points, it is relevant for the analysis of trajectories away from critical points.} to consider the reduced system~(\ref{eq:DS-Eq1})–(\ref{eq:DS-Eq4}). In particular, Eq.~(\ref{eq:DS-Eq4}) admits three distinct branches, which we discuss in turn.

\begin{tcolorbox}[
    breakable,
    enhanced,
    colback   = white,
    colframe  = gray,
    arc       = 4pt,
    boxrule   = 0.6pt,
    left      = 6pt,
    right     = 6pt,
    top       = 6pt,
    bottom    = 6pt,
    title     = {Branches of $\alpha'=0$},
    fonttitle = \bfseries
]
\begin{itemize}

\item[i)] $x_1 = 0$ $(\alpha \neq \lambda)$: this represents a
genuinely new fixed point that has no counterpart in the pure
exponential system. Since $x_1 = 0$ means $\dot{z}_1 = 0$, the
first field is kinetically frozen and does not roll. Substituting
$x_1=0$ into Eq.~(\ref{eq:DS-Eq1}) gives
\begin{equation}
    \alpha = \frac{2\xi x_2^2}{y^2},
\end{equation}
which pins $\alpha$ to a finite value generically non-vanishing and different from $\lambda$ as long as $y, \xi \neq 0$. It should be emphasized that this does not imply the general nonexistence of kinetic, scaling, or dark-energy solutions --- $x_2$ and $y$ remain free, so $z_2$-kinetic domination and potential domination are still possible. Notably, there are no solutions satisfying the constraint \(x_1 \neq 0\), namely the standard single-field scaling solution and the geodesic fixed point.

\item[ii)] $\alpha = 0$: from the definition
$\alpha \equiv - V_{z_1}/V$, this means $V'(z_1) = 0$,
i.e.\ the field sits at a critical point of the potential, and therefore its nature, of the critical point, depends on the specific form of the potential.

\item[iii)] $\Gamma = 1$ $(\alpha \neq 0,\; x_1 \neq 0)$: the potential locally mimics a pure
exponential, admitting scaling solutions analogous to those of the
single-field exponential system.  New fixed points correspond to the zeros of the function $\Gamma(\alpha)-1$, which exists only if the function $\alpha(z)$ is invertible. In other words, this corresponds to make the system of equations (\ref{eq:General-DS}) autonomous. From the expression for $\alpha(z)$ in Eq. (\ref{eq:General-Alpha&Gamma}), one can find the inverse function by finding the roots of the equation
\begin{equation}
    (z_1-B)^n+\frac{n}{\alpha-\lambda}(z_1-B)^{n-1}+A=0.
\end{equation}

\end{itemize}
\end{tcolorbox}

\subsection{Case $n=0$: \ac{EXP}}
\label{subsec:exp-pot}

\begin{figure}[t]
    \centering
    \begin{subfigure}[b]{0.495\textwidth}
        \centering
        \includegraphics[width=\textwidth]{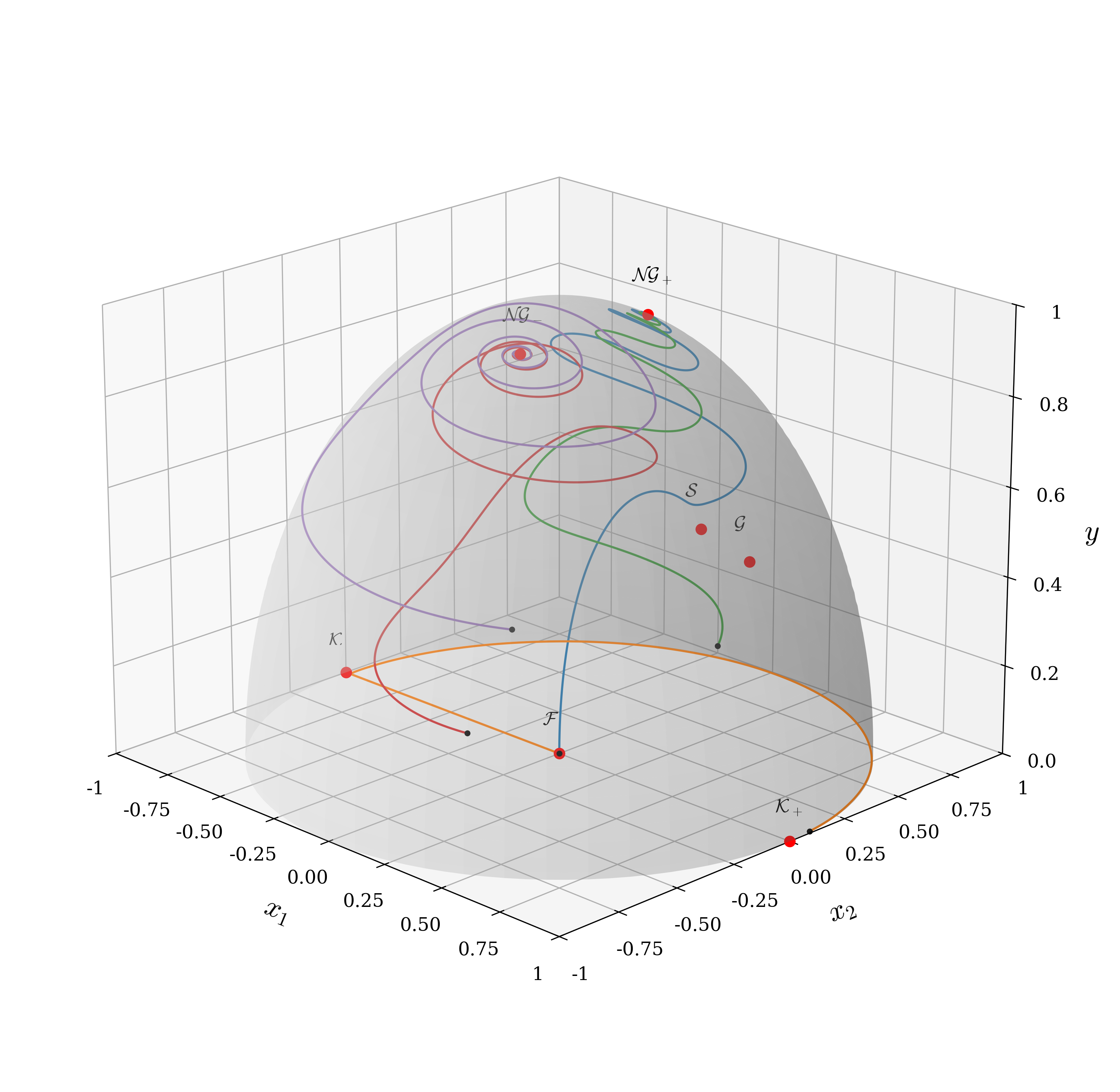}
        \caption{Example 1: $\xi=10$.}
        \label{fig:TrajectoryE1}
    \end{subfigure}
    \hfill
    \begin{subfigure}[b]{0.495\textwidth}
        \centering
        \includegraphics[width=\textwidth]{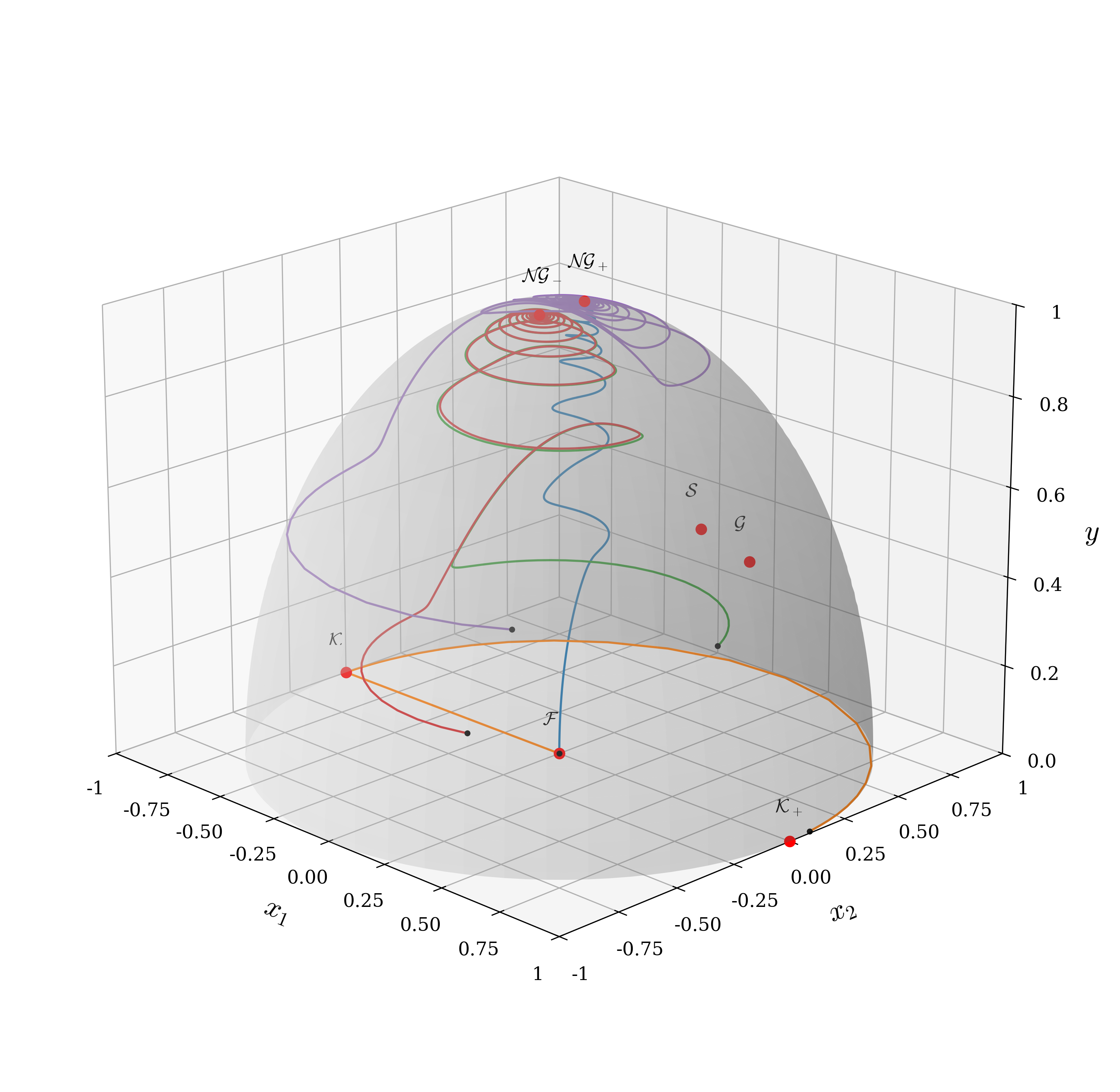}
        \caption{Example 2: $\xi=100$.}
        \label{fig:TrajectoryE2}
    \end{subfigure}
    \caption{Phase-space trajectories for the \ac{EXP} case. The coloured lines represent trajectories corresponding to different initial conditions, indicated by the black points, while the fixed points are shown in red. For this hierarchy of parameters, the fixed point $\mathcal{NG}$ acts as the late-time attractor. In both cases $\gamma=1$ and $\lambda=2$.}
    \label{fig:Trajectories}
\end{figure}

\begin{figure}[t!]
    \centering
    \begin{subfigure}[b]{0.495\textwidth}
        \centering
            \includegraphics[width=\textwidth]{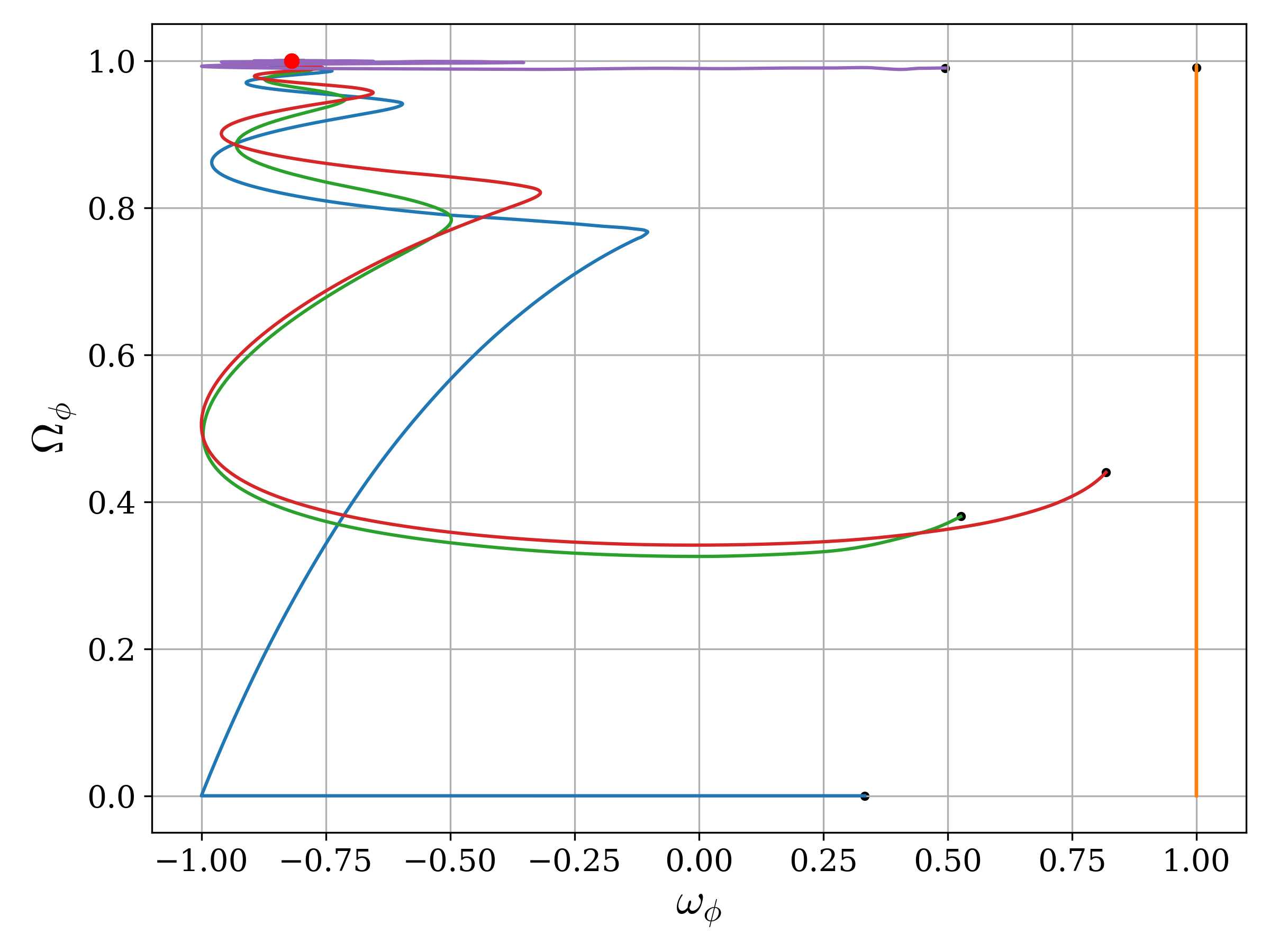}
        \caption{Example 1: $\xi=10$.}
        \label{fig:OmegavsomegaE1}
    \end{subfigure}
    \hfill
    \begin{subfigure}[b]{0.495\textwidth}
        \centering
        \includegraphics[width=\textwidth]{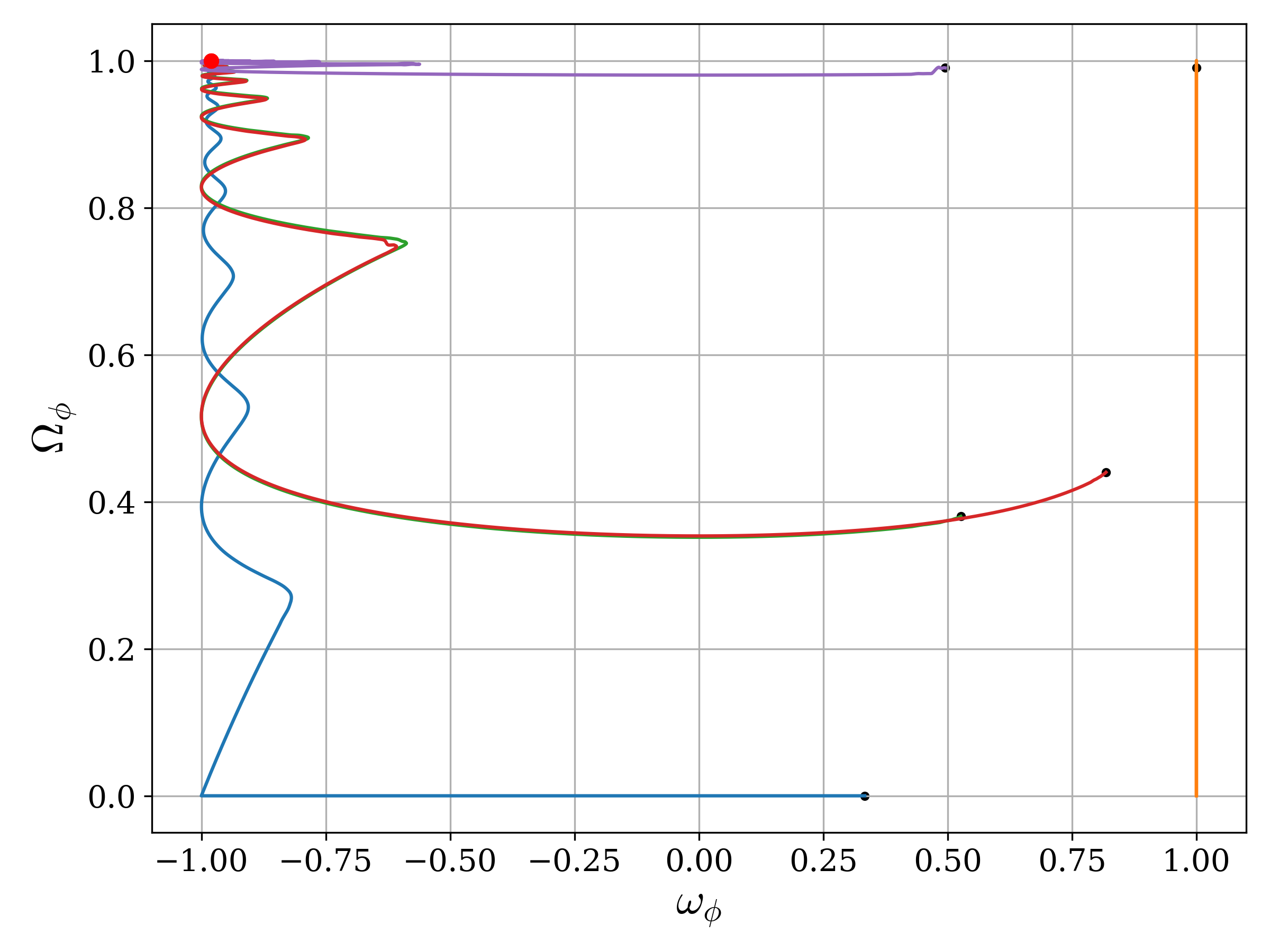}
        \caption{Example 2: $\xi=100$.}
        \label{fig:OmegavsomegaE2}
    \end{subfigure}
    \caption{Evolution of the scalar energy density, $\Omega_\phi$, and the \ac{EOS} parameter, $w_\phi$, for the different initial conditions considered in Fig.~\ref{fig:Trajectories} for an \ac{EXP} potential. In both cases $\gamma=1$ and $\lambda=2$.}
    \label{fig:Omegas}
\end{figure}

In this section, we consider $n=0$ in Eq.~(\ref{eq:AS-General-Pot}), and the auxiliary functions are simply expressed as
\begin{equation}
    V(z_1)= V_0 e^{-\lambda z_1}, \quad \alpha=\lambda \quad \text{and}\quad \Gamma=1.
\label{eq:EXP-potential}
\end{equation}
This reduces the dynamical system to only three dimensions, and the critical points are found to be the solution of the constraint: $x_1'=x_2'=y'=0$. Given Eq.~(\ref{Eq:constraint}), we impose $x_1, x_2 \in [-1,1]$ and $y\in (0,1]$. As previously found in \cite{Cicoli:2020cfj}, the fixed points contain the single-field solutions corresponding to kinetic domination ($\mathcal{K}_{\pm}$), fluid domination ($\mathcal{F}$), and the scaling solution ($\mathcal{S}$) where the scalar tracks the fluid energy density at a geodesic point ($\mathcal{G}$) named after their corresponding field space trajectory. Additionally, the two-field solution is found evolving in a non-geodesic ($\mathcal{NG}$) trajectory in the field space. In Figure \ref{fig:Trajectories}, we plot the phase-space trajectory for several initial conditions showing how the non-geodesic point $\mathcal{NG}$ corresponds to an attractor with an \ac{EOS} that allows dark energy when $\xi \gg \lambda$. This can also be understood analytically as in  \cite{Krippendorf:2018tei} by using the ansatz
\begin{equation}
    \phi_1(t)= \frac{2 M_P}{\lambda}\ln\left(\frac{t}{t_0}\right) - C_1 \quad \text{and} \quad a(t)= \left(\frac{t}{t_0}\right)^{D},
\end{equation}
for $C_1$ (in mass units) and $D$ some constants. This turns out to be $\dot{\phi}_2(t)= K\; a(t)^{-3} W^{-2}$, with $K$ as a constant, and therefore
\begin{equation}
    \phi_2(t)=
    e^{-2 \xi C_1/M_P} M_P^2 t \; \left(\frac{t}{t_0}\right)^{-1+ \frac{2 \xi}{\lambda}} \frac{ \lambda}{2 \xi}+ C_2,
\end{equation}
with $C_2$ an integration constant with mass units, and with the exponent in the scale factor as $D= (\lambda+2\xi)/3\lambda$ coming from the solution in flat space $a(t)=(t/t_0)^{{2/3( \omega_\phi+1)}}$ and the resulting \ac{EOS} for the scalar sector,
\begin{equation}
    -1\leq\omega_\phi= \frac{\lambda-2\xi}{\lambda+2 \xi}\leq 1, \quad \text{for} \quad \lambda>0,
\end{equation}
which is precisely the \ac{EOS} in the fixed point $\mathcal{NG}$ found in \cite{Cicoli:2020cfj}. The fact that we assumed $\lambda >0$ implies that the \ac{EOS} does not allow a phantom regime. In addition,  identification $\lambda = 2\xi$ reproduces the matter-dominated regime ($\omega = 0$), while the limit $\lambda \gg 1$ corresponds to kination domination. Likewise, $\lambda = 4\xi$ yields the radiation-dominated equation of state ($\omega = 1/3$). 
Finally, the condition $\xi \gg \lambda$ highlights the necessity of a curved target space to obtain a viable dark-energy-like solution ($\omega = -1$), irrespective of the steepness of the potential. Nevertheless, for a solution to be viable and describe dark energy, in addition to requiring $\omega_\phi \simeq -1$, it is also necessary that $\Omega_\phi \simeq 0.7$ which would immediately rule out all the fixed points. However, as pointed out in \cite{Cicoli:2020cfj}, the key idea is to have a transient phase passing through matter domination and eventually culminating in the stable fixed points that allow for accelerated expansion. In Figure \ref{fig:Omegas}, we plot the evolution of the system in the plane $(\omega_\phi, \Omega_\phi)$ for different values of the parameters. Notice that when the hierarchy of the parameters is large, the system can evolve from an equation of state that transitions from matter domination to dark energy.


\subsection{Case $n=1$: \ac{CL}}
\label{subsec:CL-pot}

\begin{figure}[t]
    \centering
    \begin{subfigure}[b]{0.495\textwidth}
        \centering
        \includegraphics[width=\textwidth]{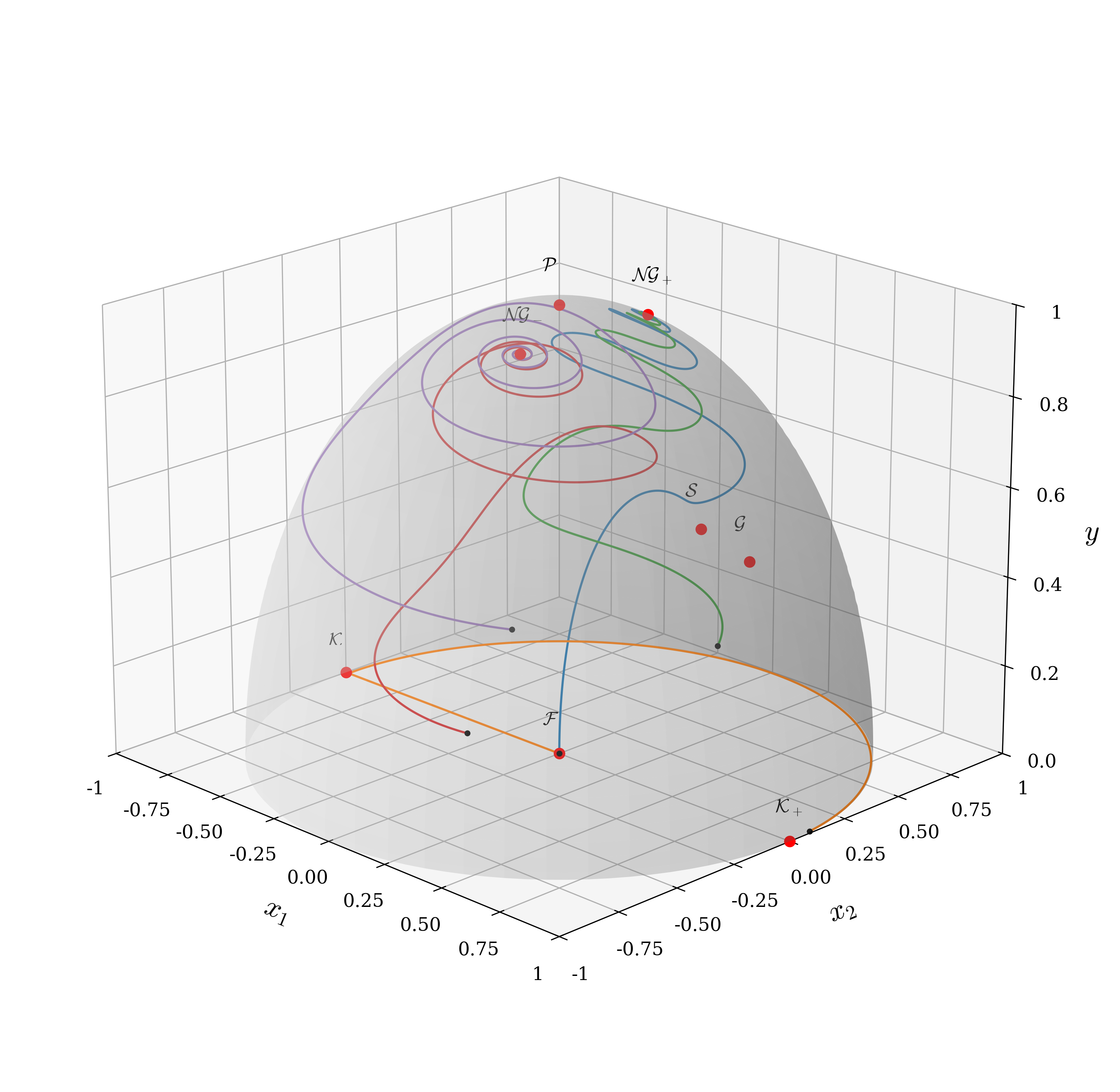}
        \caption{Example 1:  $\xi=10$ and $C=0.1$.}
        \label{fig:TrajectoryE1-CL}
    \end{subfigure}
    \hfill
    \begin{subfigure}[b]{0.495\textwidth}
        \centering
        \includegraphics[width=\textwidth]{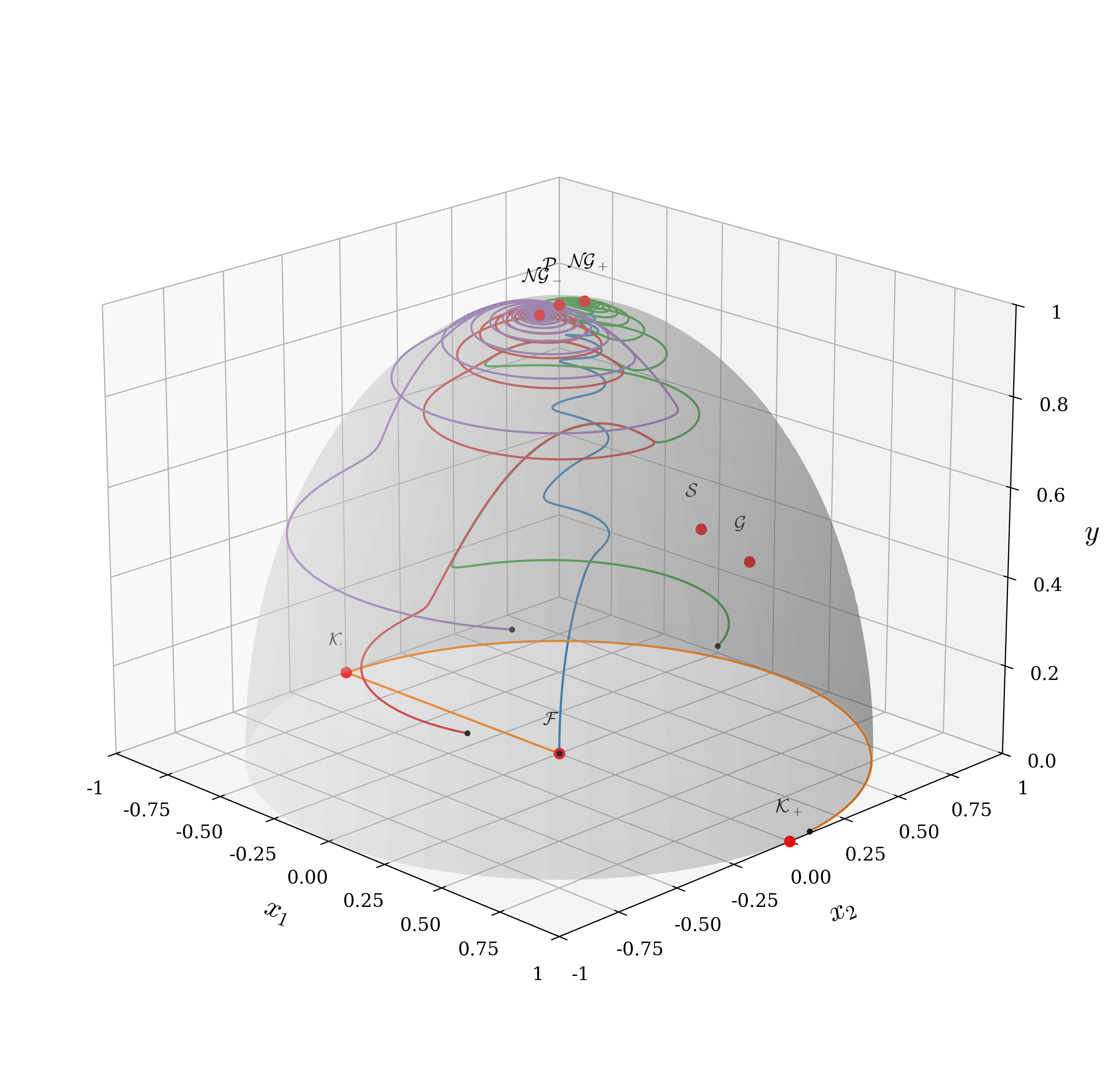}
        \caption{Example 2: $\xi=100$ and $C=0.1$.}
        \label{fig:TrajectoryE2-CL}
    \end{subfigure}
    \caption{Phase-space trajectories for the \ac{CL} case. The coloured lines represent trajectories corresponding to different initial conditions, indicated by the black points, while the fixed points are shown in red. For this parameter hierarchy, the fixed point $\mathcal{NG}$ act as the late-time dark energy attractor. In both cases $\gamma=1$ and $\lambda=2$.}
    \label{fig:Trajectories-CL}
\end{figure}

\begin{figure}[t!]
    \centering
    \begin{subfigure}[b]{0.495\textwidth}
        \centering
            \includegraphics[width=\textwidth]{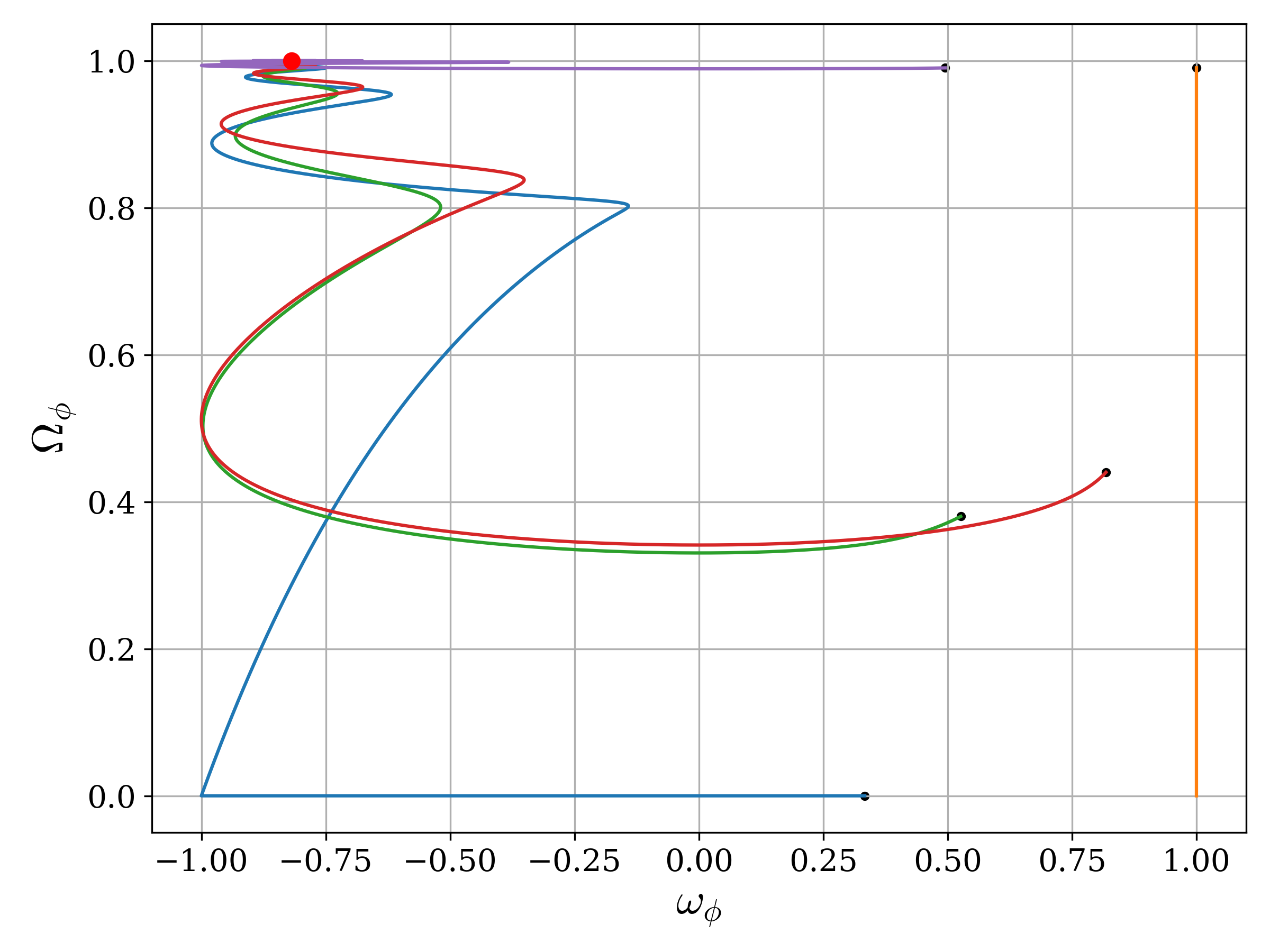}
        \caption{Example 1: $\xi=10$ and $C=0.1$.}
        \label{fig:OmegavsomegaE1-CL}
    \end{subfigure}
    \hfill
    \begin{subfigure}[b]{0.495\textwidth}
        \centering
        \includegraphics[width=\textwidth]{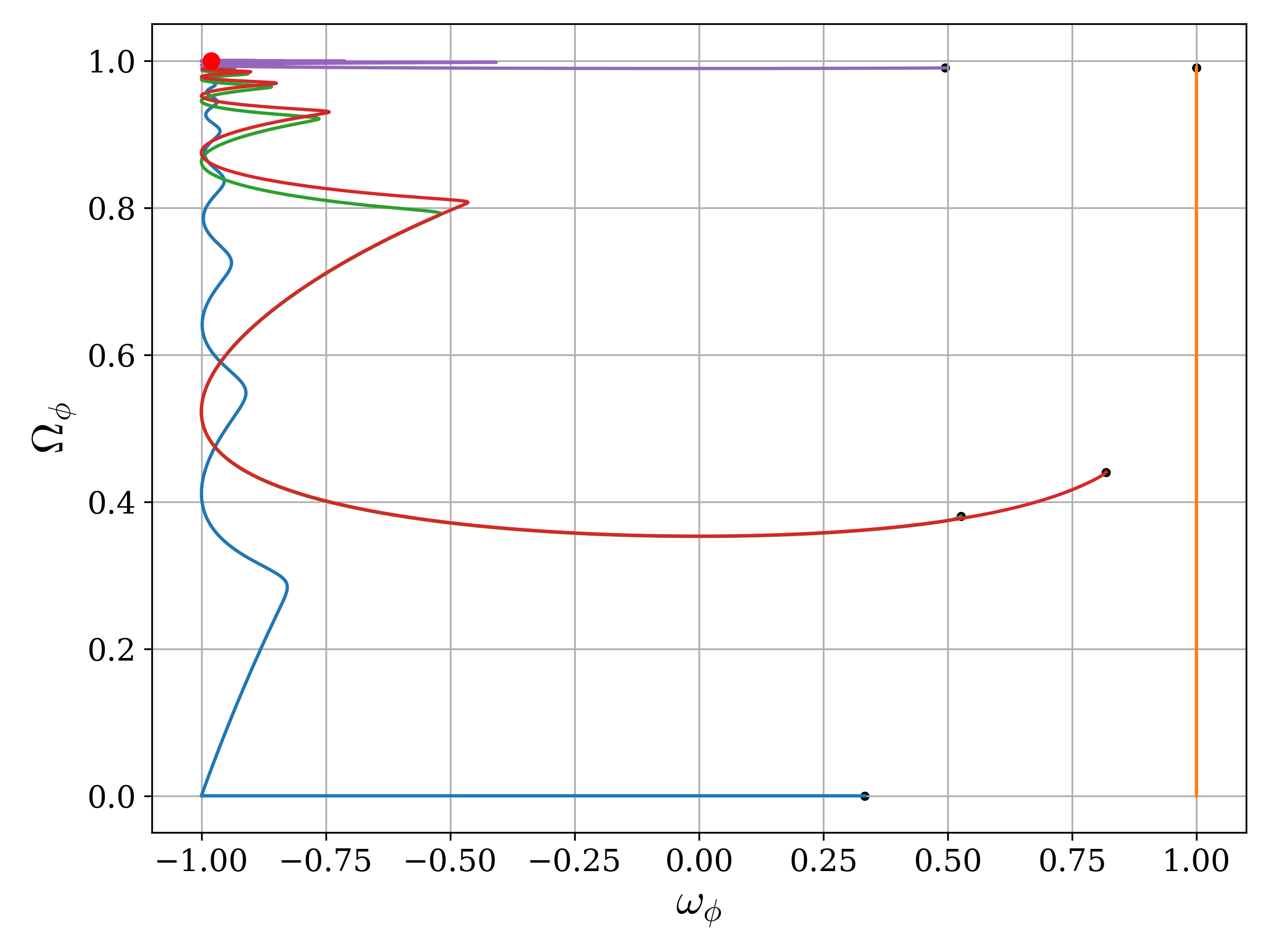}
        \caption{Example 2:  $\xi=100$ and $C=0.1$.}
        \label{fig:OmegavsomegaE2-CL}
    \end{subfigure}
    \caption{Evolution of the scalar energy density, $\Omega_\phi$, and the \ac{EOS} parameter, $w_\phi$, for the different initial conditions considered in Fig.~\ref{fig:Trajectories-CL} for an \ac{CL} potential. In both cases $\gamma=1$ and $\lambda=2$.}
    \label{fig:Omegas-CL}
\end{figure}

If we define $C=A-B$, the general form of the potential reduces to $V(z_1)= V_0 (z_1+ C)e^{-\lambda z_1}$. For an appropriate redefinition of $V_0$, we can re-express it as
\begin{equation}
     V(z_1)= V_0 (1+ C z_1 ) e^{-\lambda z_1}. 
\end{equation}
This potential was proposed for the single field scenario by \ac{CL} in \cite{Clemson:2008ua} and further studied from a numerical perspective in \cite{Pantazis:2016nky}. It has a critical point at $z_{1*} = 1/\lambda-1/C$ for $\lambda,C>0$. Evaluating the second derivative at $z_{1*}$:
\begin{equation}
    V''\big|_{z_{1*}}= -V_0\,\lambda C\, e^{-\lambda z_{1*}} < 0,
\end{equation}
this critical point is always a maximum for $\lambda, C > 0$. Notice that although $V'(z_{1*}) = 0$ implies $\alpha(z_{1*}) = 0$, the branch $\alpha = 0$ is not an independent branch of $\alpha' = 0$ for the CL potential. To see that, we compute
\begin{equation}
    \alpha= \lambda -\frac{C }{1+Cz_1} \quad \text{and} \quad \Gamma-1= -\frac{C^2}{[-C+ \lambda(1+Cz_1)]^2}.
\end{equation}  
If we use the expression for $\alpha$ to obtain $1+Cz_1= C/(\lambda-\alpha)$, and plugging this into the expression for $\Gamma-1$, we get
\begin{equation}
    \Gamma-1= -\frac{(\alpha-\lambda)^2}{\alpha^2},
\end{equation}
and therefore the equation for $\alpha'$ becomes $\alpha' = \sqrt{6}\,(\lambda - \alpha)^2\, x_1$, which vanishes at $\alpha = 0$ only if simultaneously $x_1 = 0$, since $(\lambda - 0)^2 = \lambda^2 \neq 0$. Therefore, for the CL potential, the maximum $z_{1*}$ is captured within branch~i) rather than constituting an independent branch. Moreover, the system of equations is already autonomous and the fixed points are listed in Table \ref{Tab:Physical_points_CL}. Note that in contrast to the pure \ac{EXP} case, a new fixed point appears --- and it allows for dark energy --- when the potential energy dominates $(\hat{\mathcal{P}})$ and corresponds to when the field is at the maximum of the potential with $\alpha=0$ but the fields are frozen so there is no dynamics. In Figure \ref{fig:Trajectories-CL} we plot the trajectories for different initial conditions and in Figure \ref{fig:Omegas-CL} we show the evolution of the density and \ac{EOS} for the respective trajectories. 

\begin{table}[t]
\centering
\renewcommand{\arraystretch}{2.2} 
\begin{adjustbox}{width=\textwidth}
\begin{tabular}{
    >{\centering\arraybackslash}m{1cm}   
    >{\centering\arraybackslash}m{2cm}     
    >{\centering\arraybackslash}m{2.8cm}   
    >{\centering\arraybackslash}m{2.5 cm}   
    >{\centering\arraybackslash}m{1cm}   
    >{\centering\arraybackslash}m{1 cm}   
    >{\centering\arraybackslash}m{2.2cm}   
    >{\centering\arraybackslash}m{5 cm}   
}
\toprule
\rowcolor{lightgray}
\textbf{Points} & $x_1$ & $x_2$ & $y$ & $\alpha$ & $\Omega_\phi$ & $\omega_\phi$ & \textbf{Additional conditions} \\
\midrule
\midrule

$\mathcal{P}$ 
& $0$ & $0$ & $1$ 
& $0$
& $1$ & $-1$ & None \\

$\mathcal{K}_{\pm}$ 
& $\pm1$ & $0$ & $0$ 
& $\lambda$
& $1$ & $1$ & None \\

$\mathcal{F}$ 
& $0$ & $0$ & $0$ 
& $\alpha$
& $0$ & indeterminate & None \\

$\mathcal{S}$ 
& $\dfrac{\sqrt{\tfrac{3}{2}}\gamma}{\lambda}$ 
& $0$ 
& $\dfrac{\sqrt{\tfrac{3}{2}\gamma(2-\gamma)}}{\lambda}$ 
& $\lambda$
& $\dfrac{3\gamma}{\lambda^2}$ 
& $-1+\gamma$ 
& $\lambda^2 \geq 3\gamma$ \\

$\mathcal{G}$ 
& $\dfrac{\lambda}{\sqrt{6}}$ 
& $0$ 
& $\sqrt{1 - \tfrac{\lambda^2}{6}}$ 
& $\lambda$
& $1$ 
& $-1 + \tfrac{\lambda^2}{3}$ 
& $0 < \lambda \leq \sqrt{6}$ \\

$\mathcal{NG}_{\pm}$ 
& $\dfrac{\sqrt{6}}{\lambda + 2\xi}$ 
& $\dfrac{\pm\sqrt{-6 + \lambda^2 + 2\lambda\xi}}{\lambda + 2\xi}$ 
& $\dfrac{\sqrt{2\xi}}{\sqrt{\lambda + 2\xi}}$
& $\lambda$
& $1$ 
& $\dfrac{\lambda - 2\xi}{\lambda + 2\xi}$ 
& $ \sqrt{6 + \xi^2} - \xi \leq  \lambda \leq 2 \xi$ \\

\bottomrule
\bottomrule
\end{tabular}
\end{adjustbox}

\caption{Physical fixed points of the two-field dynamical system with a \ac{CL} potential and a barotropic fluid.}
\label{Tab:Physical_points_CL}
\end{table}

\subsection{Case $n=2$: \ac{AS}}
\label{subsec:AS-pot}

This case is particularly motivated in string theory \cite{Burgess:2022nbx} when the dilaton's potential mimics the \ac{AS} form\footnote{The single-field scenario has been studied in \cite{Albrecht:1999rm, dePutter:2008wt} and the Bayesian analysis with different datasets in \cite{Barnard:2007ta, Park:2014poa, Adil:2022hkj}. Recently, in \cite{Smith:2025grk}, the background evolution together with the angular and matter power spectra was also investigated for this potential but with no Bayesian analysis. In the latter work, they provided a bound for $\lambda$ to satisfy solar-system screening constraints for specific $\xi$ in Yoga scenarios \cite{Burgess:2021obw}.}, that is
\begin{equation}
    V(z_1)= V_0 \left[A+ (z_1 -B)^2\right]e^{-\lambda z_1},
\end{equation}
with the critical points at $z_{1\pm}= \frac{1+B \lambda \pm \sqrt{1-A\lambda^2}}{\lambda}$, where the condition $A \lambda^2\leq1$ has to be satisfied to have real critical points, and where $z_{1-}$ and $z_{1+}$ correspond to a local minimum and maximum, respectively. Evaluating
the second derivative at each root gives
\begin{equation}
    V''\big|_{z_{1\pm}}
   = \mp\,2V_0\sqrt{1-\lambda^2 A}\;e^{-\lambda z_{1\pm}},
\end{equation}
so $V''(z_{1+}) < 0$ (local maximum) and $V''(z_{1 -}) > 0$ (local minimum). Unlike the CL case, the branch $\alpha = 0$ is here a genuinely independent branch of
$\alpha' = 0$, since for the AS potential
\begin{equation}
    \alpha'= \sqrt{6}\,x_1\,\alpha^2
    \left(\Gamma - 1\right),
\end{equation}
in which $\alpha^2$ appears as an explicit factor, so $\alpha = 0$
forces $\alpha' = 0$ regardless of the values of $x_1$ and $\Gamma$. Therefore, the branch $\alpha = 0$ generically contains two distinct fixed points for the AS potential --- one at the maximum ($z_{1+}$) and one at the minimum ($z_{1-}$) of the potential. The two critical points merge into a single inflexion point when $\lambda^2 A = 1$, and neither exists when $\lambda^2 A > 1$. Let us now explore the branch ~iii) and compute
\begin{equation}
    \alpha(z_1)=\lambda  -\frac{2 (z_1-B)}{A+(z_1-B)^2} \quad \text{and} \quad \Gamma = 1+ \frac{2 (A-(z_1-B)^2)}{[-2(z_1-B)+ (A+(z_1-B)^2)\lambda]^2},
\label{eq:alphanGamma-AS}
\end{equation}
from which it is straightforward to see that $\Gamma =1$ implies $A - (z_1-B)^2 = 0$, i.e., \ $ z_1 - B = \pm\sqrt{A}$. Plugging the condition $\Gamma = 1$ into Eq.~\eqref{eq:alphanGamma-AS} yields two distinct values of $\alpha$, namely
\begin{equation}
    \alpha_1 \equiv \lambda - \frac{1}{\sqrt{A}} \quad \text{and} \quad
    \alpha_2 \equiv \lambda + \frac{1}{\sqrt{A}}.
\end{equation}
Since $\lambda > 0$ and $A > 0$, one has $\alpha_2 > 0$ unconditionally, while
the sign of $\alpha_1$ depends on the parameter combination: $\alpha_1 \leq 0$
when $A\lambda^2 \leq 1$, and $\alpha_1 > 0$ when $A\lambda^2 > 1$.
At these fixed points $\Gamma = 1$ and $\alpha = \alpha_{1,2}$ are both constants, so the dynamics effectively reduces to the three-dimensional autonomous system of the \ac{EXP} potential with the replacement $\lambda \to \alpha_{1,2}$, and all fixed-point results of that case carry over directly. Two non-geodesic fixed points $\mathcal{NG}^{1,2}$ emerge from the two possibilities of $\alpha$, and their equation of state reads
\begin{equation}
    w_\phi = \frac{\alpha_{1,2} - 2\xi}{\alpha_{1,2} + 2\xi},
\end{equation}
whose denominator must be strictly positive for the $y$-coordinate of the fixed point to be real. The coordinate $x_2$ is real and physical only when the following existence condition is satisfied:
\begin{equation}
\label{eq:reality-condition}
    \mathcal{R}(\alpha) \equiv \alpha^2 + 2\xi\,\alpha - 6 \geq 0.
\end{equation}
The roots of $\mathcal{R}(\alpha) = 0$ are $\alpha_{\pm}^{\rm roots} = -\xi \pm \sqrt{\xi^2 + 6}$, and since 
$\mathcal{R}(\alpha)$ is an upward-opening parabola, condition~\eqref{eq:reality-condition} holds in two disjoint branches,
\begin{equation}
\label{eq:branches}
    \alpha \geq \alpha_+^{\rm root} \equiv -\xi + \sqrt{\xi^2+6} > 0 
    \qquad \text{or} \qquad 
    \alpha \leq \alpha_-^{\rm root} \equiv -\xi - \sqrt{\xi^2+6} < -2\xi<0,
\end{equation}
where both inequalities follow directly from $\sqrt{\xi^2 + 6} > \xi > 0$. A phantom equation of state requires a positive denominator, $\alpha + 2\xi > 0$, together with a sufficiently negative numerator, which reduces to the condition $-2\xi < \alpha < 0$. We refer to this interval as the \emph{phantom window} $\mathfrak{P} \equiv (-2\xi,\,0)$. From Eq.~\eqref{eq:branches} one reads immediately that $\alpha_-^{\rm roots} < -2\xi$ and $\alpha_+^{\rm roots} > 0$, so the entire window $\mathfrak{P}$ is contained strictly between the two roots of $\mathcal{R}$,
   \begin{equation}
    \alpha_-^{\rm roots} \;<\; -2\xi \;<\; \mathfrak{P} \;<\; 0 
    \;<\; \alpha_+^{\rm roots}.
    \end{equation}
   so $\mathcal{R}(\mathfrak{P}) < 0$ and therefore unphysical. We now analyse each case in turn:

\begin{itemize}
    \item \textit{Case $\mathcal{NG}^2$:}~since $\alpha_2 > 0$ always, only the positive branch of Eq.~\eqref{eq:branches} is relevant, that is, $\alpha_2 \geq \alpha_+^{\text{root}}$, and the denominator $\alpha_2 + 2\xi > 0$ is automatically satisfied. Furthermore, since $\lambda > 0$ implies $\alpha_2 > 0$, a phantom equation of state $w_{\phi} < -1$ would require $\alpha_2 < 0$, which is an immediate contradiction. Hence $w_{\phi} \in (-1,\,1)$ for all allowed parameter values. The existence conditions in this case are therefore
    \begin{equation}
   \lambda + \frac{1}{\sqrt{A}} \geq - \xi+\sqrt{\xi^2+6}   \quad \text{and} \quad A\lambda^2 \leq 1.
    \label{eq:AS-existence}
    \end{equation}
   
    \item \textit{Case $\mathcal{NG}^1$:}~the physical status of this point depends
    on the regime. When $A\lambda^2 \leq 1$ one has $\alpha_1 \leq 0$, the positive
    branch of Eq.~\eqref{eq:branches} is immediately excluded, and the remaining
    possibility $\alpha_1 \leq \alpha_-^{\text{root}}$ reduces to $\alpha_1 < -2\xi$,
    which violates the requirement that the denominator of $w_\phi$ be positive.
    Hence $\mathcal{NG}^1$does not exist as a physical fixed point when
    $A\lambda^2 \leq 1$: the same condition that guarantees the potential has two
    real extrema locks $\alpha_1$ into a regime that is simultaneously incompatible
    with the reality of $x_2$, the positivity of the denominator of $w_\phi$, or
    both. When $A\lambda^2 > 1$, however, $\alpha_1$ becomes strictly positive,
    the positive branch of Eq.~\eqref{eq:branches} becomes available, and
    $\mathcal{NG}^1$ is a physical fixed point with $w_\phi = (\alpha_1 -
    2\xi)/(\alpha_1 + 2\xi) \in (-1,\,1)$, where the non-phantom bound follows
    immediately from $\alpha_1 > 0$. The existence conditions in this case are
    \begin{equation}
        \lambda - \frac{1}{\sqrt{A}} \geq -\xi + \sqrt{\xi^2+6}
        \quad \text{and} \quad A\lambda^2 > 1.
    \label{eq:AS-NG1-existence}
    \end{equation}
\end{itemize}
Therefore, the phantom divide is never crossed at the level of the fixed-point
structure of the \ac{AS} model: $\mathcal{NG}^2$ satisfies $w_\phi > -1$ since
$\alpha_2 > 0$ unconditionally; $\mathcal{NG}^1$ satisfies $w_\phi > -1$ whenever
it is physical ($\alpha_1 > 0$, i.e.\ $A\lambda^2 > 1$); and $\mathcal{P}$ sits
exactly at $w_\phi = -1$ and exists only when $A\lambda^2 \leq 1$. As a concrete
illustration, consider $A = 0.001$, $\lambda = 8$, and $\xi = 20$. One finds
$\alpha_1 = 8 - 1/\sqrt{0.001} \approx -23.6$, which naively yields $w_{\phi}
\approx -3.9 < -1$. However, $\alpha_1 \approx -23.6$ lies inside $\mathfrak{P} =
(-40,\,0)$, strictly between $\alpha_-^{\rm roots} \approx -40.1$ and
$\alpha_+^{\rm roots} \approx 0.15$, where $\mathcal{R}(\alpha_1) < 0$ and $x_2^{2}
< 0$. The apparent phantom behaviour is therefore an artefact of neglecting the
physicality constraints; the system instead flows to $\mathcal{P}$ with $w_\phi =
-1$.

Analogously to the \ac{EXP} case, the equation of state at the 
$\mathcal{NG}^2$ fixed point spans a wide range of cosmological epochs 
depending on the relative magnitude of $\alpha_2$ and $\xi$. In the 
asymptotic limit $\alpha_2 \ll \xi$, one has $w_\phi \to -1$, recovering 
dark-energy-like domination, whereas in the opposite limit $\alpha_2 \gg \xi$, 
one finds $w_\phi \to 1$, corresponding to a kination-dominated era. The 
exact conditions $\alpha_2 = 2\xi$ and $\alpha_2 = 4\xi$ yield $w_\phi = 0$ 
and $w_\phi = 1/3$, reproducing matter- and radiation-dominated epochs, 
respectively. The same discussion applies to $\mathcal{NG}^1$ in the regime
$A\lambda^2 > 1$, with the replacement $\alpha_2 \to \alpha_1$.

\begin{figure}[t]
    \centering
    \begin{subfigure}[b]{0.495\textwidth}
        \centering
        \includegraphics[width=\textwidth]{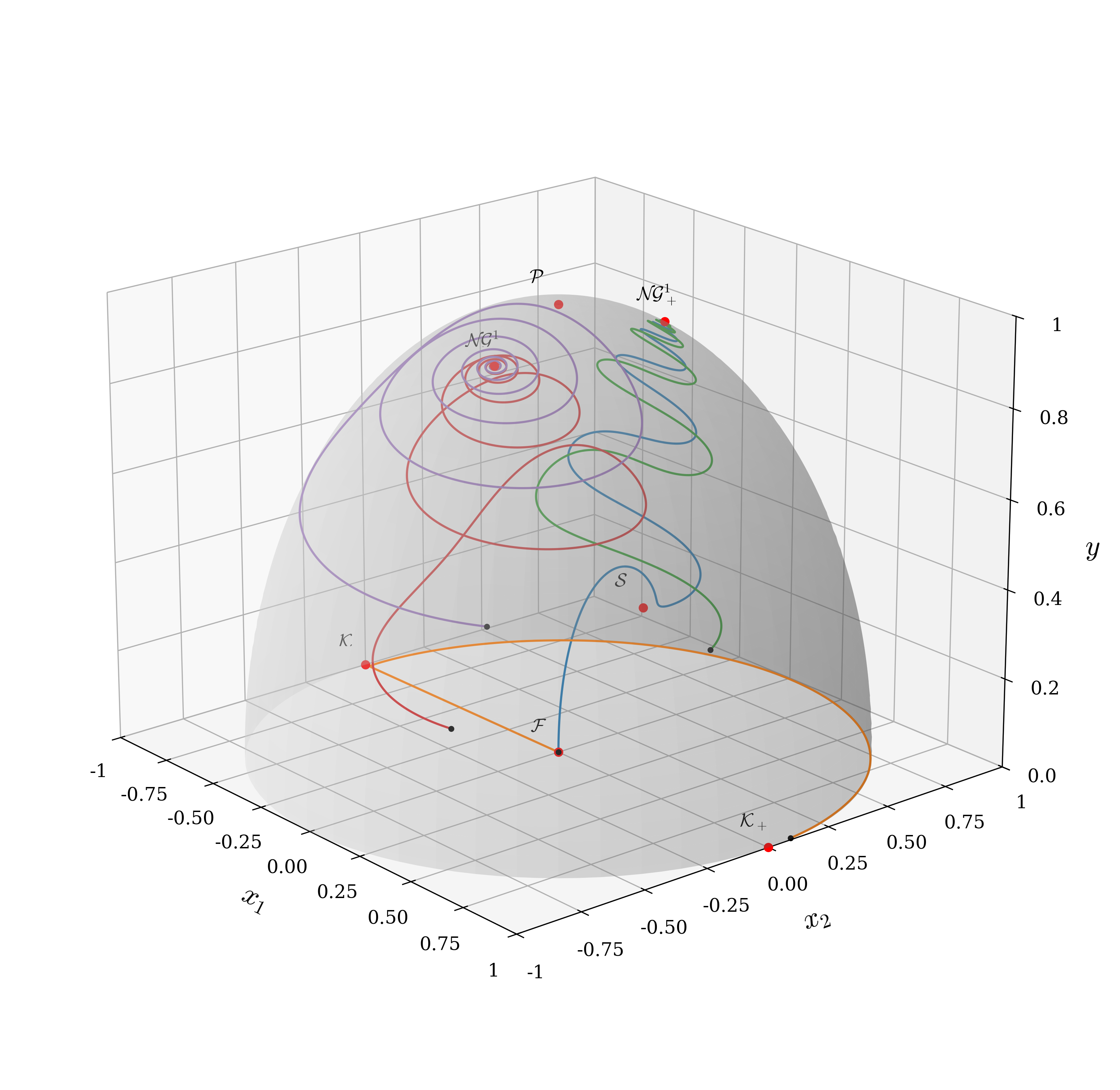}
        \caption{Example 1: $\xi=10$ and $A=1$.}
        \label{fig:TrajectoryE1-AS}
    \end{subfigure}
    \hfill
    \begin{subfigure}[b]{0.495\textwidth}
        \centering
        \includegraphics[width=\textwidth]{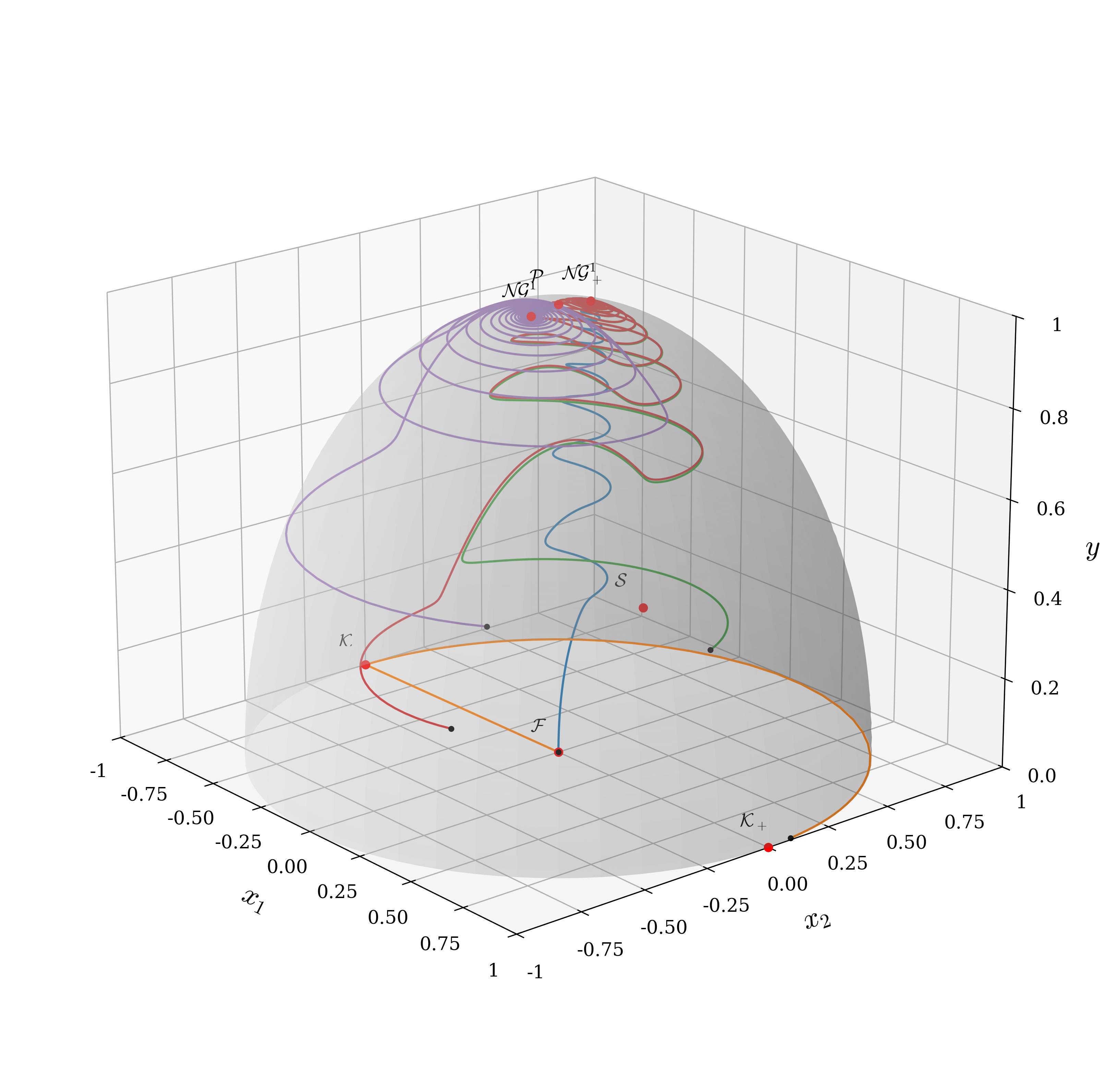}
        \caption{Example 2: $\xi=100$ and $A=1$.}
        \label{fig:TrajectoryE2-AS}
    \end{subfigure}
    \caption{Phase-space trajectories for the \ac{AS} case. The coloured lines represent trajectories corresponding to different initial conditions, indicated by the black points, while the fixed points are shown in red. For this parameter hierarchy, the fixed points $\mathcal{NG}^{1,2}$ act as late-time dark energy attractors. In both cases $\gamma=1$ and $\lambda=2$.}
    \label{fig:Trajectories-AS}
\end{figure}

\begin{figure}[h!]
    \centering
    \begin{subfigure}[b]{0.495\textwidth}
        \centering
            \includegraphics[width=\textwidth]{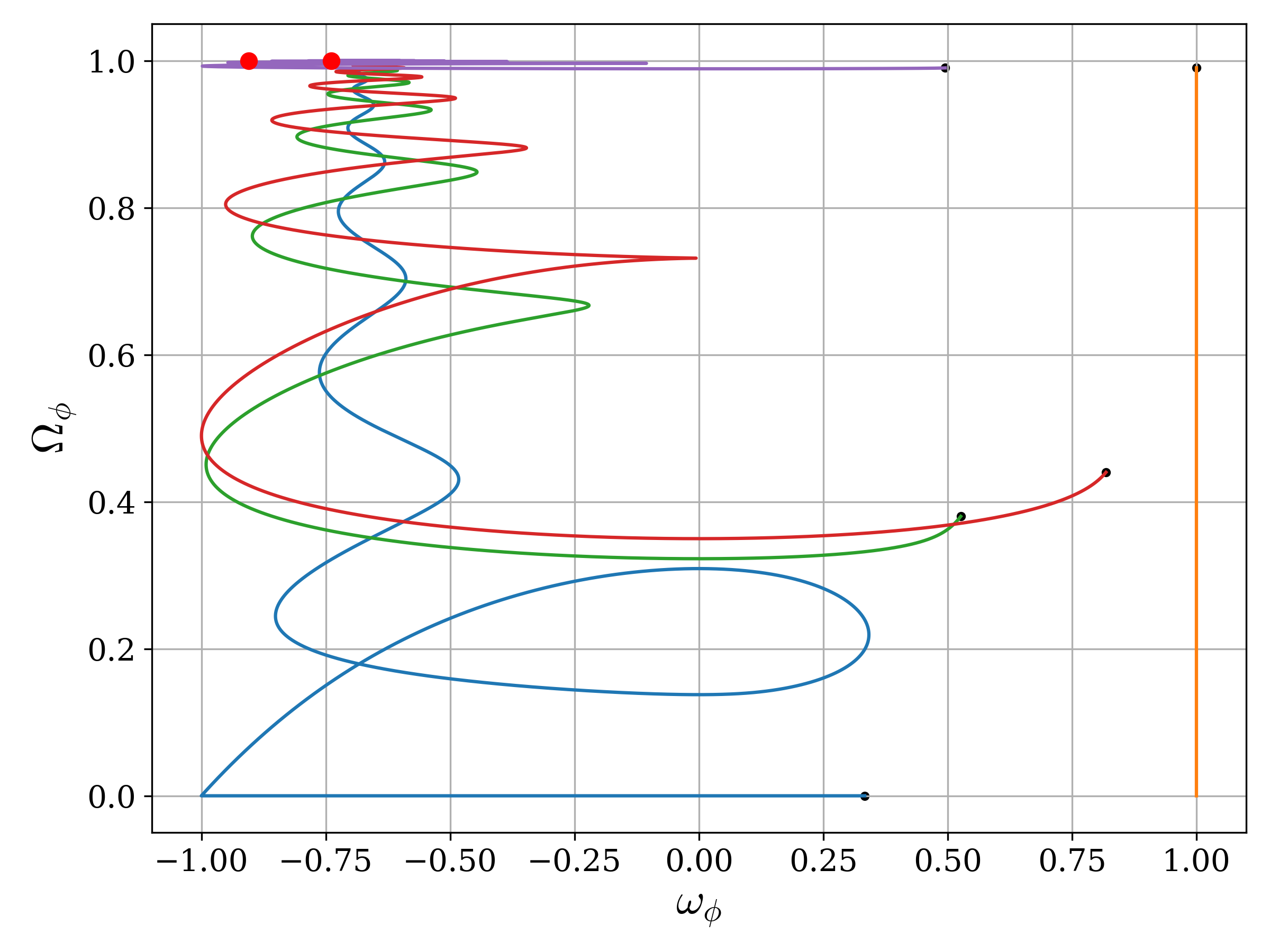}
        \caption{Example 1: $\xi=10$ and $A=1$.}
        \label{fig:OmegavsomegaE1-AS}
    \end{subfigure}
    \hfill
    \begin{subfigure}[b]{0.495\textwidth}
        \centering
        \includegraphics[width=\textwidth]{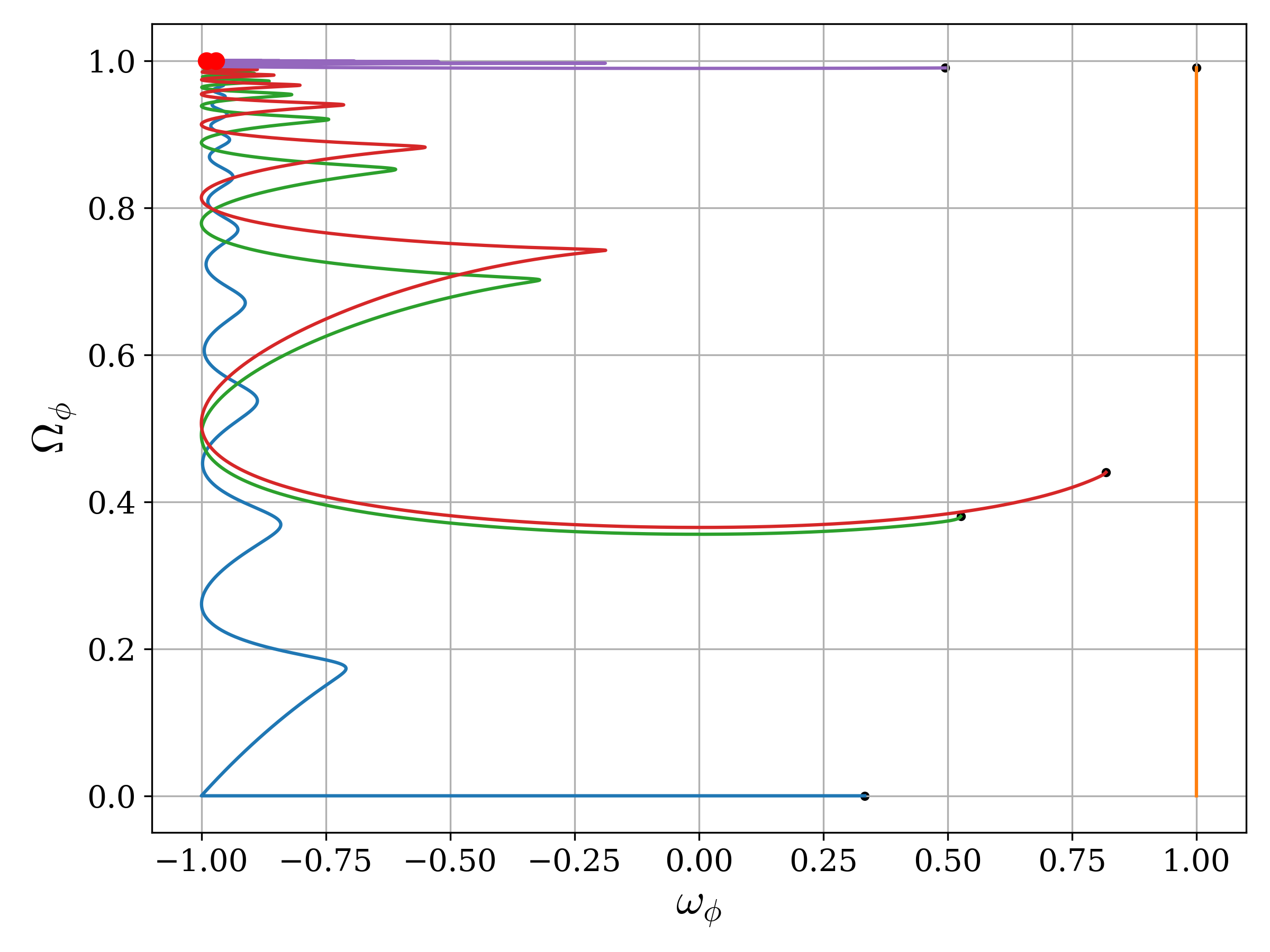}
        \caption{Example 2: $\xi=100$ and $A=1$.}
        \label{fig:OmegavsomegaE2-AS}
    \end{subfigure}
    \caption{Evolution of the scalar energy density, $\Omega_\phi$, and the \ac{EOS} parameter, $w_\phi$, for the different initial conditions considered in Fig.~\ref{fig:Trajectories-AS} for an \ac{AS} potential. In both cases $\gamma=1$ and $\lambda=2$.}
    \label{fig:Omegas-AS}
\end{figure}

In Table~\ref{Tab:Physical_points_AS}, we listed all the physical fixed points and their existence conditions. In Figure \ref{fig:Trajectories-AS} we plot the phase-space trajectory for several initial conditions. Since now $A$ plays a role in the equation of state, the hierarchy between $\lambda$ and $\xi$ is no longer strict: for $A\lambda^2 > 1$ the attractor is $\mathcal{NG}^1$ with $w_\phi = (\alpha_1 - 2\xi)/(\alpha_1 + 2\xi)$, whereas for $A\lambda^2 \leq 1$ the system flows to $\mathcal{P}$ with $w_\phi = -1$, since $\mathcal{NG}^1$ is unphysical in that regime. In Fig. \ref{fig:TrajectoryE1-AS} we show how the dark energy point $\mathcal{NG}^{2}$ acts as an attractor for $A\sim \xi \sim \mathcal{O}(10)$ and $\lambda\sim\mathcal{O}(1)$. Complementing the study, and to compare with the \ac{EXP} and \ac{CL} case,  in Figure \ref{fig:Omegas-AS}, we plot the evolution of the dynamical system in the plane $(\omega_\phi, \Omega_\phi)$ for different values of the parameters. 

\begin{table}[t]
\centering
\renewcommand{\arraystretch}{2.4}
\begin{adjustbox}{width=\textwidth, totalheight=\textheight, keepaspectratio}
\begin{tabular}{ccccccc>{\centering\arraybackslash}m{3.5cm}}
\toprule
\rowcolor{lightgray}
\textbf{Points} & $x_1$ & $x_2$ & $y$ & $\alpha$ & $\Omega_\phi$ & $\omega_\phi$ & \textbf{Additional conditions} \\
\midrule\midrule
$\mathcal{P}$
  & $0$ & $0$ & $1$
  & $0$ & $1$ & $-1$
  & $A\lambda^2\leq 1$ \\
$\mathcal{K}_{\pm}$
  & $\pm1$ & $0$ & $0$
  & $0$ & $1$ & $1$
  & None \\
$\mathcal{K}^{1,2}_{\pm}$
  & $\pm1$ & $0$ & $0$
  & $\alpha_1,\alpha_2$ & $1$ & $1$
  & None \\
$\mathcal{F}$
  & $0$ & $0$ & $0$
  & $\alpha$ & $0$ & indeterminate
  & $A(\alpha-\lambda)^2\leq 1$ \\
$\mathcal{F}^{1,2}$
  & $0$ & $0$ & $0$
  & $\alpha_1,\alpha_2$ & $0$ & indeterminate
  & None \\
$\mathcal{S}$
  & $\frac{\sqrt{3/2}\,\gamma}{\pm 1/\sqrt{A}+\lambda}$
  & $0$
  & $\frac{\sqrt{\frac{3}{2}\gamma(2-\gamma)}}{\pm 1/\sqrt{A}+\lambda}$
  & $\frac{\pm1}{\sqrt{A}}+\lambda$
  & $\frac{3\gamma}{\left(\mp\tfrac{1}{\sqrt{A}}+\lambda\right)^2}$
  & $-1+\gamma$
  & $ \frac{3A \gamma}{(1+\sqrt{A}\lambda)^2}\leq 1$ \\
$\mathcal{G}_\pm$
  & $\frac{\pm 1/\sqrt{A}+\lambda}{\sqrt{6}}$
  & $0$
  & $\sqrt{1 - \frac{1}{6}\!\left(\frac{\pm1}{\sqrt{A}}+\lambda\right)^{\!2}}$
  & $\frac{\pm1}{\sqrt{A}}+\lambda$
  & $1$
  & $-1 + \frac{\left(\frac{\pm1}{\sqrt{A}}+\lambda\right)^2}{3}$
  & $\frac{\pm1}{\sqrt{A}}+\lambda\leq\sqrt{6}$ \\
$\mathcal{NG}^{1}_{\pm}$
  & $\frac{\sqrt{6}}{\alpha_1+2\xi}$
  & $\frac{\pm\sqrt{-6+\alpha_1^2+2\alpha_1\xi}}{\alpha_1+2\xi}$
  & $\frac{\sqrt{2\xi}}{\sqrt{\alpha_1+2\xi}}$
  & $\alpha_1$
  & $1$
  & $\frac{\alpha_1-2\xi}{\alpha_1+2\xi}$
  & $\sqrt{6+\xi^2}-\xi \leq\alpha_1 \land A\lambda^2 > 1$ \\
$\mathcal{NG}^{2}_{\pm}$
  & $\frac{\sqrt{6}}{\alpha_2+2\xi}$
  & $\frac{\pm\sqrt{-6+\alpha_2^2+2\alpha_2\xi}}{\alpha_2+2\xi}$
  & $\frac{\sqrt{2\xi}}{\sqrt{\alpha_2+2\xi}}$
  & $\alpha_2$
  & $1$
  & $\frac{\alpha_2-2\xi}{\alpha_2+2\xi}$
  & $ \sqrt{6+\xi^2}-\xi \leq\alpha_2 \land A\lambda^2\leq1$  \\
\bottomrule\bottomrule
\end{tabular}
\end{adjustbox}
\caption{Physical fixed points of the two-field dynamical system for the \ac{AS} potential and a barotropic fluid.}
\label{Tab:Physical_points_AS}
\end{table}

\section{Numerical results}

To obtain the cosmological evolution independently of the dynamical systems analysis, we solve the Klein--Gordon equations for the $n=0, 1, 2$ powers in the scalar-field potential. This enables us to obtain the evolution of key cosmological quantities, including the EoS parameter, the scalar-field energy density, and the Hubble expansion rate. 
For the numerical solutions, one needs to specify the initial conditions of the fields and their derivatives, which were determined by using a shooting algorithm that is 
calibrated to ensure that the scalar-field energy density today matches the required dark energy abundance at the present epoch. Consequently, this approach provides an independent exploration of the model parameter space and serves as a consistency check of the results obtained from the dynamical systems analysis.

\begin{figure}
    \centering
    \makebox[15cm][c]{
   \includegraphics[trim = 10mm  0mm 0mm 0mm, clip, width=22.cm, height=5.cm]{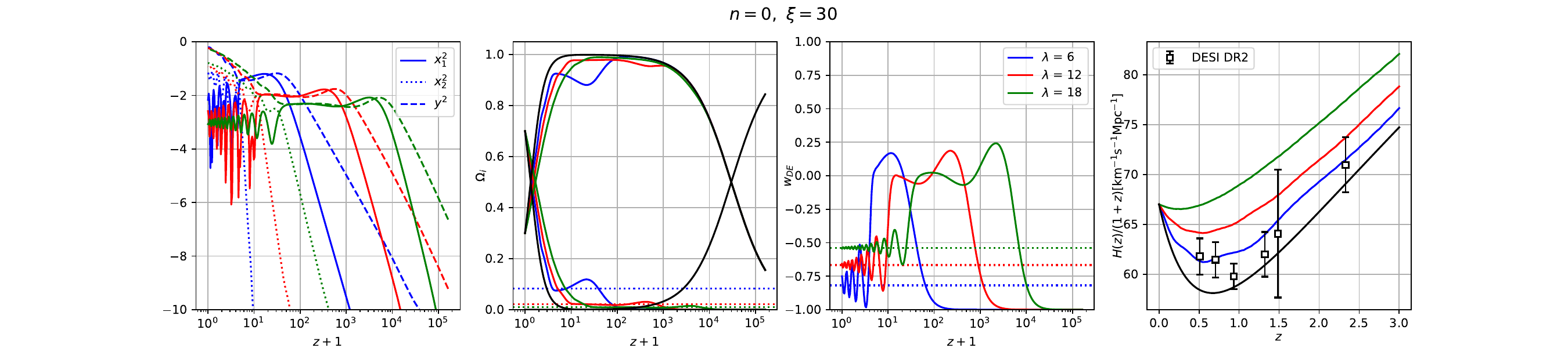}}
   \makebox[15cm][c]{
    \includegraphics[trim = 10mm  0mm 0mm 0mm, clip, width=22.cm, height=5.cm]{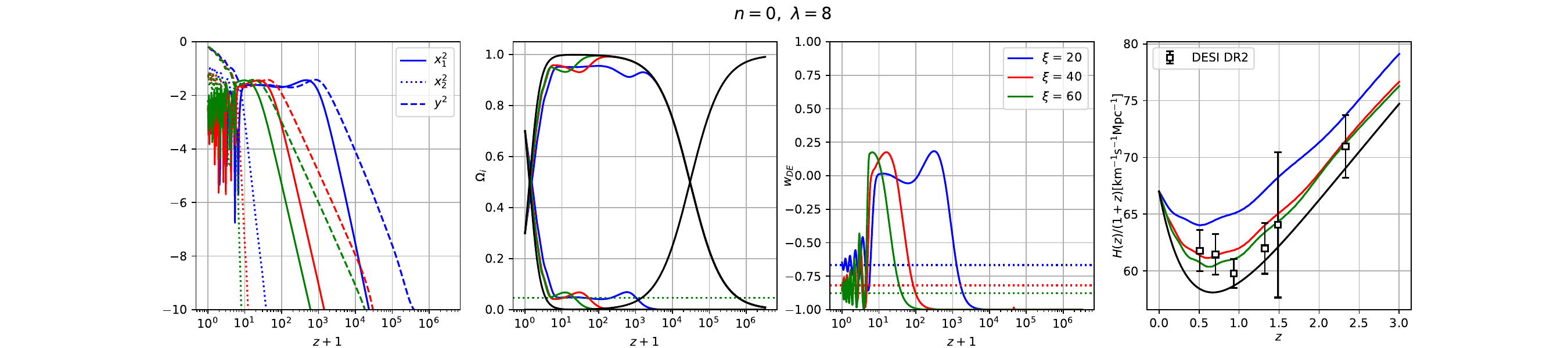}}
    \makebox[15cm][c]{
   \includegraphics[trim = 10mm  0mm 0mm 0mm, clip, width=22.cm, height=5.cm]{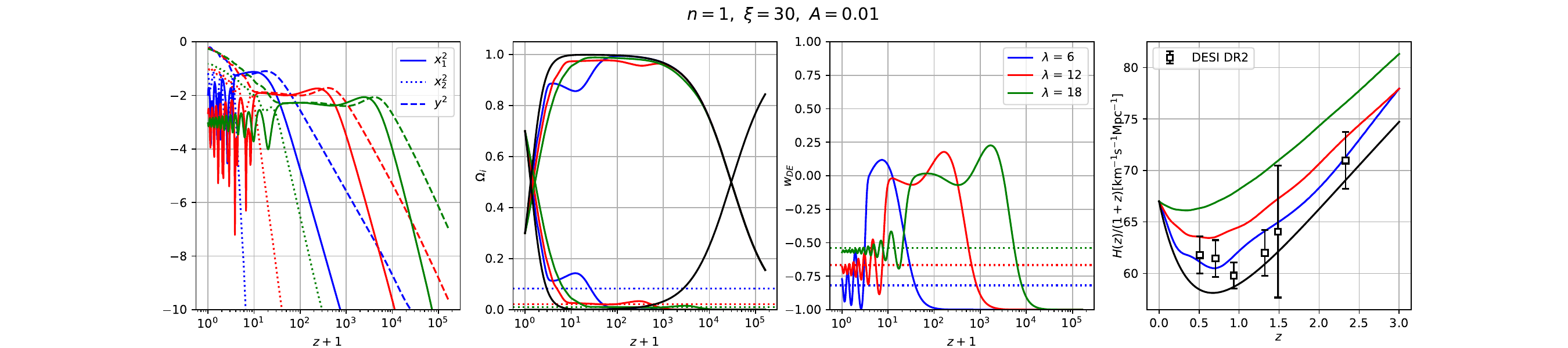}}
    \caption{Numerical evolution for $n=0$ and $n=1$. The top and middle panels  correspond to the purely exponential limit ($n=0$): the top panel fixes $\xi$ and varies $\lambda$, while the midel panel fixes $\lambda$ and varies $\xi$. 
   The bottom panel corresponds to the \ac{CL} potential ($n=1$), varying $\lambda$.
    From left to right, the columns show the kinetic and potential contributions, the density parameters, the axio-dilaton equation of state, and the Hubble parameter; where the data points and error bars correspond to values of $H(z)$ inferred from the DESI DR2 BAO measurements..}
   \label{fig:n0-varying-parameters}
\end{figure}

\begin{figure}
    \centering
    \makebox[15cm][c]{
   \includegraphics[trim = 10mm  0mm 0mm 0mm, clip, width=22.cm, height=5.cm]{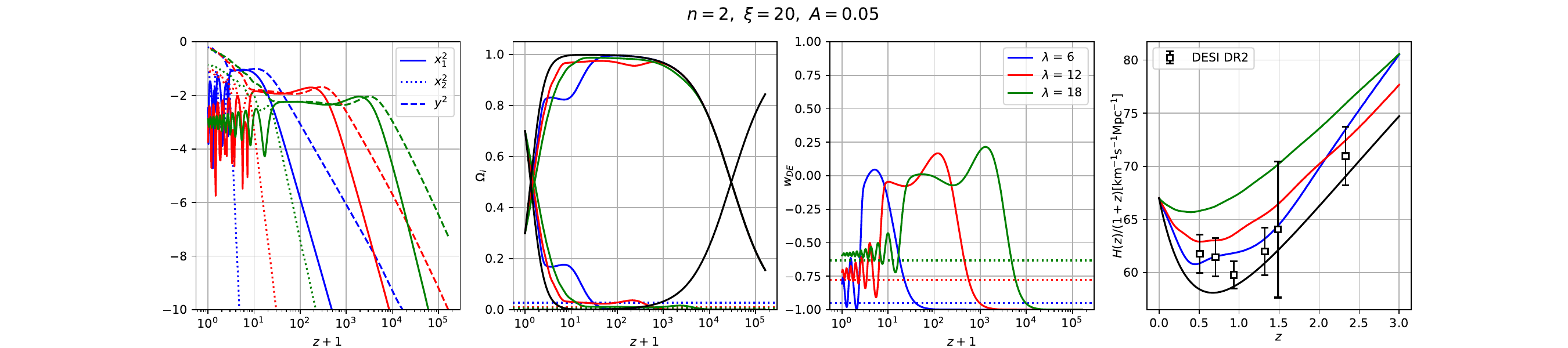}}
   \makebox[15cm][c]{
    \includegraphics[trim = 10mm  0mm 0mm 0mm, clip, width=22.cm, height=5.cm]{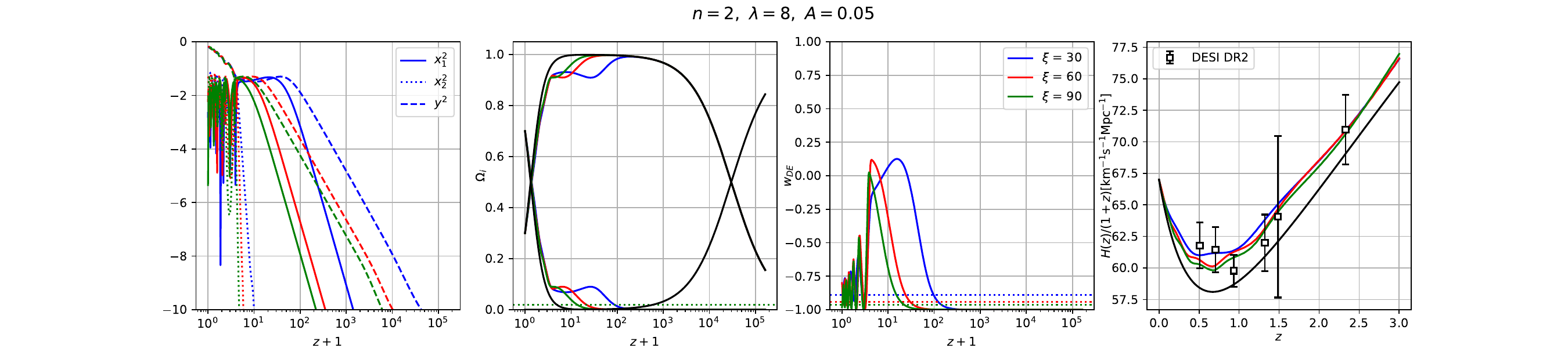}}
    \makebox[15cm][c]{
    \includegraphics[trim = 10mm  0mm 0mm 0mm, clip, width=22.cm, height=5.cm]{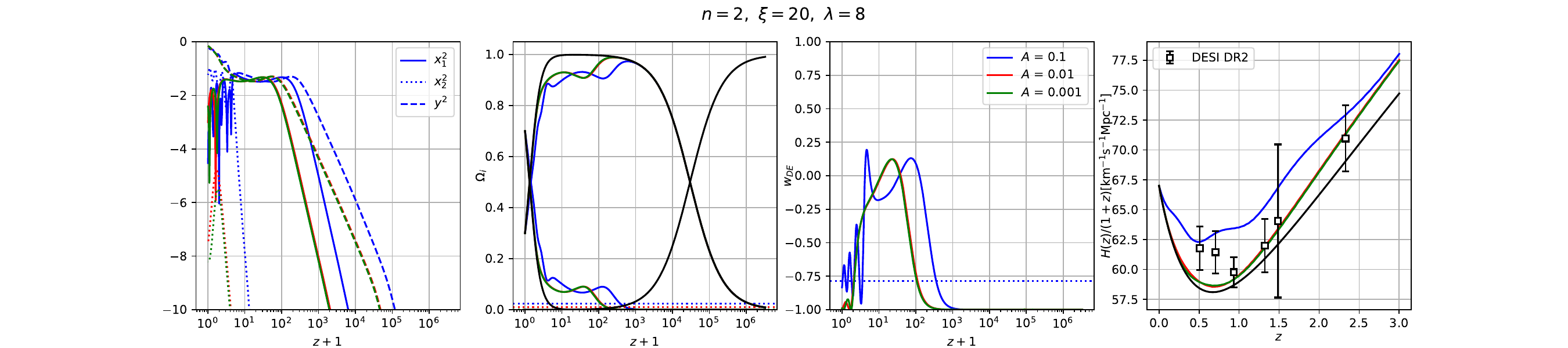}}
        \makebox[15cm][c]{
   \includegraphics[trim = 10mm  0mm 0mm 0mm, clip, width=22.cm, height=5.cm]{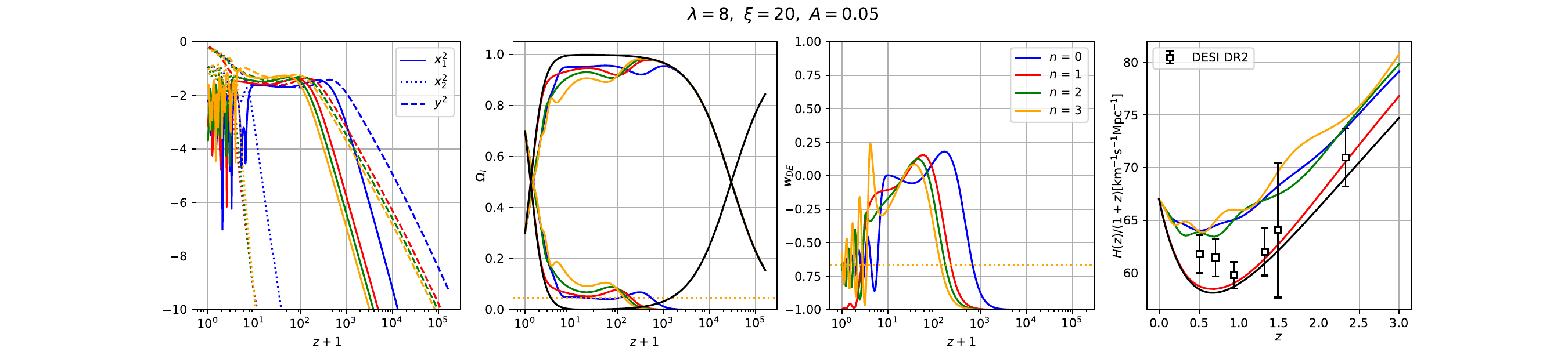}}
    \caption{Numerical evolution for the \ac{AS} potential. The top three panels correspond to the case $n=2$, varying $\lambda$, $\xi$, and $A$, respectively. The bottom panel compares the evolution for different values of the exponent $n$, with $\lambda$, $\xi$, and $A$ held fixed. The columns are similarly ordered as in Fig.~\ref{fig:n0-varying-parameters}.}
   \label{fig:n2-varying-parameters}
\end{figure}

In Figs.~\ref{fig:n0-varying-parameters} and \ref{fig:n2-varying-parameters} we show, from left to right, the redshift evolution of the potential and kinetic terms of the fields, the corresponding density parameters, the axio-dilaton equation of state, and the Hubble parameter, for $n=0, 1, 2$, considering different values of the scalar field parameters, as shown in the coloured labels.

The case $n=0$, shown in the upper and middle panels of Fig.~\ref{fig:n0-varying-parameters}, corresponds to the exponential-potential limit. In this case, $\lambda$ mainly controls the effective steepness of the potential, while $\xi$ controls the strength of the kinetic mixing between the fields. As a result, the late-time acceleration is determined by their interplay: in both panels, the  \ac{EOS} approaches values closer to $-1$ as the hierarchy $\xi \gg \lambda$ becomes more pronounced. By contrast, the early-time scalar contribution is controlled primarily by $\lambda$, with larger values of $\lambda$ leading to a smaller scalar fraction, as expected from the scaling behaviour of the exponential potential.

The case $n=1$, shown in the bottom panels of Fig.~\ref{fig:n0-varying-parameters}, corresponds to the \ac{CL} potential discussed above.  Although the polynomial prefactor introduces an additional parameter, the fixed points in Table~\ref{Tab:Physical_points_CL} show that the relevant late-time quantities are still governed mainly by $\lambda$ and $\xi$. The overall behaviour is therefore similar to the exponential case: increasing $\lambda$ reduces the early dark energy contribution. However, for a fixed value of $\xi$, the \ac{EOS} deviates from $-1$.

It is worth noting that, in all the examples shown in Fig.~\ref{fig:n0-varying-parameters}, the equation-of-state parameter approaches but never exactly reaches  $w=-1$. As shown in the left panels, the kinetic contributions at late times ($z+1=1$) are not negligible when compared with the potential, and then the \ac{EOS} slightly deviates from -1.

The scenario $n=2$ is shown in first three panels of Fig.~\ref{fig:n2-varying-parameters}. In this case, the polynomial prefactor introduces two extra parameters, although $B$ can be absorbed by a field redefinition and does not affect the dynamics. By contrast, the parameter $A$ enters the critical-point structure through the effective slopes $\alpha_{1,2}=\lambda\mp1/\sqrt{A}$, as shown in Table~\ref{Tab:Physical_points_AS}. As a result, both the early scalar contribution and the late-time behaviour depend on the combined values of $A$, $\lambda$, and $\xi$. 
The first three panels illustrate this dependence by varying one parameter at a time while keeping the other two fixed. The approach to a cosmological-constant-like regime is controlled by the hierarchy $\alpha_{1,2}\ll\xi$, so smaller effective slopes relative to $\xi$ drive the \ac{EOS} closer to $-1$. In the first and second panels, where $A$ is fixed, the early scalar fraction is mainly controlled by $\lambda$, while changing $\xi$ primarily affects the late-time approach to acceleration. In the third panel, varying $A$ changes the effective slopes directly, and therefore modifies both the early contribution and the late-time equation of state.

Finally, the bottom panel of Fig.~\ref{fig:n2-varying-parameters} compares the background evolution for different values of the exponent $n$, while keeping the remaining parameters fixed. This comparison highlights that only the \ac{AS} potential ($n=2$) is sensitive to the parameter $A$, allowing its equation of state to approach values closer to $-1$. By contrast, the remaining potentials, which do not depend on $A$, exhibit nearly identical background evolution for the same values of the common parameters.

\section{Bayesian analysis}

\subsection{Data sets and methodology}

To constrain the cosmological and model parameters, we perform a Bayesian joint likelihood analysis using the SimpleMC code \cite{simplemc, aubourg2015cosmological}, combining \ac{SNe Ia}, \ac{BAO} observations and \ac{PLK} 2018 distance-prior information\footnote{See~\cite{Padilla:2019mgi} for a cosmological Bayesian inference review.}.
\begin{itemize}
\item \ac{SNe Ia} (\ac{BP}): We use the compressed (binned) version of the Pantheon compilation, consisting of 40 binned measurements derived from the original sample of 1,048 Type Ia supernovae \cite{Pan-STARRS1:2017jku}. We employ the corresponding covariance matrix, including statistical and systematic uncertainties, and marginalise over the supernova absolute-magnitude normalisation.

    \item \ac{BAO} (DESI \ac{BAO}): We include \ac{BAO} distance measurements from DESI DR2 \cite{DESI:2025zgx}, obtained from the clustering of several tracers including the \ac{BGS}, \ac{LRG}, \ac{ELG}, quasars (QSO), and the Ly$\alpha$ forest. These measurements cover a broad range of redshifts and provide constraints on the cosmic expansion history through distance measurements relative to the sound horizon at the drag epoch.

    \item \ac{PLK}: We use the Planck 2018 distance priors \cite{Chen:2018dbv}, which compress the \ac{CMB} information into a set of shift parameters. These measurements constrain the integrated expansion history up to recombination and provide an early-Universe calibration of the cosmic distance scale.
\end{itemize}

The standard cosmological parameters, namely the present-day matter density parameter $\Omega_m$, the normalised Hubble constant $h$ and the present-day physical baryon density parameter $\Omega_bh^2$, along with the model parameters $\lambda, \xi, A$ are treated as free parameters with uniformly distributed priors reported in Tab. \ref{tab:priors}.
The exponent $n$ is fixed separately to $n=0$, $n=1$, and $n=2$ in the three main model runs.
\begin{table}[ht]
\centering
\begin{tabular}{cc}
\hline\hline
Parameter & Prior \\
\hline
$\Omega_m$ & $\mathcal{U} [0.1,\,0.5]$ \\
$\Omega_{b}h^2$ & $\mathcal{U} [0.01,\,0.05]$ \\
$H_0$ & $\mathcal{U} [50,\,90]$ \\
$\lambda$ & $\mathcal{U}[1,\,  20]$\\
$\log_{10}\xi$ & $\mathcal{U}[1,\, 3]$ \\
$\log_{10}A$ & $\mathcal{U}[-3,\, 0]$ \\
\hline\hline
\end{tabular}
\caption{Uniform priors adopted for the cosmological and model parameters used throughout the analysis.}
\label{tab:priors}
\end{table}

\subsection{Results}

\begin{table}[htbp]
\centering
\renewcommand{\arraystretch}{1.25}
\begin{adjustbox}{width=\textwidth}
\begin{tabular}{lcccccc}
\toprule
& \multicolumn{2}{c}{$n=0$} & \multicolumn{2}{c}{$n=1$} & \multicolumn{2}{c}{$n=2$} \\
\cmidrule(lr){2-3} \cmidrule(lr){4-5} \cmidrule(lr){6-7}
\textbf{Parameter}
& \textbf{DESI \ac{BAO}}
& \textbf{DESI \ac{BAO}  + \ac{BP} + \ac{PLK}}
& \textbf{DESI \ac{BAO} }
& \textbf{DESI \ac{BAO}  + \ac{BP} + \ac{PLK}}
& \textbf{DESI \ac{BAO} }
& \textbf{DESI \ac{BAO} + \ac{BP} + \ac{PLK}} \\
\midrule
$\Omega_{b}h^2$
& $0.0220\pm 0.0003$
& $0.0225\pm 0.0001$
&$0.0220^{+0.0003}_{-0.0003}$
&$0.0225\pm 0.0001$
&$0.0220\pm 0.0003$
&$0.0225\pm 0.0001$ \\
$\Omega_m$
& $0.294^{+0.013}_{-0.011}$
& $0.3093^{+0.0051}_{-0.013}$
& $0.293 \pm 0.013$
& $0.3048^{+0.0053}_{-0.0065}$
& $0.294 \pm 0.014$
& $0.3099^{+0.0059}_{-0.014}$ \\
$h$
& $0.662^{+0.026}_{-0.017}$
& $0.676^{+0.013}_{-0.0044}$
& $0.668^{+0.023}_{-0.016}$
& $0.6807^{+0.0059}_{-0.0046}$
& $0.665^{+0.022}_{-0.015}$
& $0.675^{+0.014}_{-0.0053}$ \\
$\lambda$
& --
& $>6.90$
& $>5.13$
& $>8.57$
& $>5.76$
& $>9.05$ \\
$\log_{10}\xi$
& $>1.71$
& $>2.57$
& $>1.32$
& $>2.51$
& $>1.51$
& $>2.55$ \\
$\log_{10}A$
& --
& --
& --
& --
& --
& -- \\
\midrule
$\chi^2_{\rm min}$
& 5.8
& 50.4
& 6.2
& 50.4
& 4.2
& 50.4\\
\bottomrule
\end{tabular}
\end{adjustbox}
\caption{Marginalized constraints on the cosmological and model parameters for the three polynomial exponents. Error bars denote 68\% confidence limits, while lower and upper bounds are reported at the 95\% confidence level for both the DESI \ac{BAO} alone and the combined DESI \ac{BAO} + \ac{BP} + \ac{PLK} data set.}
\label{tab:nested_summary}
\end{table}

\begin{figure}[h]
    \centering
    \makebox[12cm][c]{
    \includegraphics[width=16cm]{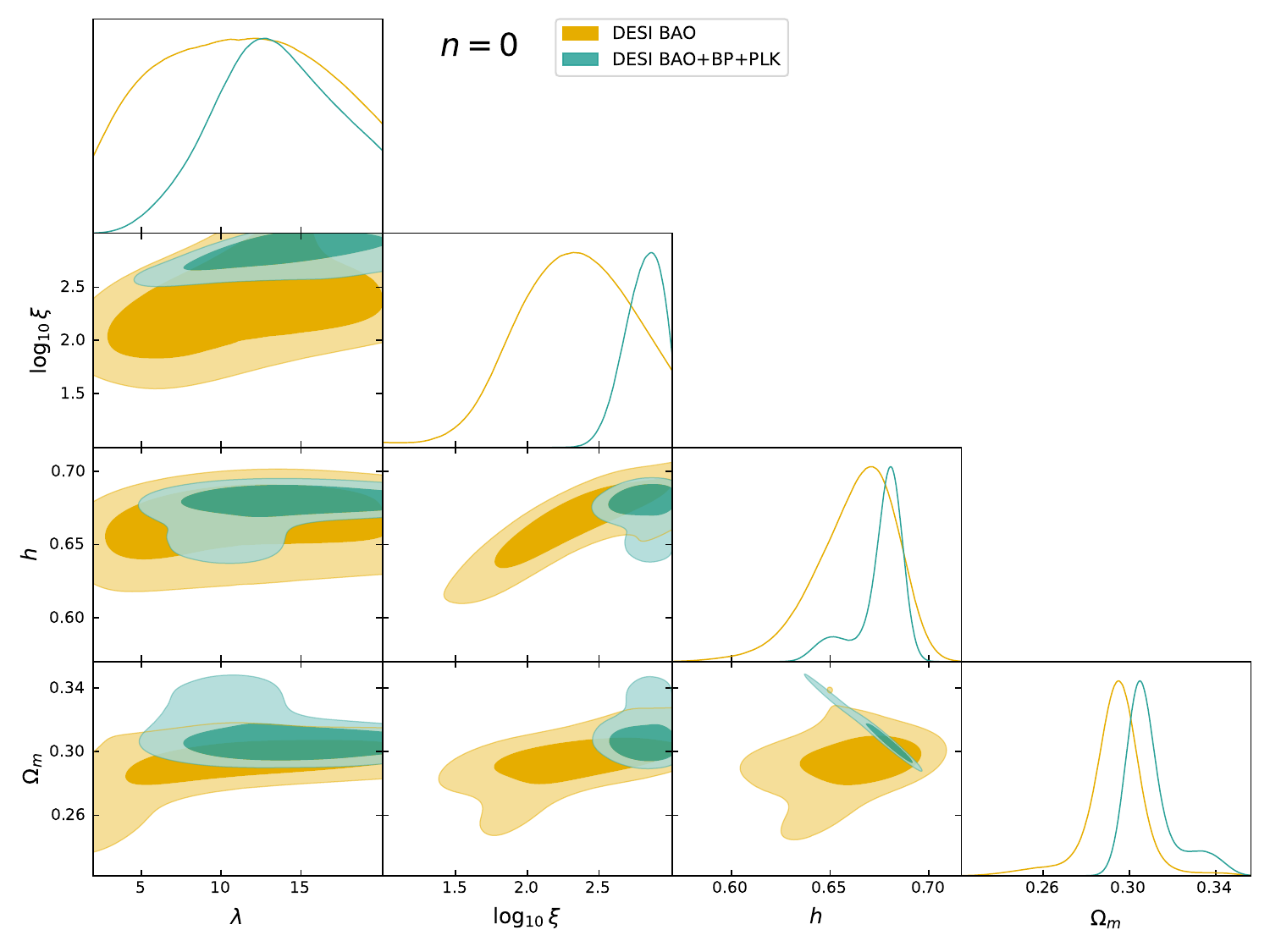}}
   \caption{Marginalised posterior distributions for the model with $n=0$, corresponding to the purely exponential potential. The contours show the 68\% and 95\% confidence regions for DESI \ac{BAO}  and DESI \ac{BAO}  + \ac{BP} + \ac{PLK}, according to the labels.}
   \label{fig:posterior-n0}
\end{figure}

\begin{figure}[h]
    \centering
    \makebox[12cm][c]{
    \includegraphics[width=16.cm]{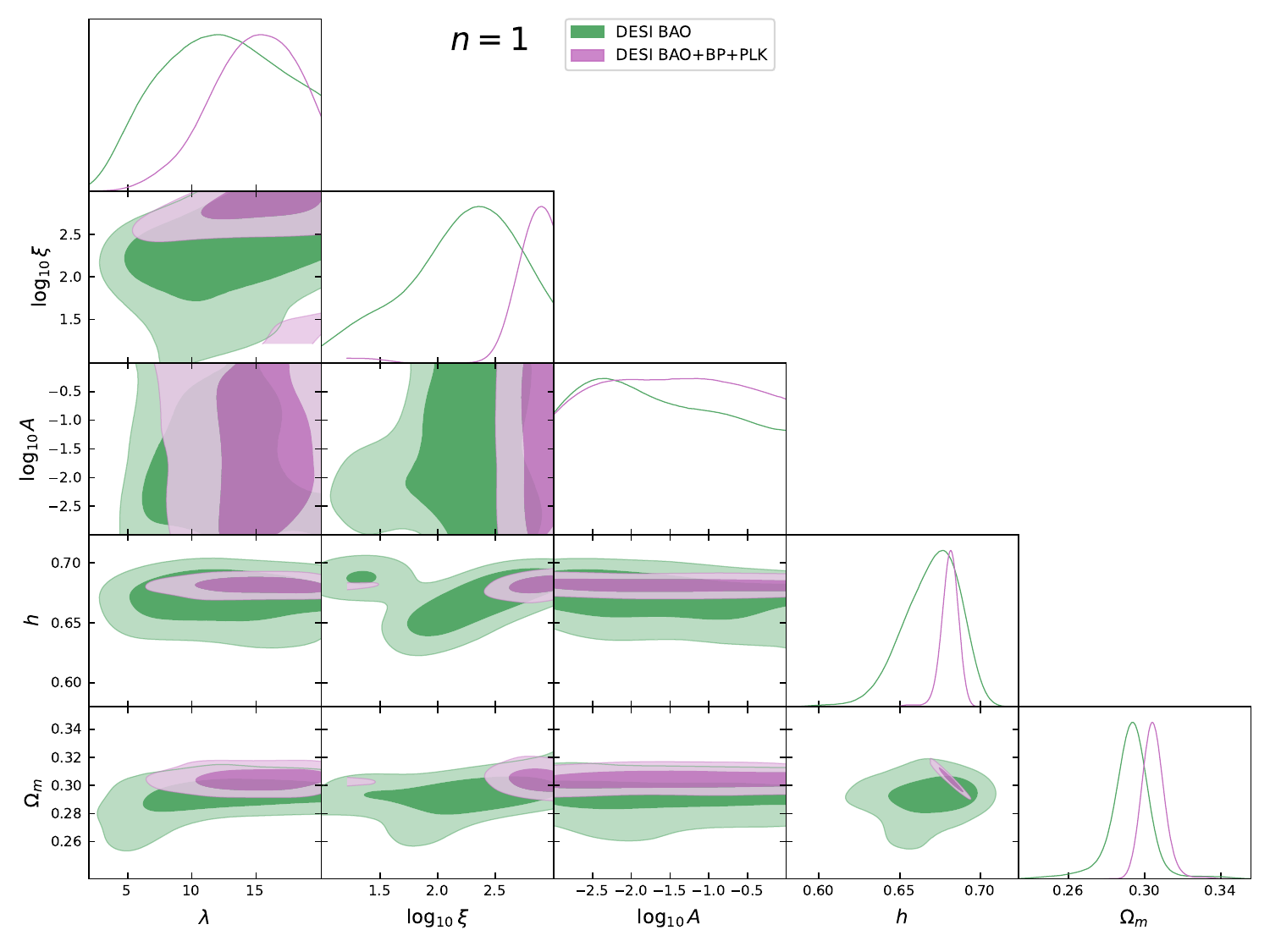}
    }
   \caption{Marginalised posterior distributions for the model with $n=1$, corresponding to the \ac{CL} potential. The contours show the 1-$\sigma$ and 2-$\sigma$ confidence regions for DESI \ac{BAO} and DESI \ac{BAO} DR2 + \ac{BP} + \ac{PLK}, according to the labels.}
   \label{fig:posterior-n1}
\end{figure}

\begin{figure}[h]
    \centering
    \makebox[12cm][c]{
    \includegraphics[width=16.cm]{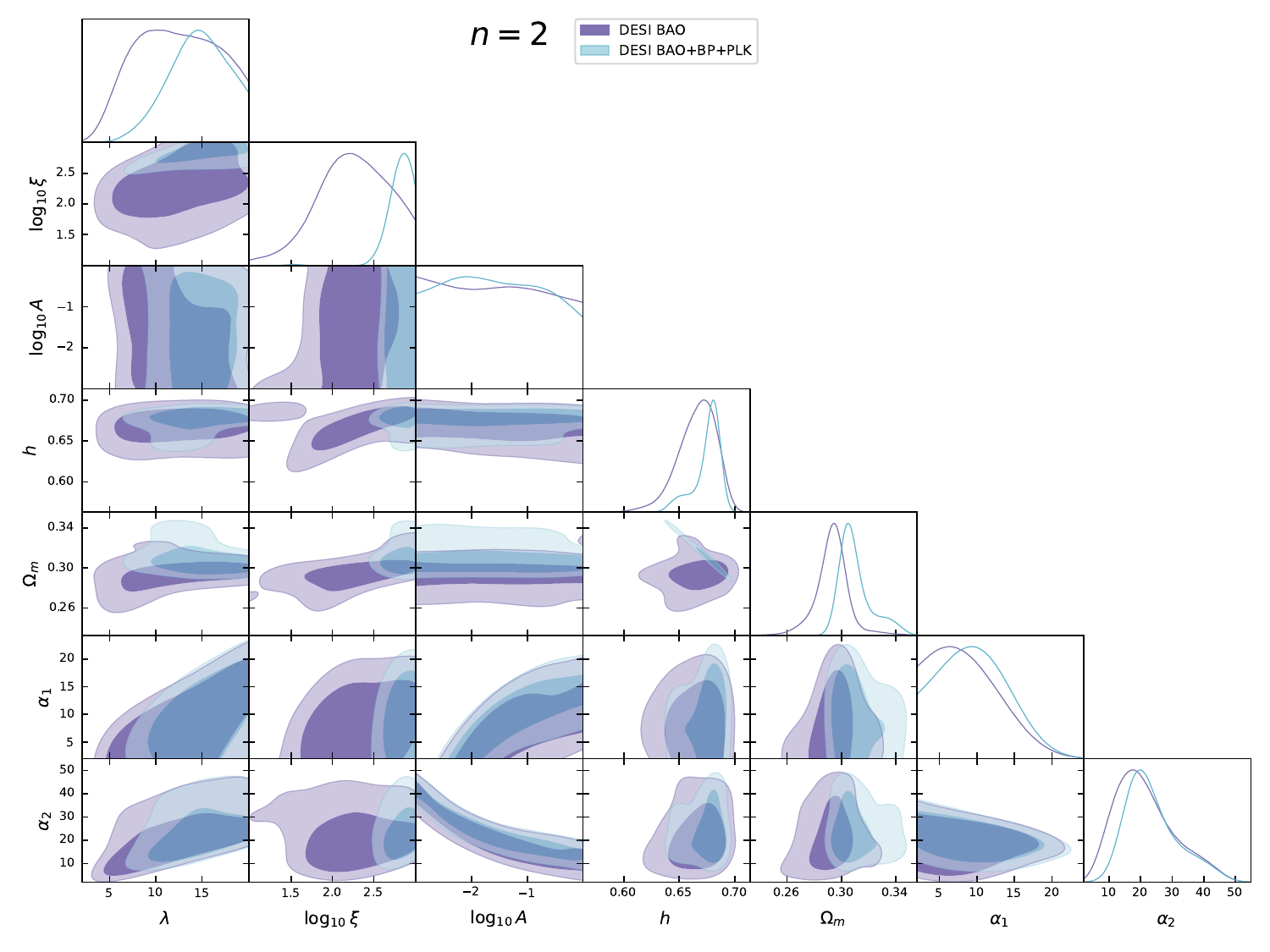}
    }
   \caption{Marginalised posterior distributions for the model with $n=2$, corresponding to the \ac{AS} potential. The contours show the 1-$\sigma$ and 2-$\sigma$ confidence regions for DESI \ac{BAO} and DESI \ac{BAO}  + \ac{BP} + \ac{PLK}, according to the labels.}
   \label{fig:posterior-n2}
\end{figure}

The confidence limits for the main parameters and the three fixed polynomial cases ($n=0$, 1, and 2) are shown in Figs.~\ref{fig:posterior-n0}--\ref{fig:posterior-n2}, while the corresponding mean values and uncertainties are summarised in Tab.~\ref{tab:nested_summary}. 
For each fixed value of $n$, we report results for DESI \ac{BAO} dataset alone, and for  DESI \ac{BAO} + \ac{BP} + \ac{PLK} dataset. The former constrains the late-time expansion history through geometrical distance measurements, whereas the latter also incorporates supernova and \ac{CMB} distance-prior information, thereby providing stronger sensitivity to the integrated expansion history.  
The results obtained by treating $n$ as a free parameter are presented in Appendix~\ref{app:nfree}, where the $\Lambda$CDM constraints are also included in Tab.~\ref{tab:nsummary} for completeness.

Regarding the standard cosmological parameters, the constraints show only mild variations across different $n-$models and datasets, with $\Omega_bh^2$ remaining around $0.022$ and $h$ approximately in the range $0.66$--$0.69$. A more noticeable dataset dependence is found for $\Omega_m$, which shifts towards larger values when \ac{BP} + \ac{PLK} are included, a trend also observed in the DESI $\Lambda$CDM analysis \cite{DESI:2025zgx}.

\begin{figure}
    \centering
    \makebox[15cm][c]{
   \includegraphics[trim = 5mm  0mm 5mm 0mm, clip, width=5.cm, height=5.cm]{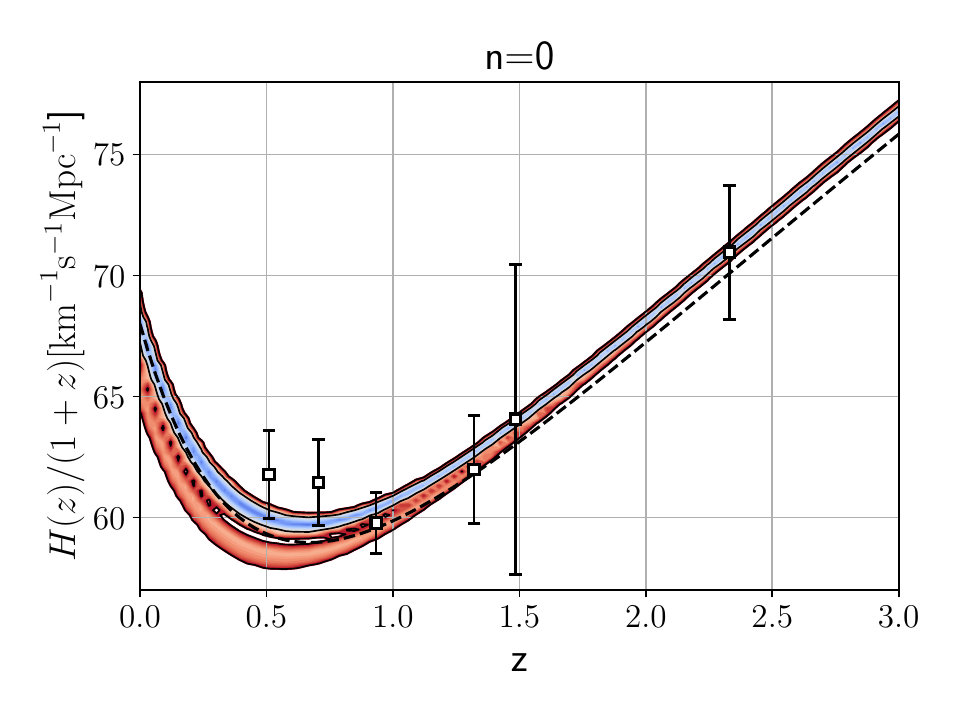} 
    \includegraphics[trim = 5mm  0mm 5mm 0mm, clip, width=5.cm, height=5.cm]{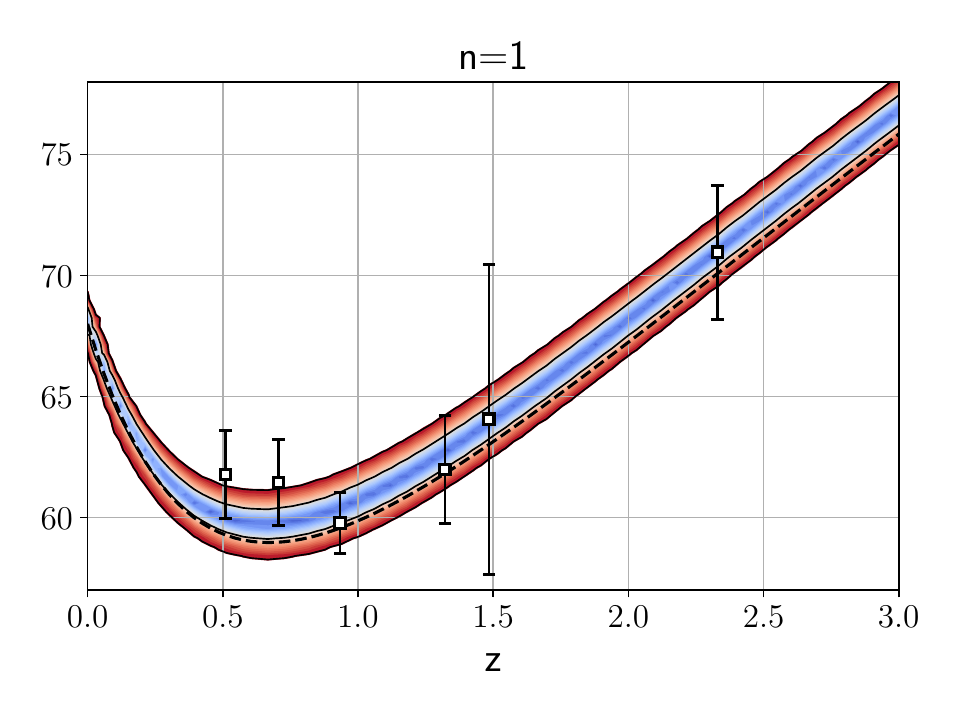} 
     \includegraphics[trim = 5mm  0mm 5mm 0mm, clip, width=5.cm, height=5.cm]{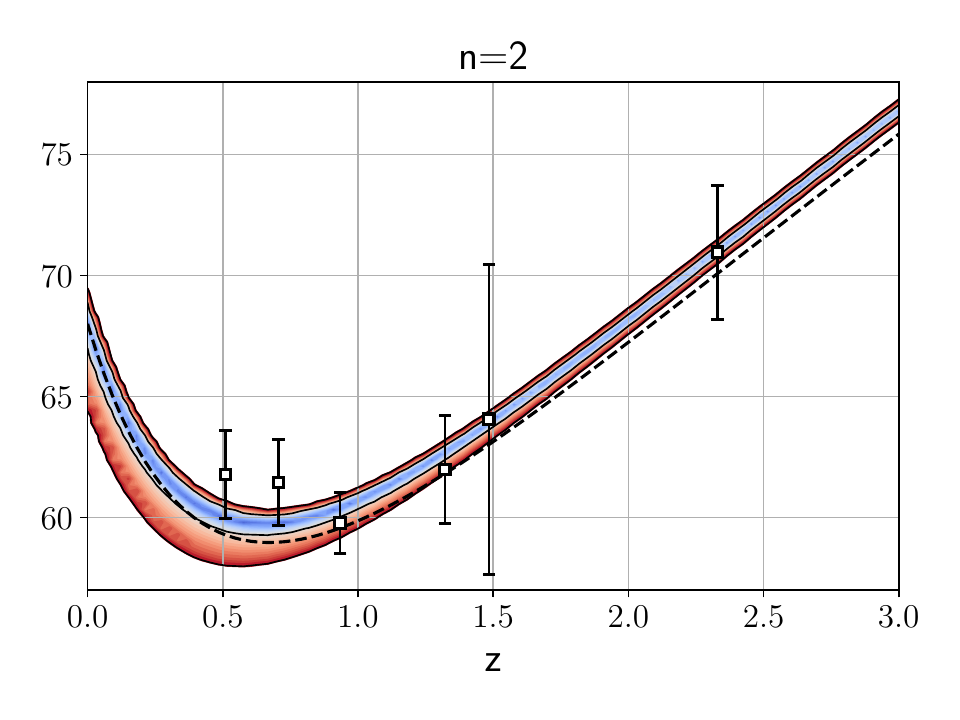} 
      \includegraphics[trim = 5mm  0mm 5mm 0mm, clip, width=5.cm, height=5.cm]{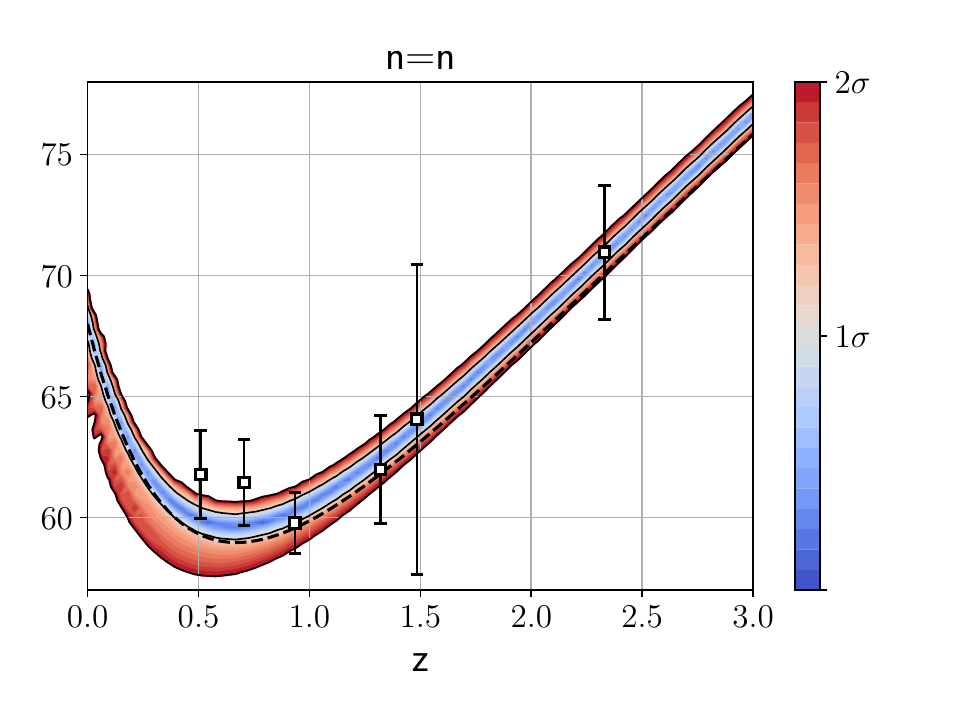}} \\
   
    \makebox[15cm][c]{
   \includegraphics[trim = 5mm  0mm 5mm 0mm, clip, width=5.cm, height=5.cm]{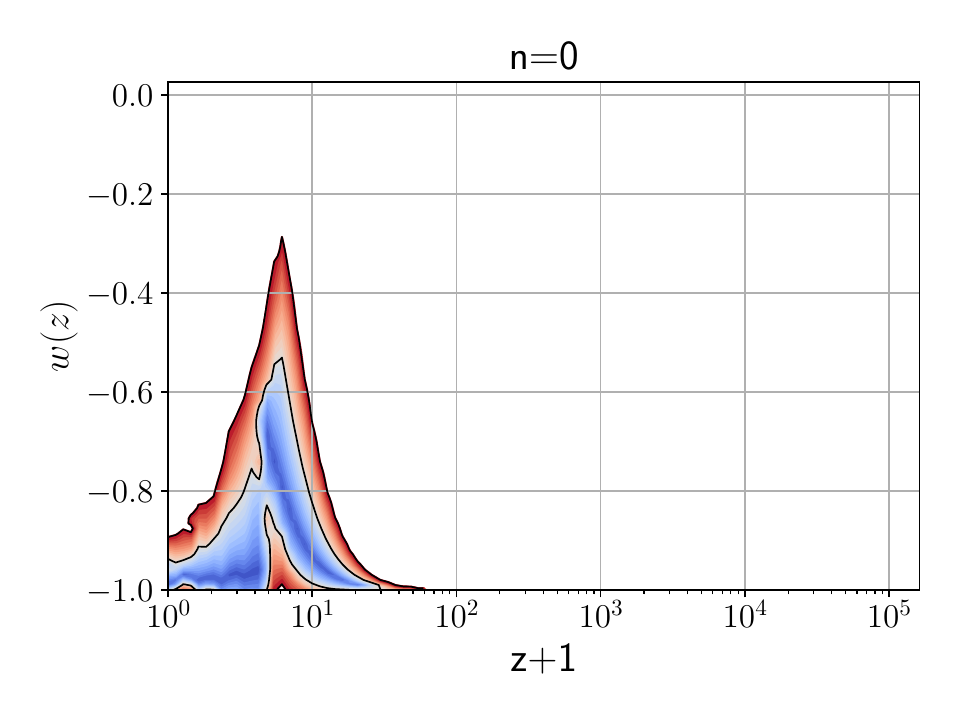}
    \includegraphics[trim = 5mm  0mm 5mm 0mm, clip, width=5.cm, height=5.cm]{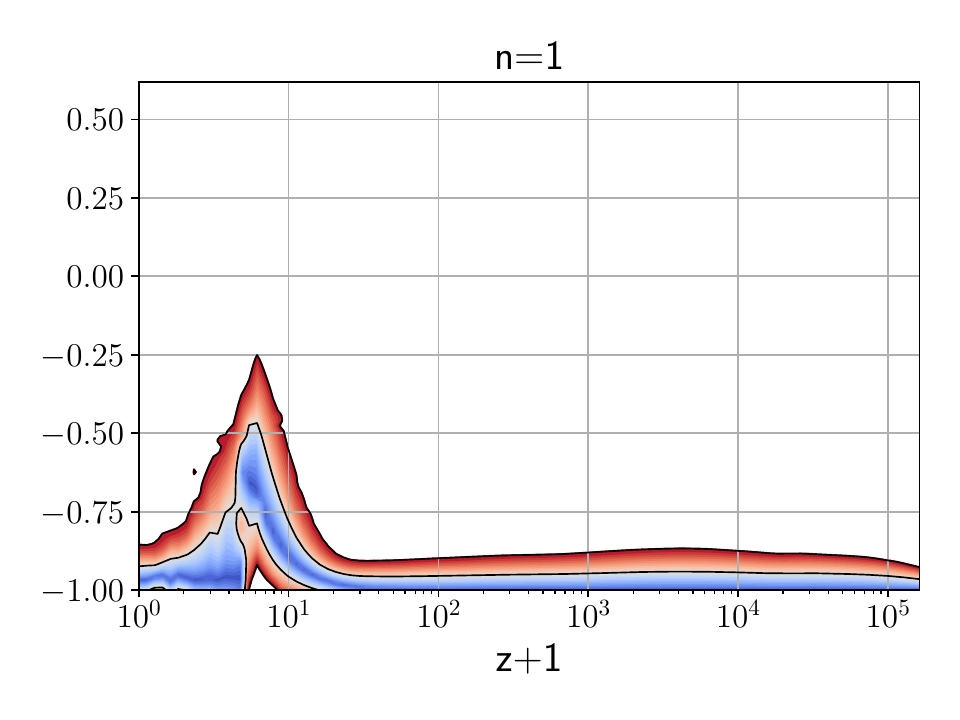}
    \includegraphics[trim = 5mm  0mm 5mm 0mm, clip, width=5.cm, height=5.cm]{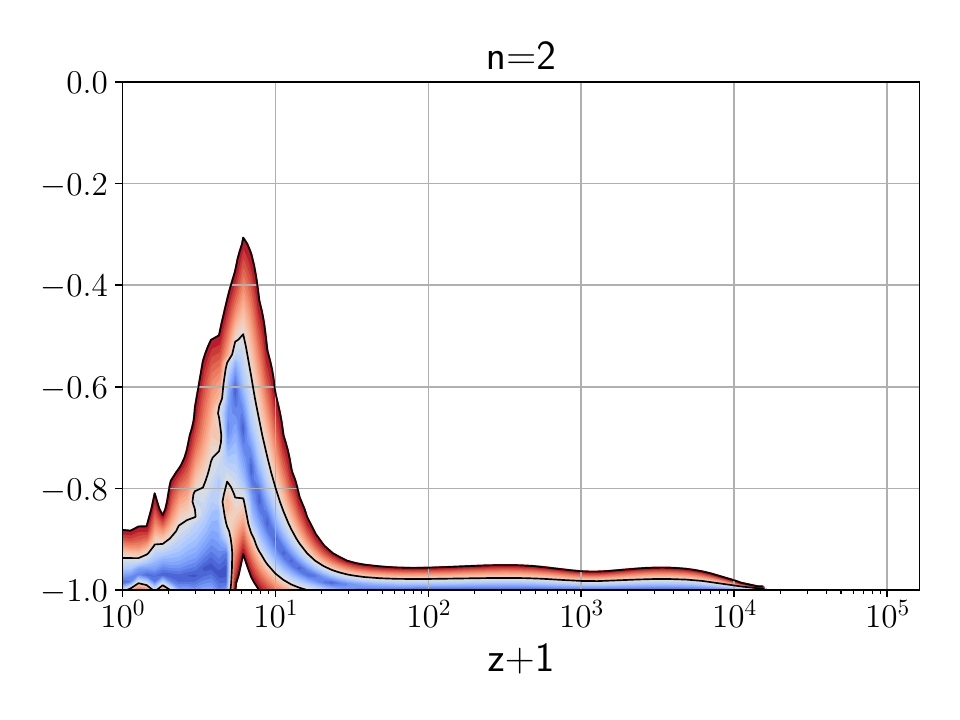}
    \includegraphics[trim = 5mm  0mm 5mm 0mm, clip, width=5.cm, height=5.cm]{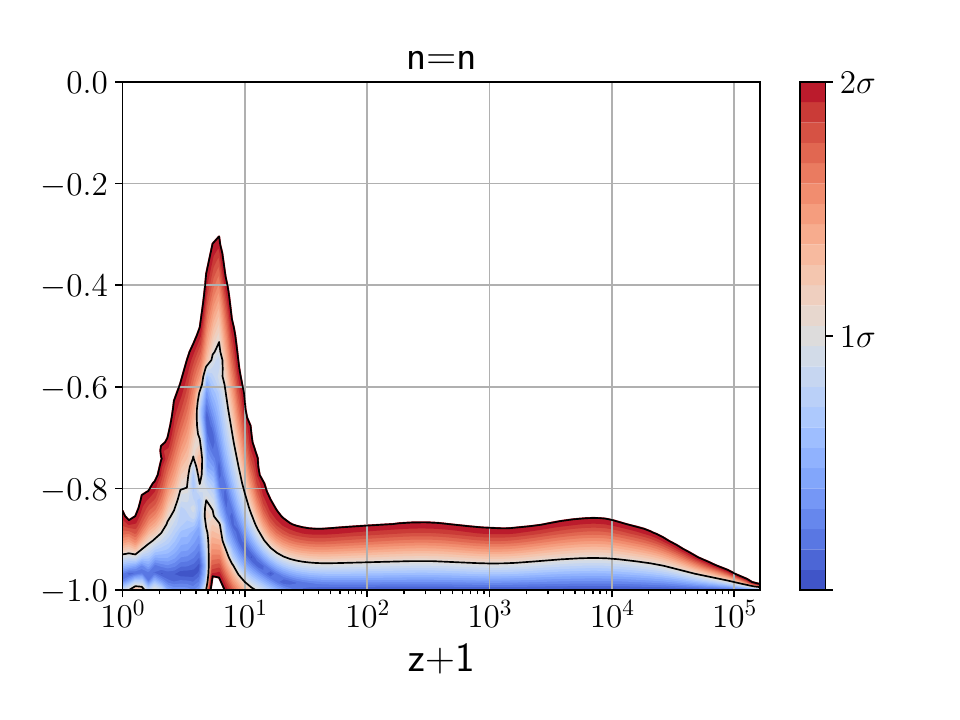}}
    \caption{Reconstructed expansion histories obtained from the DESI \ac{BAO} + \ac{BP} + \ac{PLK} posterior samples. From left to right, the columns correspond to $n=0$, $n=1$, $n=2$, and $n$--free. The top row shows the Hubble rate, $H(z)/(1+z)$, and the bottom row shows the dark-energy equation of state, $w(z)$. The shaded regions indicate the 68\% and 95\% credible intervals.}
   \label{fig:n0-fgivenx}
\end{figure}

Focusing first on the DESI \ac{BAO} constraints, it is worth noting that these observations alone have limited sensitivity to the model-specific parameters and therefore do not significantly constrain them. However, once the \ac{BP} and \ac{PLK} data sets are added to the dataset, the allowed parameter space becomes more restricted. In particular, independent of the $n-$power, Figs.~\ref{fig:posterior-n0}--\ref{fig:posterior-n2} show a correlation between $\lambda$ and $\xi$: larger values of $\lambda$ require correspondingly larger values of $\xi$ to preserve the hierarchy needed to obtain a late-time accelerated expansion, as pointed by the dynamical system analysis and numerical exploration.

For the exponential case, Fig.~\ref{fig:posterior-n0} shows the 68\% and 95\% confidence regions for the $n=0$ case. At the 95\% confidence level, the combined DESI \ac{BAO} + \ac{BP} + \ac{PLK} data constrain the potential and coupling parameters to $\lambda > 6.9$ and $\xi > 10^{2.57}$. Although the condition for obtaining a late-time equation of state close to $-1$ implies $\lambda \ll 2\xi$, allowing in principle for much smaller values of $\lambda$, the data impose a lower bound on this parameter, which is mainly driven by the contribution of the scalar field to the early dark energy fraction, which increases as $\lambda$ decreases.

We next consider the simplest polynomial deformation of the exponential potential, obtained by introducing a linear prefactor ($n=1$), corresponding to the \ac{CL} case. Table~\ref{tab:nested_summary} reports $\lambda>8.57$ and $\log_{10}\xi>2.51$ at 95\% confidence for DESI BAO + BP + PLK, with the associated posterior distributions shown in Fig.~\ref{fig:posterior-n1}. Although this prefactor modifies the form of the potential, its constant $A$ can be absorbed into a shift of the scalar field and, therefore, does not introduce an additional degree of freedom into the background evolution. Consequently, the constraints remain close to those of the purely exponential model, indicating that the linear prefactor has only a limited effect on the preferred cosmological parameter space.

In contrast, the \ac{AS} case ($n=2$) introduces $A$ as a genuine degree of freedom through the effective slopes $\alpha_{1,2}=\lambda\mp1/\sqrt{A}$, which modulates the early scalar fraction and the late-time attractor. For this model, Tab.~\ref{tab:nested_summary} shows that the combined DESI BAO + BP + PLK data constrain $\lambda>9.05$ and $\log_{10}\xi>2.55$ at the 95\% confidence level. With DESI BAO alone, this model also yields the smallest minimum $\chi^2$ value among the three fixed-$n$ cases, $\chi^2_{\min}=4.2$, compared with $5.8$ for $n=0$ and $6.2$ for $n=1$. As shown in Fig.~\ref{fig:posterior-n2}, the posterior of $A$ remains broad, whereas the derived slopes $\alpha_1$ and $\alpha_2$ are more tightly constrained. Thus, the observations primarily restrict the combinations of potential parameters entering the dynamics, while the bound on $\lambda$ remains comparable to those obtained for $n=0$ and $n=1$.

To examine how these parameter constraints translate into the cosmological evolution, we reconstruct the expansion history from the DESI BAO + BP + PLK posterior samples using \texttt{fgivenx}. Fig.~\ref{fig:n0-fgivenx} presents the reconstructed equation of state and Hubble rate, $H(z)/(1+z)$, for the three fixed values $n=0,1,2$ and $n$-free. Despite the different powers of the potential, all the models predict remarkably similar background evolutions; the reconstructed Hubble rate  remains close to the $\Lambda$CDM expectation over the full redshift range, with only small differences in the width of the credible regions among the models. Likewise, the reconstructed dark-energy equation of state approaches $w_{\rm DE}\simeq-1$ at both early and late time, while allowing a transient departure from a cosmological-constant behaviour at intermediate redshifts, approximately $2 \lesssim 1+z \lesssim 50$. Although the $n=0$ model remains closer to $\Lambda$CDM throughout the evolution, the polynomial extensions admit slightly broader excursions from $w_{\rm DE}=-1$ without producing appreciable changes in the expansion history.

Overall, the explored scalar field potentials yield consistent cosmological constraints, with no clear evidence that one model provides a significantly better fit than the others, including the case in which  $n$ is treated as a free parameter, presented in Appendix~\ref{app:nfree}. 
This behaviour is also reflected in the minimum $\chi^2$ values reported in the bottom rows of Tabs.~\ref{tab:nested_summary} and \ref{tab:nsummary} which remain very similar among the different axio-dilaton models for a given dataset. For reference, in Tab.~\ref{tab:nsummary} (Appendix~\ref{app:nfree}) we also report the corresponding $\Lambda \rm CDM$ values. It is worth noting that, when DESI BAO data are considered alone, the scalar--field models reach lower minimum values, with $\chi^2_{\min}$ ranging from $4.2$ to $6.2$, compared with $10.28$ for $\Lambda$CDM. This difference should not, however, be interpreted as evidence for a better statistical description of the data, since the axio-dilaton scenarios contain several additional free parameters and therefore have substantially greater fitting flexibility. Moreover, the difference becomes much smaller once information from additional cosmological scales is included. For the full DESI BAO+BP+PLK combination, for example, $\chi^2_{\min}$ changes only from $51.6$ in $\Lambda$CDM to approximately $50.4$ in the axio-dilaton models. The minimum-$\chi^2$ values, therefore, indicate that the different axio-dilaton realisations can accommodate the current data.

\section{Conclusions}

This article presented a systematic cosmological analysis of the axio-dilaton system, adopting a generalised Albrecht–Skordis dilaton potential motivated by recent proposals addressing the dark-energy problem and by string-compactification scenarios incorporating logarithmic corrections.

We confronted the models with DESI \ac{BAO} DR2 measurements, the binned
Pantheon supernova sample, and the Planck 2018 distance-prior information. For all
three fixed values $n=0,1,2$, the combined data reveal a clear correlation
between the potential slope $\lambda$ and the field-space curvature parameter
$\xi$: steeper potentials require stronger kinetic coupling to maintain a
viable late-time accelerated evolution. At the 95\% credible level, the
combined data yield $\lambda>6.90$ and $\log_{10}\xi>2.57$ for $n=0$,
$\lambda>8.57$ and $\log_{10}\xi>2.51$ for $n=1$, and $\lambda>9.05$ and
$\log_{10}\xi>2.55$ for $n=2$. By contrast, the parameter $A$ remains only
weakly constrained, indicating that present background observations are more
sensitive to the effective slope and kinetic coupling than to the detailed
shape of the polynomial correction.

Treating $n$ as a continuous free parameter leads to the same overall
picture. Its posterior distribution extends across the adopted prior range,
so the current data do not select a preferred polynomial exponent. The bounds
$\lambda>8.50$ and $\log_{10}\xi>2.49$ obtained from the combined data are
consistent with those of the fixed-$n$ analyses. Moreover, the reconstructed
background histories are remarkably similar among all the models. The
normalised Hubble rate remains close to the $\Lambda$CDM prediction over the
redshift interval considered, whereas the reconstructed dark-energy equation
of state permits a transient departure from $-1$ at intermediate redshifts
before returning to a cosmological-constant-like behaviour.

Although the axio--dilaton models attain slightly smaller minimum $\chi^2$
values than $\Lambda$CDM, particularly for DESI \ac{BAO} alone, this reduction
is not sufficient to establish a statistical preference because the scalar
models contain additional parameters. For the full data combination, the
difference is small, with $\chi^2_{\min}\simeq 50.4$ for the axio--dilaton
realisations and $51.6$ for $\Lambda$CDM. We therefore conclude that the
models provide viable descriptions of the current background data, but 
the available observations neither distinguish among the different
polynomial potentials nor favour them decisively over $\Lambda$CDM. 

Our work complements the studies of \cite{Smith:2025grk,Smith:2025uaq}, which also investigate the axio-dilaton system. In addition to presenting a complete dynamical systems analysis, we explore a broader class of models by considering different values of the polynomial degree $n$ and the kinetic-coupling parameter $\xi$.

A more comprehensive assessment will require the inclusion of cosmological perturbations and structure-growth observables. It would also be interesting to consider the addition of an axion potential or interactions between dark energy and dark matter, which could provide a route to an effective phantom regime along the lines of \cite{Das:2005yj,Huey:2004qv,Smith:2025uaq,Bedroya:2025fwh}. These questions will be addressed in a forthcoming publication, which is currently in preparation.

\acknowledgments

We thank Cliff Burgess, Anne Davis, Carsten van de Bruck, Gonzalo Villa and Ivonne Zavala for useful conversations. FQ research is funded by Tankeem under grant  ADHPG-AD457 to NYUAD. MR-H thanks St. John's College, NYUAD and Ibero Puebla for their hospitality. The work of MR-H has been partially supported by Cambridge Trust, SECIHTI, DAMTP and St. Edmund’s College. JAV acknowledges support from UNAM-DGAPA-PAPIIT IN109126, IN110325, HTC project LANCAD-UNAM-DGTIC-477 and C\'{a}tedra de Investigaci\'{o}n Marcos Moshinsky. GG-A acknowledges the postdoctoral fellowship from SECIHTI.

\appendix
\section{Stability Analysis}
\label{sec:stability}

The local stability of a fixed point is determined by linearising the
autonomous system around it and computing the eigenvalues $\mu_i$ of the
Jacobian $M_{ij}=\partial h_i/\partial x_j$ evaluated at that point.  A
fixed point is called \emph{hyperbolic} if none of the $\mu_i$ have zero
real part; for hyperbolic points the Hartman--Grobman theorem guarantees
that the linearised flow faithfully reproduces the qualitative behaviour of
the full non-linear system in a neighbourhood of the fixed
point~\cite[Sec.~2.2]{Bahamonde:2017ize}.  One then classifies the point
as a \emph{stable attractor} if $\operatorname{Re}(\mu_i)<0$ for all $i$,
an \emph{unstable repeller} if $\operatorname{Re}(\mu_i)>0$ for all $i$,
or a \emph{saddle} if the eigenvalues have real parts of both signs.

When one or more eigenvalues have zero real part the fixed point is
\emph{non-hyperbolic} and linear stability theory is
inconclusive~\cite[Sec.~2.2]{Bahamonde:2017ize}: the eigenvalue spectrum
alone cannot decide whether nearby trajectories approach, recede from, or
orbit the fixed point in the associated directions.  The
remaining eigenvalues with non-zero real parts still govern attraction or
repulsion in their own directions; only the centre directions require
further analysis.  In the \ac{CL} model, where the centre direction
corresponds to a decoupled ordinary differential equation for $\alpha$, we analyse it directly
by integrating the reduced equation near the fixed point.  In the \ac{AS}
model the zero eigenvalue at Branch~(iii) points arises from a qualitatively
different vanishing structure in $h_4$ and is not further analysed here;
the stability conclusions for those points rest on the three non-zero
eigenvalues inherited from the \ac{EXP} model.

The three models share a common structure for the first three equations
of the autonomous system.  In the variables $(x_1,x_2,y,\alpha)$, the system (\ref{eq:General-DS}) can be expressed more explicitly as:
\begin{align}
h_1\equiv &\; x_1' = 3x_1(x_1^2+x_2^2-1)
        +\sqrt{\tfrac{3}{2}}\,(\alpha y^2-2\xi x_2^2)
        -\tfrac{3}{2}\gamma\, x_1(x_1^2+x_2^2+y^2-1),
        \label{eq:h1}\\[4pt]
h_2\equiv &\; x_2' = 3x_2(x_1^2+x_2^2-1)
        +\sqrt{6}\,\xi\, x_1 x_2
        -\tfrac{3}{2}\gamma\, x_2(x_1^2+x_2^2+y^2-1),
        \label{eq:h2}\\[4pt]
h_3\equiv &\; y' = -\sqrt{\tfrac{3}{2}}\,\alpha x_1 y
        +3y(x_1^2+x_2^2)
        -\tfrac{3}{2}\gamma\, y(x_1^2+x_2^2+y^2-1),
        \label{eq:h3}\\[4pt]
h_4\equiv &\; \alpha'= f(\alpha,x_1),
        \label{eq:h4}
\end{align}
where the slope evolution $f(\alpha,x_1)$ is the only model-dependent
ingredient:
\begin{equation}
f(\alpha,x_1) =
\begin{cases}
0 & \text{EXP: }\alpha\equiv\lambda\text{ is constant (3D system)},\\[6pt]
\sqrt{6}\,(\lambda-\alpha)^2\,x_1 & \text{CL},\\[6pt]
\sqrt{6}\,(1-\Gamma)\,\alpha^2\,x_1\,,\quad
\Gamma\equiv\dfrac{V V_{zz}}{(V_z)^2} & \text{AS}.
\end{cases}
\label{eq:h4-models}
\end{equation}
The equations $h_1$, $h_2$, $h_3$ are identical across all three
models once $\alpha$ is treated as a dynamical variable; the entire
model hierarchy is encoded in $h_4$ alone.  The variable $\alpha$
enters $h_1$ and $h_3$ through the terms $\sqrt{3/2}\,\alpha y^2$ and
$-\sqrt{3/2}\,\alpha x_1 y$ respectively, so $\alpha$ feeds back into
the $(x_1,x_2,y)$ subsector, and $x_1$ drives $\alpha$ through $h_4$.
This mutual coupling between $x_1$ and $\alpha$ is the origin of the
off-diagonal Jacobian entries $J_{14}$ and $J_{41}$ that appear
throughout the stability analysis below.
\subsection{Case $n=0$ (\ac{EXP})}
\label{subsec:stab_exp}

For the two-field exponential potential $V=V_0 e^{-\lambda\phi_1/M_P}$ with
field-space curvature parameter $\xi$, the three-dimensional autonomous
system $(x_1,x_2,y)$ admits the six fixed points found
in~\cite{Cicoli:2020cfj}. The exact eigenvalues are collected in Table~\ref{tab:stab_exp}.  The $\mathcal{NG}$
point is the only fixed point that simultaneously satisfies $\Omega_\phi=1$
and $\omega_\phi\approx-1$ (requiring $\xi\gg\lambda$) while being stable.
This is the key result of~\cite{Cicoli:2020cfj}: a large field-space
curvature $\xi$ can drive the universe toward $\omega_\phi\approx-1$ even
for a steep potential ($\lambda\gg1$), without violating the swampland de
Sitter conjectures.

\begin{table}[t]
\centering
\renewcommand{\arraystretch}{2.2}
\begin{adjustbox}{width=\textwidth}
\begin{tabular}{
    >{\centering\arraybackslash}m{1.2cm}
    >{\centering\arraybackslash}m{3cm}
    >{\centering\arraybackslash}m{3cm}
    >{\centering\arraybackslash}m{3cm}
    >{\centering\arraybackslash}m{4.5cm}
}
\toprule
\rowcolor{lightgray}
\textbf{Point} & $\mu_1$ & $\mu_2$ & $\mu_3$ & \textbf{Stability} \\
\midrule\midrule
$\mathcal{K}_+$ & $\sqrt{6}\,\xi$ & $\tfrac{1}{2}(6-\sqrt{6}\lambda)$ & $3(2-\gamma)$ & Repeller \\
$\mathcal{K}_-$ & $-\sqrt{6}\,\xi$ & $\tfrac{1}{2}(6+\sqrt{6}\lambda)$ & $3(2-\gamma)$ & Saddle \\
$\mathcal{F}$   & $\tfrac{3}{2}(\gamma-2)$ & $\tfrac{3}{2}(\gamma-2)$ & $\tfrac{3}{2}\gamma$ & Saddle \\
$\mathcal{S}$   & $-\tfrac{3}{2}+\tfrac{3\xi}{\lambda}$
      & $-\tfrac{3}{4}\!\left(1+\tfrac{\sqrt{24-7\lambda^2}}{\lambda}\right)$
      & $-\tfrac{3}{4}\!\left(1-\tfrac{\sqrt{24-7\lambda^2}}{\lambda}\right)$
      & $\lambda>2\xi$,\; $\lambda>\sqrt{3}$ \\
$\mathcal{G}$   & $\tfrac{1}{2}(\lambda^2-6)$
      & $\xi\lambda+\tfrac{\lambda^2}{2}-3$
      & $\lambda^2-3\gamma$
      & $\lambda<\sqrt{3}\ \land\ \lambda<\sqrt{\xi^2+6}-\xi$ \\
$\mathcal{NG}_\pm$  & $-3\gamma$
      & $\tfrac{1}{2}\!\left(-3+\sqrt{\Delta}\right)$
      & $\tfrac{1}{2}\!\left(-3-\sqrt{\Delta}\right)$
      & $\sqrt{\xi^2+6}-\xi\leq\lambda\leq2\xi$ \\
\bottomrule\bottomrule
\end{tabular}
\end{adjustbox}
\caption{Eigenvalues of the stability matrix for the exponential-potential
model~\cite{Cicoli:2020cfj}.  The discriminant is
$\Delta=3(27-8\xi\lambda-4\lambda^2)$.  Points $\mathcal{K}_\pm$,
$\mathcal{F}$ are always saddles or repellers; $\mathcal{S}$,
$\mathcal{G}$, $\mathcal{NG}_\pm$ are conditionally stable.  For
$0<\gamma<2$ and any $\xi,\lambda>0$, at most one of $\mathcal{S}$,
$\mathcal{G}$, $\mathcal{NG}_\pm$ is stable at a time.}
\label{tab:stab_exp}
\end{table}

\subsection{Case $n=1$ (\ac{CL})}
\label{subsec:stab_CL}

The CL model adds a fourth dynamical variable $\alpha$ with evolution
equation $h_4=\sqrt{6}(\lambda-\alpha)^2x_1$.  We determine the stability
of all fixed points via the $4\times4$ Jacobian of the autonomous system
$(x_1,x_2,y,\alpha)$.  Recall that $h_4=0$ factorises as
\begin{equation}
  \sqrt{6}(\lambda-\alpha)^2x_1=0
  \;\Longleftrightarrow\;
  \alpha^\star=\lambda
  \quad\text{or}\quad
  x_1^\star=0,
  \label{eq:CL-branches}
\end{equation}
defining two distinct branches.  Crucially, when $x_1^\star=0$, equation
$h_4=0$ is satisfied for \emph{every} value of $\alpha^\star$; it is the
residual equations $h_1=h_2=h_3=0$ that fix (or leave free) the value of
$\alpha^\star$ in Branch~B.

\begin{itemize}

\item \textbf{Branch A — fixed points at $\alpha^\star=\lambda$:}

To understand the Jacobian structure, we compute the fourth row and fourth
column explicitly.  The fourth row comes from differentiating
$h_4=\sqrt{6}(\lambda-\alpha)^2x_1$:
\begin{equation}
  J_{41} = \frac{\partial h_4}{\partial x_1}\bigg|_{\alpha^\star=\lambda}
         = \sqrt{6}(\lambda-\alpha)^2\big|_{\alpha^\star=\lambda} = 0,
  \qquad
  J_{44} = \frac{\partial h_4}{\partial\alpha}\bigg|_{\alpha^\star=\lambda}
         = -2\sqrt{6}(\lambda-\alpha)x_1\big|_{\alpha^\star=\lambda} = 0,
\end{equation}
and $J_{42}=J_{43}=0$ since $h_4$ contains neither $x_2$ nor $y$.  Hence
the entire fourth row vanishes.  The fourth column comes from
differentiating $h_1,h_2,h_3$ with respect to $\alpha$.  In $h_1$ the
only $\alpha$-dependent term is $\sqrt{3/2}\,\alpha y^2$; in $h_2$ there
is no $\alpha$; in $h_3$ the only $\alpha$-dependent term is
$-\sqrt{3/2}\,\alpha x_1 y$:
\begin{equation}
  J_{14} = \frac{\partial h_1}{\partial\alpha}\bigg|_\star
         = \sqrt{\tfrac{3}{2}}\,y^{\star2},
  \qquad
  J_{24} = \frac{\partial h_2}{\partial\alpha}\bigg|_\star = 0,
  \qquad
  J_{34} = \frac{\partial h_3}{\partial\alpha}\bigg|_\star
         = -\sqrt{\tfrac{3}{2}}\,x_1^\star y^\star.
\end{equation}

The $4\times4$ Jacobian therefore takes the upper block-triangular form
\begin{equation}
  J_{4D}\big|_{\alpha^\star=\lambda}
  =\begin{pmatrix}
     J_{3D}^{\rm EXP} & \mathbf{c} \\
     \mathbf{0}^\top  & 0
   \end{pmatrix},
  \qquad
  \mathbf{c}
  =\begin{pmatrix}
     \sqrt{3/2}\,y^{\star2}\\[2pt]
     0\\[2pt]
     -\sqrt{3/2}\,x_1^\star y^\star
   \end{pmatrix},
\end{equation}
where $J_{3D}^{\rm EXP}$ is the $3\times3$ Jacobian of the exponential
model evaluated at the same $(x_1^\star,x_2^\star,y^\star)$.  For any
upper block-triangular matrix $\bigl(\begin{smallmatrix}A&\mathbf{c}\\
\mathbf{0}^\top&d\end{smallmatrix}\bigr)$ the determinant factorises as
$\det(A)\cdot d$, so
\begin{equation}
  \det(J_{4D}-\mu I_4)
  = \det(J_{3D}^{\rm EXP}-\mu I_3)\cdot(-\mu) = 0,
\end{equation}
giving
\begin{equation}
  \operatorname{spec}(J_{4D})
  = \operatorname{spec}(J_{3D}^{\rm EXP})\cup\{0\}.
\end{equation}
The three non-trivial eigenvalues and their stability conditions are
therefore identical to the exponential case (Table~\ref{tab:stab_exp}),
with $\lambda=\alpha^\star$.

The zero eigenvalue is a \emph{centre direction} in $\alpha$: linear
stability analysis is inconclusive in this direction since a vanishing
eigenvalue means perturbations in $\alpha$ are neither amplified nor damped
at linear order~\cite[Sec.~2.2]{Bahamonde:2017ize}.  To determine the
actual long-time behaviour, we analyse the $\alpha$ equation directly.  Define $u\equiv\lambda-\alpha$, so that
$\alpha'=\sqrt{6}(\lambda-\alpha)^2x_1$ becomes
\begin{equation}
  u' = -\sqrt{6}\,(\lambda-\alpha)^2\,x_1 = -\sqrt{6}\,x_1^\star\,u^2
  + \mathcal{O}(u^2\,\delta x_1),
  \label{eq:u-ode}
\end{equation}
where at leading order near the fixed point we replace $x_1\approx
x_1^\star$.  Equation~\eqref{eq:u-ode} is separable and integrates
exactly:
\begin{equation}
  u(N) = \frac{1}{\sqrt{6}\,x_1^\star(N-N_0)+u_0^{-1}}
  \;\xrightarrow{N\to\infty}\;
  \frac{1}{\sqrt{6}\,x_1^\star\,N},
  \label{eq:u-solution}
\end{equation}
showing that $\lambda-\alpha\sim N^{-1}$: an algebraic (power-law)
decay rather than the exponential decay $\sim e^{-|\mu|N}$ associated with
a strictly negative eigenvalue~\cite{Copeland:1997et}.  The convergence $\alpha\to\lambda$ is
therefore guaranteed for any fixed point with $x_1^\star>0$ (i.e.\
$\mathcal{S}$, $\mathcal{G}$, $\mathcal{NG}_\pm$), confirming these
remain genuine late-time attractors despite the zero eigenvalue.  At
$\mathcal{F}$, where $x_1^\star=0$, equation~\eqref{eq:u-ode} gives
$u'=0$, so $\alpha$ is frozen; but $\mathcal{F}$ is always a saddle due to
$\mu_3=\frac{3}{2}\gamma>0$ and is not an attractor regardless.

\item \textbf{Branch B — fixed points at $x_1^\star=0$:}

Substituting $x_1^\star=0$ into $h_1=h_2=h_3=0$ gives
\begin{align}
  h_1\big|_{x_1=0} &= \sqrt{\tfrac{3}{2}}\!\left(\alpha^\star y^{\star2}
    - 2\xi x_2^{\star2}\right) = 0, \label{eq:CL-B-h1}\\
  h_2\big|_{x_1=0} &= x_2^\star\!\left[3(x_2^{\star2}-1)
    -\tfrac{3}{2}\gamma(x_2^{\star2}+y^{\star2}-1)\right] = 0,
    \label{eq:CL-B-h2}\\
  h_3\big|_{x_1=0} &= y^\star\!\left[3x_2^{\star2}
    -\tfrac{3}{2}\gamma(x_2^{\star2}+y^{\star2}-1)\right] = 0.
    \label{eq:CL-B-h3}
\end{align}
Suppose $x_2^\star\neq0$ and $y^\star\neq0$; dividing
\eqref{eq:CL-B-h2} by $x_2^\star$ and \eqref{eq:CL-B-h3} by
$y^\star$ and subtracting gives
\begin{equation}
  3(x_2^{\star2}-1) - 3x_2^{\star2} = -3 = 0,
\end{equation}
a contradiction.  Hence $x_2^\star=0$.  With $x_2^\star=0$,
equation~\eqref{eq:CL-B-h2} is satisfied identically for any
$y^\star$, while equation~\eqref{eq:CL-B-h3} reduces to
\begin{equation}
  -\tfrac{3}{2}\gamma\,y^\star(y^{\star2}-1) = 0.
  \label{eq:CL-B-h3-reduced}
\end{equation}
For $\gamma\neq0$ this requires $y^\star=0$ or $y^{\star2}=1$,
i.e.\ $y^\star=1$ (taking the physical non-negative root).
With $x_2^\star=0$, equation~\eqref{eq:CL-B-h1} further simplifies to
\begin{equation}
  \sqrt{\tfrac{3}{2}}\,\alpha^\star y^{\star2} = 0.
  \label{eq:CL-B-reduced}
\end{equation}
The two roots of \eqref{eq:CL-B-h3-reduced} lead to qualitatively
different constraints on $\alpha^\star$ from \eqref{eq:CL-B-reduced}.

\begin{itemize}

  \item \textbf{Sub-case B1 — $y^\star=0$: continuous family
    $\mathcal{F}(\alpha^\star)$:}

  When $y^\star=0$, equation~\eqref{eq:CL-B-h3-reduced} is satisfied
  trivially, and equation~\eqref{eq:CL-B-reduced} holds for
  \emph{every} value of $\alpha^\star$.  This yields a continuous
  one-parameter family of fixed points
  \begin{equation}
    \mathcal{F}(\alpha^\star) = (0,\,0,\,0,\,\alpha^\star),
    \qquad \alpha^\star\in\mathbb{R}.
    \label{eq:CL-F-family}
  \end{equation}
  To compute the Jacobian at a general member of this family, we
  evaluate all 16 partial derivatives of $(h_1,h_2,h_3,h_4)$ at
  $(0,0,0,\alpha^\star)$.  The key diagonal entries are
\begin{align}
  J_{11} &= \left[3(x_1^2+x_2^2-1)+6x_1^2
            -\tfrac{3}{2}\gamma(x_1^2+x_2^2+y^2-1)-3\gamma x_1^2
            \right]_{(0,0,0,\alpha^\star)} \notag\\
         &= -3+\tfrac{3}{2}\gamma = \tfrac{3}{2}(\gamma-2), \\[6pt]
  J_{22} &= \left[3(x_1^2+x_2^2-1)+6x_2^2
            -\tfrac{3}{2}\gamma(x_1^2+x_2^2+y^2-1)-3\gamma x_2^2
            \right]_{(0,0,0,\alpha^\star)} \notag\\
         &= \tfrac{3}{2}(\gamma-2), \\[6pt]
  J_{33} &= \left[-\sqrt{\tfrac{3}{2}}\,\alpha x_1+3(x_1^2+x_2^2)
            -\tfrac{3}{2}\gamma(x_1^2+x_2^2+y^2-1)-3\gamma y^2
            \right]_{(0,0,0,\alpha^\star)} \notag\\
         &= \tfrac{3}{2}\gamma.
\end{align}
  All off-diagonal entries in the first three rows vanish.  The only non-zero off-diagonal entry in
  the entire matrix is in the fourth row:
  \begin{equation}
    J_{41} = \left[\sqrt{6}(\lambda-\alpha)^2\right]_{(0,0,0,\alpha^\star)}
           = \sqrt{6}(\lambda-\alpha^\star)^2,
  \end{equation}
  while $J_{42}=J_{43}=0$ and $J_{44}=-2\sqrt{6}(\lambda-\alpha)x_1
  \big|_{x_1^\star=0}=0$.  The full Jacobian is therefore
  \begin{equation}
    J_{\mathcal{F}} =
    \begin{pmatrix}
      \tfrac{3}{2}(\gamma-2) & 0 & 0 & 0 \\[4pt]
      0 & \tfrac{3}{2}(\gamma-2) & 0 & 0 \\[4pt]
      0 & 0 & \tfrac{3}{2}\gamma & 0 \\[4pt]
      \sqrt{6}(\lambda-\alpha^\star)^2 & 0 & 0 & 0
    \end{pmatrix}.
    \label{eq:CL-F-Jacobian}
  \end{equation}
  This matrix is \emph{lower triangular}: every entry above the
  diagonal is zero.  For a lower triangular matrix the characteristic
  polynomial is the product of the diagonal entries minus $\mu$,
  regardless of the sub-diagonal entry $J_{41}$:
  \begin{equation}
    \det(J_{\mathcal{F}}-\mu I)
    = \left(\tfrac{3}{2}(\gamma-2)-\mu\right)^2
      \left(\tfrac{3}{2}\gamma-\mu\right)(-\mu) = 0.
  \end{equation}
  The four eigenvalues are
  \begin{equation}
    \mu_{1,2}=\tfrac{3}{2}(\gamma-2),\quad
    \mu_3=\tfrac{3}{2}\gamma,\quad
    \mu_4=0,
    \label{eq:CL-F-eigs}
  \end{equation}
  which are \emph{independent of $\alpha^\star$}: every point in the
  family has the same stability character.  Since $\mu_3=\frac{3}{2}
  \gamma>0$ for all $\gamma>0$, each $\mathcal{F}(\alpha^\star)$ is
  always a saddle.  The sub-diagonal entry $J_{41}=\sqrt{6}(\lambda-
  \alpha^\star)^2$ does not affect the eigenvalues; it governs the
  eigenvector geometry but not the stability type.  Note that setting
  $\alpha^\star=\lambda$ recovers the Branch-A $\mathcal{F}$ point with
  the same eigenvalues, confirming consistency between the two branches.

  \item \textbf{Sub-case B2 — $y^\star=1$: isolated de Sitter point
    $\hat{\mathcal{P}}$:}

  When $y^\star=1$, equation~\eqref{eq:CL-B-reduced} becomes
  $\sqrt{3/2}\,\alpha^\star=0$, which forces $\alpha^\star=0$.  This
  gives the single isolated point
  \begin{equation}
    \hat{\mathcal{P}} = (0,\,0,\,1,\,0).
  \end{equation}
  Unlike the family~\eqref{eq:CL-F-family}, here $\alpha^\star=0$ is
  not a choice but is uniquely determined by the fixed-point conditions;
  it corresponds to the field sitting at a potential extremum
  ($V_{z_1}=0$).  We now compute all 16 entries of the Jacobian at
  $\hat{\mathcal{P}}=(0,0,1,0)$.\\ 

  \textit{-Row 1} (from $h_1$):
  \begin{align}
    J_{11} &= \left[3(x_1^2+x_2^2-1)+6x_1^2
              -\tfrac{3}{2}\gamma(x_1^2+x_2^2+y^2-1)-3\gamma x_1^2
              \right]_{\hat{\mathcal{P}}}
           = 3(-1) - \tfrac{3}{2}\gamma(0) = -3,\\
    J_{12} &= 0,\quad
    J_{13} = 0,\quad  J_{14} = \left[\sqrt{\tfrac{3}{2}}\,y^2\right]_{\hat{\mathcal{P}}} = \sqrt{\tfrac{3}{2}}.
  \end{align}

  \textit{-Row 2} (from $h_2$, which contains no $\alpha$):
  \begin{equation}
    J_{21}=0,\quad
    J_{22}=3(0{-}1)-\tfrac{3}{2}\gamma(0)=-3,\quad
    J_{23}=0,\quad
    J_{24}=0.
  \end{equation}

  \textit{-Row 3} (from $h_3$):
  \begin{align}
    J_{33} &= \left[-\sqrt{\tfrac{3}{2}}\,\alpha x_1
              +3(x_1^2+x_2^2)
              -\tfrac{3}{2}\gamma\bigl[(x_1^2+x_2^2+y^2-1)+2y^2\bigr]
             \right]_{\hat{\mathcal{P}}}
           =-\tfrac{3}{2}\gamma(2)=-3\gamma,\\
    J_{31}&=0, \quad
    J_{32}=0,\quad
    J_{34}=0.
  \end{align}

  \textit{-Row 4} (from $h_4=\sqrt{6}(\lambda-\alpha)^2x_1$):
  \begin{equation*}
    J_{41}=\left[\sqrt{6}(\lambda-\alpha)^2\right]_{\hat{\mathcal{P}}}=\sqrt{6}\lambda^2,\quad J_{42}=0, \quad J_{43}=0, \quad J_{44} =0.
  \end{equation*}

  Assembling all entries, the $4\times4$ Jacobian at $\hat{\mathcal{P}}$
  is
  \begin{equation}
    J_{\hat{\mathcal{P}}} =
    \begin{pmatrix}
      -3 & 0 & 0 & \sqrt{3/2} \\[4pt]
       0 & -3 & 0 & 0 \\[4pt]
       0 & 0 & -3\gamma & 0 \\[4pt]
      \sqrt{6}\,\lambda^2 & 0 & 0 & 0
    \end{pmatrix}.
    \label{eq:Jhat}
  \end{equation}
  
  To compute the characteristic polynomial we expand
  $\det(J_{\hat{\mathcal{P}}}-\mu I)$ by cofactors along column~2
  (which has a single non-zero entry $-3-\mu$ at row~2), then along row~2 of the resulting $3\times3$ minor (single non-zero entry
  $-3\gamma-\mu$ at its second column), and finally evaluate the
  remaining $2\times2$ determinant:
  \begin{equation}
    \det(J_{\hat{\mathcal{P}}}-\mu I)
    = (-3-\mu)(-3\gamma-\mu)
      \underbrace{\left[(-3-\mu)(-\mu)
      - \sqrt{\tfrac{3}{2}}\cdot\sqrt{6}\,\lambda^2\right]}_{=\,\mu^2+3\mu
      -3\lambda^2} = 0.
  \end{equation}
   The full characteristic polynomial is therefore
  \begin{equation}
    (-3-\mu)(-3\gamma-\mu)(\mu^2+3\mu-3\lambda^2)=0.
    \label{eq:CL-P-charpoly}
  \end{equation}
  Setting each factor to zero gives: $\mu_1=-3$ and $\mu_2=-3\gamma$, both negative for $0<\gamma<2$.  The quadratic factor has roots
  \begin{equation}
    \mu_\pm = \frac{-3\pm\sqrt{9+12\lambda^2}}{2}.
    \label{eq:CL-P-roots}
  \end{equation}
  Since $\sqrt{9+12\lambda^2}>3$ for all $\lambda>0$ (squaring:
  $9+12\lambda^2>9\iff12\lambda^2>0$, which holds for any
  $\lambda\neq0$), we have $\mu_+>0$ always.  Therefore
  $\hat{\mathcal{P}}$ is always a saddle for every $\lambda>0$.

  \end{itemize}

\end{itemize}

The stability structure of the \ac{CL} model is summarised in
Table~\ref{tab:stab_CL}.  Apart from point $\hat{\mathcal{P}}$, the stability of the points is identical to those of the exponential model since the \ac{CL}-specific parameter $C$ enters neither the Branch-A eigenvalues
nor the condition for $\hat{\mathcal{P}}$.

\begin{table}[t]
\centering
\renewcommand{\arraystretch}{2.2}
\begin{adjustbox}{width=\textwidth}
\begin{tabular}{
    >{\centering\arraybackslash}m{2cm}
    >{\centering\arraybackslash}m{1.5cm}
    >{\centering\arraybackslash}m{3.5cm}
    >{\centering\arraybackslash}m{3.5cm}
    >{\centering\arraybackslash}m{3.5cm}
    >{\centering\arraybackslash}m{3cm}
}
\toprule
\rowcolor{lightgray}
\textbf{Point} & $\mu_1$ & $\mu_2$ & $\mu_3$ & $\mu_4$ & \textbf{Stability} \\
\midrule\midrule
$\hat{\mathcal{P}}$
  & $-3$ & $-3\gamma$
  & $\dfrac{-3+\sqrt{9+12\lambda^2}}{2}$
  & $\dfrac{-3-\sqrt{9+12\lambda^2}}{2}$
  & Saddle (always) \\
$\mathcal{K}_\pm$, $\mathcal{F}(\alpha^\star)$,
$\mathcal{S}$, $\mathcal{G}$, $\mathcal{NG}_\pm$
  & \multicolumn{4}{c}{EXP values (Table~\ref{tab:stab_exp}) $+\;\mu_4=0$
    (centre in $\alpha$; see text)}
  & Same conditions as EXP \\
\bottomrule\bottomrule
\end{tabular}
\end{adjustbox}
\caption{Eigenvalues and stability of the fixed points for the Clemson-Liddle model corresponding to $n=1$.}
\label{tab:stab_CL}
\end{table}

\subsection{Case $n=2$ (\ac{AS})}
\label{subsec:stab_AS}

The AS system is four-dimensional, with the slope parameter
$\alpha\equiv-M_{\rm pl}\,V_{\phi_1}/V=-V_{z_1}/V$ promoted to a
dynamical variable alongside $(x_1,x_2,y)$, evolving as
\begin{equation}
  \alpha' = \sqrt{6}\,x_1\,(1-\Gamma)\,\alpha^2,
  \label{eq:alpha-evol}
\end{equation}
where $\Gamma\equiv \frac{V\,V_{z_1 z_1}}{V_{z_1}^2}$. Fixed points require $\alpha'=0$; the right-hand side of
Eq.~\eqref{eq:alpha-evol} vanishes in three structurally distinct
ways, each giving a different class of fixed points and a different
Jacobian structure, analysed in turn below.

\begin{itemize}

\item \textbf{Branch (i) --- $x_1^\star=0$, fixed points
  $\mathcal{F}^{1,2}=(0,0,0,\alpha_i)$:}

When $x_1^\star=0$ the field is kinetically frozen and
$\alpha'=0$ is satisfied trivially, for any
value of $\alpha^\star$.  As in the CL case, $h_4=0$ places no
constraint on $\alpha^\star$; the residual equations $h_1=h_2=h_3=0$
determine it.  Substituting $x_1=0$ and using the same argument as
Section~\ref{subsec:stab_CL} gives $x_2^\star=0$, leaving
$\sqrt{3/2}\,\alpha^\star y^{\star2}=0$.  For $y^\star=0$ any
$\alpha^\star$ is permitted; the physically relevant values are
$\alpha_i=\lambda\mp1/\sqrt{A}$, at which $\Gamma(\alpha_i)=1$ and the
remaining fixed-point conditions are satisfied, giving
$\mathcal{F}^{1,2}=(0,0,0,\alpha_{1,2})$.  These are the AS
generalisations of the fluid-dominated point $\mathcal{F}$ of the EXP
model.  The case $y^\star=1$ leads instead to $\alpha^\star=0$ and the
de Sitter point $\mathcal{P}=(0,0,1,0)$, which is treated separately
in Branch~(ii) because its Jacobian requires a different approach.

We now evaluate all 16 partial derivatives of $(h_1,h_2,h_3,h_4)$ at
$(0,0,0,\alpha_i)$.

\textit{-Rows 1--3} (from $h_1,h_2,h_3$): the computation is identical
to the CL family $\mathcal{F}(\alpha^\star)$ in Sub-case~B1
(Section~\ref{subsec:stab_CL}), with $\alpha^\star\to\alpha_i$.  All
off-diagonal entries vanish because every term in rows 1--3 carries
$x_2^\star=0$ or $y^\star=0$ as a factor, and
$J_{14}=\sqrt{3/2}\,y^{\star2}=0$.  The diagonal entries are
\begin{equation}
  J_{11} = J_{22} = \tfrac{3}{2}(\gamma-2), \qquad
  J_{33} = \tfrac{3}{2}\gamma.
\end{equation}

\textit{-Row 4} (from $h_4=\sqrt{6}\,x_1(1-\Gamma)\alpha^2$):
\begin{align}
  J_{41} &= \frac{\partial h_4}{\partial x_1}\bigg|_\star
           = \sqrt{6}\,\underbrace{(1-\Gamma(\alpha_i))}_{=\,0}\,
             \alpha_i^2 = 0, \\
  J_{42} &= J_{43} = 0 \quad (h_4\text{ contains no }x_2,y), \\
  J_{44} &= \frac{\partial h_4}{\partial\alpha}\bigg|_\star
         = \sqrt{6}\,x_1^\star\left[-\frac{\partial\Gamma}{\partial\alpha}\,(\alpha^\star)^2
           + 2\underbrace{(1-\Gamma(\alpha^\star))}_{=\,0}\,\alpha^\star\right]= 0 .
\end{align}

\textit{-Column 4} (from $\partial h_i/\partial\alpha$ at $y^\star=0$,
$x_1^\star=0$):
\begin{equation}
  J_{14} = \sqrt{\tfrac{3}{2}}\,y^{\star2} = 0, \quad
  J_{24} = 0, \quad
  J_{34} = -\sqrt{\tfrac{3}{2}}\,x_1^\star y^\star = 0.
\end{equation}

Both the fourth row and the fourth column vanish entirely, so the
$4\times4$ Jacobian is fully block-diagonal:
\begin{equation}
  J_{4D}\big|_{\mathcal{F}^{1,2}} =
  \begin{pmatrix}
    \tfrac{3}{2}(\gamma-2) & 0 & 0 & 0 \\[4pt]
    0 & \tfrac{3}{2}(\gamma-2) & 0 & 0 \\[4pt]
    0 & 0 & \tfrac{3}{2}\gamma & 0 \\[4pt]
    0 & 0 & 0 & 0
  \end{pmatrix}
  =
  \begin{pmatrix} J_{3D}(\alpha_i) & 0 \\ 0 & 0 \end{pmatrix}.
  \label{eq:AS-block-F}
\end{equation}
For a block-diagonal matrix the determinant factorises as the product
of the block determinants:
\begin{equation}
  \det(J_{4D}-\mu I_4)
  = \det(J_{3D}(\alpha_i)-\mu I_3)\cdot(-\mu) = 0,
  \qquad
  \operatorname{spec}(J_{4D})
  = \operatorname{spec}(J_{3D}(\alpha_i))\cup\{0\}.
\end{equation}
The three non-trivial eigenvalues are those of the EXP model at the
$\mathcal{F}$ coordinates with $\lambda\to\alpha_i$
(Table~\ref{tab:stab_exp}):
\begin{equation}
  \mu_{1,2} = \tfrac{3}{2}(\gamma-2) < 0,
  \qquad
  \mu_3 = \tfrac{3}{2}\gamma > 0,
  \qquad
  \mu_4 = 0.
\end{equation}
Since $\mu_3>0$ for all $\gamma>0$, both $\mathcal{F}^1$ and
$\mathcal{F}^2$ are \emph{always saddles}.  The zero eigenvalue
$\mu_4=0$ is a centre direction in $z_1$: with $x_1^\star=0$ the field
is frozen and $z_1$ does not evolve, so no linearised information about
the $z_1$ direction is available.

\item \textbf{Branch (ii) --- $\alpha^\star=0$, de Sitter point
  $\mathcal{P}=(0,0,1,0)$:}

When $\alpha^\star=0$ the field sits at an extremum of $V$ where
$V_{z_1}=0$.  Since $(\alpha^\star)^2=0$, the right-hand side of
Eq.~\eqref{eq:alpha-evol} vanishes for \emph{any} value of
$x_1^\star$; $h_4=0$ places no constraint on $x_1^\star$.  The value
of $x_1^\star$ is determined by the residual equations
$h_1=h_2=h_3=0$.  Substituting $\alpha^\star=0$ into $h_1=0$ and
$h_2=0$ and imposing the physical requirement $\Omega_\phi=x_1^{\star
2}+x_2^{\star2}+y^{\star2}\leq1$ uniquely selects $x_1^\star=
x_2^\star=0$, $y^\star=1$, giving the isolated de Sitter point
$\mathcal{P}=(0,0,1,0)$.

This branch is treated separately from Branch~(i) because at
$\alpha^\star=0$ the quantity $\Gamma=VV_{z_1z_1}/V_{z_1}^2$ is
ill-defined ($V_{z_1}=0$ in both numerator and denominator), so the
form~\eqref{eq:alpha-evol} cannot be linearised directly.  Stability
must instead be read off from the chain-rule form
\begin{equation}
  \alpha' = \sqrt{6}\,x_1\,\frac{d\alpha}{dz_1},
  \qquad
  \frac{d\alpha}{dz_1} = \alpha^2-\frac{V_{z_1z_1}}{V},
  \label{eq:alpha-chain}
\end{equation}
evaluated at $\alpha^\star=0$:
\begin{equation}
  \frac{d\alpha}{dz_1}\bigg|_{z_1^\star}
  = -\frac{V_{z_1z_1}}{V}\bigg|_{z_1^\star}.
\end{equation}
We now compute all 16 entries of the Jacobian at $\mathcal{P}=
(0,0,1,0)$.

\textit{Rows 1--3}: since $h_1,h_2,h_3$ are identical to the CL
equations, the computation is exactly the same as for $\hat{\mathcal
{P}}$ in Sub-case~B2 (Section~\ref{subsec:stab_CL}).  The non-zero
entries are
\begin{equation}
  J_{11}=J_{22}=-3, \quad J_{33}=-3\gamma, \quad
  J_{14}=\sqrt{\tfrac{3}{2}},
\end{equation}
and all other entries in rows 1--3 vanish by the same arguments as
Sub-case~B2.

\textit{Row 4} (from the chain-rule form
$h_4=\sqrt{6}\,x_1\,d\alpha/dz_1$):
\begin{align}
  J_{41} &= \frac{\partial(\alpha')}{\partial x_1}\bigg|_{\mathcal{P}}
           = \sqrt{6}\,\frac{d\alpha}{dz_1}\bigg|_{z_1^\star}
           = -\sqrt{6}\,\frac{V_{z_1z_1}}{V}\bigg|_{z_1^\star}, \\
  J_{42} &= J_{43} = J_{44}= 0.
\end{align}

Assembling all entries, the full $4\times4$ Jacobian at $\mathcal{P}$
is
\begin{equation}
  J_{\mathcal{P}} =
  \begin{pmatrix}
    -3 & 0 & 0 & \sqrt{3/2} \\[4pt]
     0 & -3 & 0 & 0 \\[4pt]
     0 & 0 & -3\gamma & 0 \\[4pt]
    -\sqrt{6}\,V_{z_1z_1}/V\big|_{z_1^\star} & 0 & 0 & 0
  \end{pmatrix}.
  \label{eq:JP-AS}
\end{equation}
Comparing with the CL result~\eqref{eq:Jhat}, the structure is
identical except that $J_{41}=\sqrt{6}\lambda^2$ (always positive)
is replaced by $J_{41}=-\sqrt{6}\,V_{z_1z_1}/V|_{z_1^\star}$ (whose
sign depends on the curvature of the potential at the extremum).  This
sign change is the origin of the qualitatively different stability
behaviour.

We compute $\det(J_{\mathcal{P}}-\mu I)$ by the same three-step
cofactor expansion used in Sub-case~B2. First we expand along column~2: one non-zero entry $-3-\mu$
at row~2, giving
\begin{equation}
  \det(J_{\mathcal{P}}-\mu I) = (-3-\mu)\cdot M_{22},
\end{equation}
where $M_{22}$ is the $3\times3$ minor obtained by deleting row~2 and
column~2:
\begin{equation}
  M_{22} = \det\begin{pmatrix}
    -3-\mu & 0 & \sqrt{3/2} \\[4pt]
    0 & -3\gamma-\mu & 0 \\[4pt]
    -\sqrt{6}\,V_{z_1z_1}/V & 0 & -\mu
  \end{pmatrix}.
\end{equation}

Analogously, we expand $M_{22}$ along its column~2: one non-zero
entry $-3\gamma-\mu$ at row~2, giving
\begin{equation}
  M_{22} = (-3\gamma-\mu)\cdot M_{22}^{(22)},
\end{equation}
where $M_{22}^{(22)}$ is the $2\times2$ minor of $M_{22}$ obtained
by deleting its row~2 and column~2:
\begin{equation}
  M_{22}^{(22)} = \det\begin{pmatrix}
    -3-\mu & \sqrt{3/2} \\[4pt]
    -\sqrt{6}\,V_{z_1z_1}/V & -\mu
  \end{pmatrix}= \mu^2+3\mu
    +3
      \frac{V_{z_1z_1}}{V}.
\end{equation}

Note the sign in the last term. Because $J_{41}$ is negative here, the cross-term enters with a \emph{positive} sign, giving $+3V_{z_1z_1}/V$ rather
than the $-3\lambda^2$ of the CL case. Collecting all three factors:
\begin{equation}
  \det(J_{\mathcal{P}}-\mu I)
  = (-3-\mu)(-3\gamma-\mu)
    \left(\mu^2+3\mu+3\,\frac{V_{z_1z_1}}{V}\bigg|_{z_1^\star}\right)
  = 0.
  \label{eq:AS-P-charpoly}
\end{equation}
The first two factors give $\mu_1=-3$ and $\mu_2=-3\gamma$, both
negative for $0<\gamma<2$.  Applying the quadratic formula to the
third factor with $a=1$, $b=3$, $c=3V_{z_1z_1}/V$, and defining the
discriminant
\begin{equation}
  \Delta \equiv \frac{9}{4} - 3\,\frac{V_{z_1z_1}}{V}
  \bigg|_{z_1^\star},
  \label{eq:Delta-P}
\end{equation}
the remaining two eigenvalues are
\begin{equation}
  \mu_{3,4} = -\frac{3}{2}\pm\sqrt{\Delta}.
\end{equation}
The stability depends entirely on the sign of $V_{z_1z_1}$ at the
extremum:
\begin{itemize}
  \item At a \emph{potential maximum} ($V_{z_1z_1}<0$):
    $\Delta=9/4+3|V_{z_1z_1}/V|>9/4$, so $\sqrt{\Delta}>3/2$ and
    $\mu_+>0$: $\mathcal{P}$ is a \emph{saddle}.
  \item At a \emph{potential minimum} ($V_{z_1z_1}>0$):
    $\Delta<9/4$.  If $\Delta\geq0$, both roots $\mu_\pm=
    -3/2\pm\sqrt{\Delta}$ are real and negative since
    $\sqrt{\Delta}<3/2$.  If $\Delta<0$, the roots are complex with
    $\operatorname{Re}(\mu_{3,4})=-3/2<0$.  In both sub-cases all
    four eigenvalues have negative real part: $\mathcal{P}$ is a
    \emph{stable attractor} whenever $A\lambda^2<1$.
\end{itemize}

\item \textbf{Branch (iii) --- $\Gamma=1$, $x_1^\star\neq0$:}

When $\Gamma(\alpha^\star)=1$ and $\alpha^\star\neq0$, the factor
$(1-\Gamma)=0$ kills the right-hand side of Eq.~\eqref{eq:alpha-evol}
regardless of $x_1^\star$.  The zeros of $\Gamma(z_1)-1$ give the two
slope values $\alpha_{1,2}=\lambda\mp1/\sqrt{A}$, and the fixed-point
coordinates $(x_1,x_2,y)$ are those of the EXP model with $\lambda\to
\alpha_i$, giving $\mathcal{K}^{1,2}_\pm$, $\mathcal{S}$,
$\mathcal{G}_\pm$, $\mathcal{NG}^{1,2}_\pm$.

We compute the fourth row of the Jacobian from $h_4=\sqrt{6}\,x_1
(1-\Gamma)\alpha^2$:
\begin{align}
  J_{41} &= \frac{\partial h_4}{\partial x_1}\bigg|_\star
           = \sqrt{6}\,\underbrace{(1-\Gamma(\alpha^\star))}_{=\,0}
             \,(\alpha^\star)^2 = 0, \\
  J_{42} &= J_{43} = 0 \quad (h_4\text{ contains no }x_2,y), \\
  J_{44} &= \frac{\partial h_4}{\partial\alpha}\bigg|_\star
           = -\sqrt{6}\,x_1^\star(\alpha^\star)^2
             \frac{\partial\Gamma}{\partial\alpha}\bigg|_\star=0.
\end{align}
For the specific AS potential, the value of $J_{44}$ at the Branch~
(iii) fixed points vanishes, giving a zero
fourth eigenvalue.  The Jacobian is therefore upper block-triangular,
\begin{equation}
  J_{4D}\big|_{\Gamma=1} =
  \begin{pmatrix}
    J_{3D}(\alpha_i) & \mathbf{c} \\
    \mathbf{0}^\top  & 0
  \end{pmatrix},
  \quad
  \mathbf{c} = \begin{pmatrix}
    \sqrt{3/2}\,y^{\star2} \\ 0 \\ -\sqrt{3/2}\,x_1^\star y^\star
  \end{pmatrix},
  \label{eq:AS-block-iii}
\end{equation}
and by the same block-triangular determinant factorisation as
Branch~A of the CL model:
\begin{equation}
  \operatorname{spec}(J_{4D})
  = \operatorname{spec}(J_{3D}(\alpha_i))\cup\{0\}.
\end{equation}
All Branch~(iii) fixed points inherit the EXP stability conditions
with $\lambda\to\alpha_i$ (Table~\ref{tab:stab_exp}), plus a centre
direction in $\alpha$ with $\mu_4=0$.

The non-geodesic fixed point $\mathcal{NG}^2$ has $\alpha_2=\lambda+
1/\sqrt{A}$, existing when $A\lambda^2\leq1$.  Its three EXP
eigenvalues are those of $\mathcal{NG}$ with $\lambda\to\alpha_2$:
\begin{equation}
  \mu_1=-3\gamma,\quad
  \mu_{2,3}=\frac{-3\pm\sqrt{\Delta_2}}{2},\quad
  \Delta_2=3(27-8\xi\alpha_2-4\alpha_2^2).
\end{equation}
All three have negative real part when
\begin{equation}
  \sqrt{\xi^2+6}-\xi \;\leq\; \alpha_2 \;\leq\; 2\xi.
  \label{eq:NG_stab_AS}
\end{equation}
When $A\lambda^2\leq1$, as established in
Section~\ref{subsec:AS-pot}, $\mathcal{NG}^1$ is unphysical and
$\mathcal{NG}^2$ is the sole non-geodesic attractor.  When
$A\lambda^2>1$, the potential has no real extrema, $\mathcal{P}$
ceases to exist, and $\alpha_1=\lambda-1/\sqrt{A}$ becomes strictly
positive; $\mathcal{NG}^1$ is then a physical Branch~(iii) fixed
point whose three EXP eigenvalues carry the same structure with
$\lambda\to\alpha_1$:
\begin{equation}
  \mu_1=-3\gamma,\quad
  \mu_{2,3}=\frac{-3\pm\sqrt{\Delta_1}}{2},\quad
  \Delta_1=3(27-8\xi\alpha_1-4\alpha_1^2),
\end{equation}
with negative real parts when
\begin{equation}
  \sqrt{\xi^2+6}-\xi \;\leq\; \alpha_1 \;\leq\; 2\xi.
  \label{eq:NG1_stab_AS}
\end{equation}
In this regime $\mathcal{NG}^1$ is the numerically observed late-time
attractor, consistent with it carrying the shallower effective slope.
The complete eigenvalue spectrum is given in Table~\ref{tab:stab_AS}.

\end{itemize}

\begin{table}[t]
\centering
\renewcommand{\arraystretch}{2.2}
\begin{adjustbox}{width=\textwidth}
\begin{tabular}{
    >{\centering\arraybackslash}m{2.4cm}
    >{\centering\arraybackslash}m{1.5cm}
    >{\centering\arraybackslash}m{3.2cm}
    >{\centering\arraybackslash}m{3.2cm}
    >{\centering\arraybackslash}m{1.5cm}
    >{\centering\arraybackslash}m{3.8cm}
}
\toprule
\rowcolor{lightgray}
\textbf{Point} & $\mu_1$ & $\mu_2$ & $\mu_3$ & $\mu_4$ &
\textbf{Stability} \\
\midrule\midrule
$\mathcal{P}$ ($V$ max)
  & \multicolumn{4}{c}{$-3,\quad-3\gamma,\quad
    -\tfrac{3}{2}\pm\sqrt{\Delta}$,\quad$\Delta>9/4$}
  & Saddle (always); $A\lambda^2\leq1$ \\
$\mathcal{P}$ ($V$ min)
  & \multicolumn{4}{c}{$-3,\quad-3\gamma,\quad
    -\tfrac{3}{2}\pm\sqrt{\Delta}$,\quad$\Delta<9/4$}
  & Stable attractor; $A\lambda^2\leq1$ \\
$\mathcal{K}^{1,2}_\pm$, $\mathcal{F}^{1,2}$, $\mathcal{S}$,
$\mathcal{G}_\pm$
  & \multicolumn{4}{c}{EXP values (Table~\ref{tab:stab_exp}) with
    $\lambda\to\alpha_{1,2}$, plus $\mu_4=0$}
  & Same conditions as EXP \\
$\mathcal{NG}^2_\pm$
  & $-3\gamma$
  & $\dfrac{-3+\sqrt{\Delta_2}}{2}$
  & $\dfrac{-3-\sqrt{\Delta_2}}{2}$
  & $0$
  & $\sqrt{\xi^2+6}-\xi\leq\alpha_2\leq2\xi$
    \newline and $A\lambda^2\leq1$ \\
$\mathcal{NG}^1_\pm$
  & $-3\gamma$
  & $\dfrac{-3+\sqrt{\Delta_1}}{2}$
  & $\dfrac{-3-\sqrt{\Delta_1}}{2}$
  & $0$
  & $\sqrt{\xi^2+6}-\xi\leq\alpha_1\leq2\xi$
    \newline and $A\lambda^2>1$ \\
\bottomrule\bottomrule
\end{tabular}
\end{adjustbox}
\caption{Eigenvalues and stability of the fixed points for the Albrecht-Skordis model corresponding to $n=2$.}
\label{tab:stab_AS}
\end{table}

\subsection{Comparison across models}
\label{subsec:stab_compare}

The three models are built on the same two-field framework and share
the same equations $h_1,h_2,h_3$ for $(x_1,x_2,y)$, differing only
in the slope equation $h_4$.  This common structure means that the
off-diagonal Jacobian entries $J_{14}=\sqrt{3/2}\,y^{\star2}$ and
$J_{34}=-\sqrt{3/2}\,x_1^\star y^\star$ (from $\partial h_1/\partial
\alpha$ and $\partial h_3/\partial\alpha$) are identical across all
three models at any given fixed point.  The difference in stability
behaviour is entirely encoded in $J_{41}=\partial h_4/\partial x_1$,
which measures how strongly $x_1$ drives $\alpha$ in the $(x_1,\alpha)$
phase plane.

\begin{enumerate}

\item \textbf{EXP}: three-dimensional system $(x_1,x_2,y)$ with
  constant slope $\lambda$.  There is no $\alpha$ direction, so no
  fourth row or column exists in the Jacobian.  The kinematic points
  $\mathcal{K}^\pm$ and $\mathcal{F}$ are always unstable; the
  geodesic attractor $\mathcal{G}$, the scaling solution $\mathcal{S}$,
  and the non-geodesic attractor $\mathcal{NG}$ are stable fixed points, but only $\mathcal{NG}$ can simultaneously
  achieve $\Omega_\phi=0.7$ and $\omega_\phi\approx-1$ for sufficiently
  large $\xi$.

\item \textbf{CL}: four-dimensional system $(x_1,x_2,y,\alpha)$ with
  $h_4=\sqrt{6}(\lambda-\alpha)^2x_1$.  The key Jacobian entry is
  $J_{41}=\sqrt{6}(\lambda-\alpha)^2|_\star$, which vanishes at
  $\alpha^\star=\lambda$ (Branch~A) and takes the positive value
  $\sqrt{6}\lambda^2$ at $\hat{\mathcal{P}}$ (Branch~B2).  Branch~A
  fixed points ($\mathcal{K}_\pm$, $\mathcal{F}$, $\mathcal{S}$,
  $\mathcal{G}$, $\mathcal{NG}$) carry the EXP eigenvalues plus a zero
  eigenvalue; their stability conditions are identical to EXP and our asymptotic analysis (Section~\ref{subsec:stab_CL})confirms $\alpha\to\lambda$
  algebraically as $N^{-1}$ for all points with $x_1^\star\neq0$.
  Branch~B splits into the fluid-dominated family
  $\mathcal{F}(\alpha^\star)=(0,0,0,\alpha^\star)$ (always a saddle,
  eigenvalues independent of $\alpha^\star$) and the isolated de Sitter
  point $\hat{\mathcal{P}}=(0,0,1,0)$.  At $\hat{\mathcal{P}}$ the
  positive entry $J_{41}=\sqrt{6}\lambda^2>0$ creates a positive
  feedback loop with $J_{14}=\sqrt{3/2}$, giving the product
  $J_{14}\cdot J_{41}=3\lambda^2>0$ and forcing the negative constant
  $-3\lambda^2$ in the quadratic factor of the characteristic
  polynomial~\eqref{eq:CL-P-charpoly}, which guarantees $\mu_+>0$:
  $\hat{\mathcal{P}}$ is always a saddle.

\item \textbf{AS}: four-dimensional system $(x_1,x_2,y,\alpha)$ with
  $h_4=\sqrt{6}\,x_1(1-\Gamma)\alpha^2$.  Fixed points fall into three
  branches.  Branch~(iii) points ($\Gamma=1$,
  $x_1^\star\neq0$) — $\mathcal{K}^{1,2}_\pm$, $\mathcal{S}$,
  $\mathcal{G}_\pm$, $\mathcal{NG}^{1,2}$ — have $J_{41}=0$ because
  $(1-\Gamma(\alpha^\star))=0$, giving the same upper block-triangular
  Jacobian as CL Branch~A and hence the same EXP stability conditions
  with $\lambda\to\alpha_i$, plus $\mu_4=0$
  (Table~\ref{tab:stab_AS}).  $\mathcal{F}^{1,2}$ lies at the
  intersection of branches~(i) and~(iii) and has both the fourth row
  and the fourth column vanishing, giving a fully block-diagonal
  Jacobian and the same saddle character as EXP $\mathcal{F}$.  The de
  Sitter point $\mathcal{P}=(0,0,1,0)$, existing when $A\lambda^2\leq
  1$, has $J_{41}=-\sqrt{6}\,V_{z_1z_1}/V|_{z_1^\star}$, whose sign
  is opposite to the CL entry $J_{41}=+\sqrt{6}\lambda^2$.  This sign
  reversal changes the constant term in the quadratic factor of the
  characteristic polynomial from $-3\lambda^2$ (always negative,
  forcing instability in CL) to $+3V_{z_1z_1}/V$ (positive at the
  potential minimum, forcing stability).  Consequently $\mathcal{P}$
  is a saddle at the potential maximum but a stable attractor at the
  potential minimum for all $A\lambda^2<1$, independently of $\xi$ and
  $\lambda$.  The two regimes are cleanly separated by
  $A\lambda^2=1$: for $A\lambda^2\leq1$, $\mathcal{NG}^1$ is
  unphysical and $\mathcal{NG}^2$ is the non-geodesic attractor,
  stable in the window~\eqref{eq:NG_stab_AS}; for $A\lambda^2>1$,
  $\mathcal{P}$ and $\mathcal{NG}^2$ are absent and $\mathcal{NG}^1$
  takes over as the physical non-geodesic attractor, stable in the
  analogous window~\eqref{eq:NG1_stab_AS}.

\end{enumerate}

The key phenomenological distinction is that the \ac{AS} model admits a
guaranteed de Sitter endpoint: whenever $A\lambda^2<1$, the point
$\mathcal{P}$ at the potential minimum is a stable attractor with
$\omega_\phi=-1$, independently of $\xi$ and $\lambda$. It can
coexist with the non-geodesic attractor $\mathcal{NG}^2$.  For $A\lambda^2>1$, by contrast, the potential has
no minimum, $\mathcal{P}$ is absent, and the late-time behaviour is
governed instead by $\mathcal{NG}^1$, stable in the
window~\eqref{eq:NG1_stab_AS}. By contrast, the \ac{CL} model
preserves the \ac{EXP} attractor structure exactly, with
$\hat{\mathcal{P}}$ playing only the role of an unstable transient
state.  The structural reason for this difference is entirely contained
in the sign of $J_{41}$: positive in CL (driving instability) and
negative in \ac{AS} at the potential minimum, driving stability.

\section{Bayesian results of $n-$free}
\label{app:nfree}
\begin{figure}[h]
    \centering
    \makebox[12cm][c]{
    \includegraphics[width=16.cm]{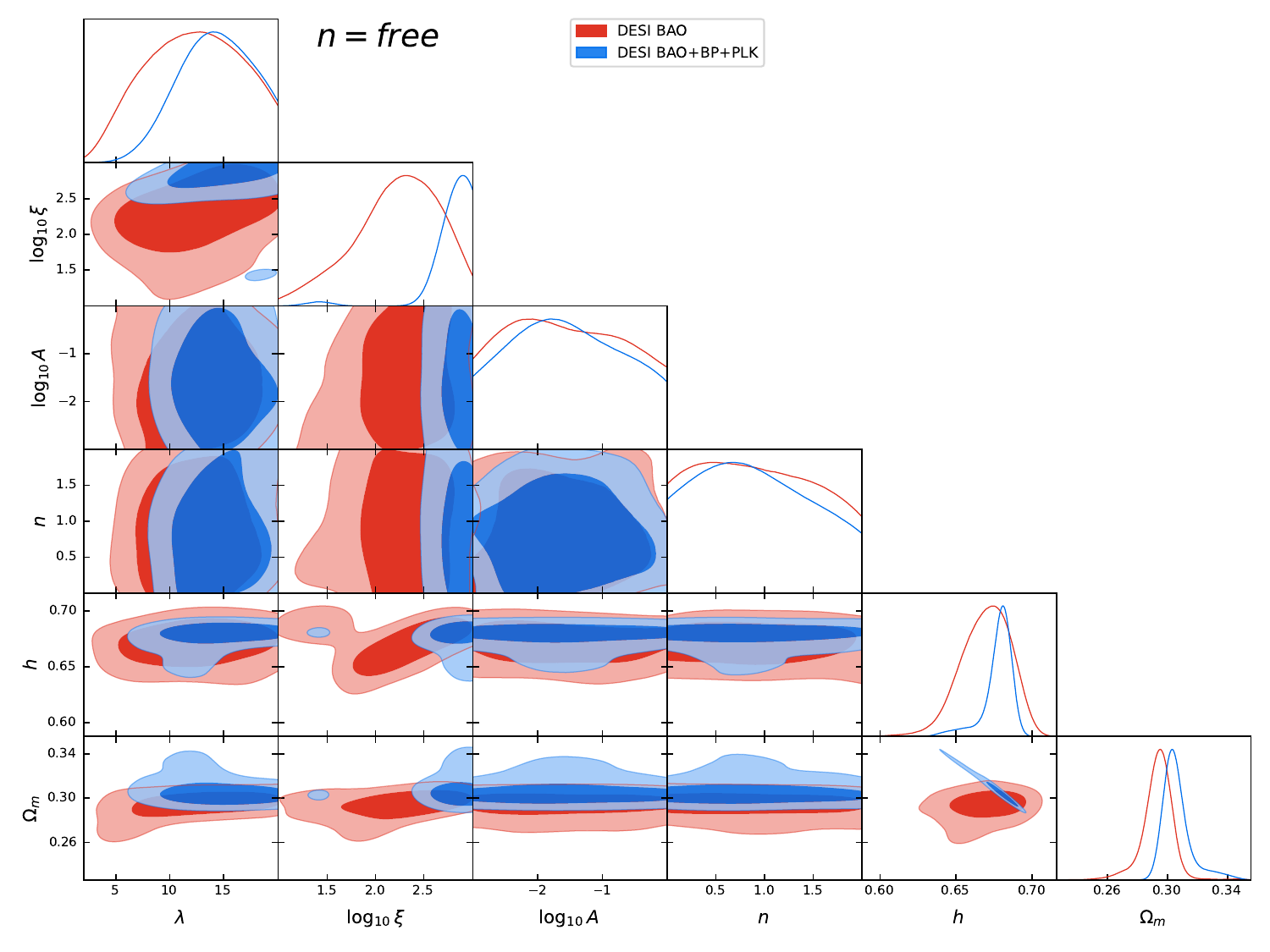}
    }
   \caption{Marginalised posterior distributions considering $n$ as free parameter. The contours show the 1-$\sigma$ and 2-$\sigma$ confidence regions for DESI BAO and DESI BAO  + BP + PLK. }
   \label{fig:posterior-nn}
\end{figure}

To investigate whether the current data show any preference for a particular value of the polynomial exponent, we repeated the Bayesian analysis by treating $n$ as a free parameter. We adopted a uniform prior, $n\in[0,2]$, while keeping the same cosmological data set and parameter priors used in the analyses with fixed values of $n$. The marginalised parameter constraints are summarised in Table~\ref{tab:nsummary}, and the corresponding posterior distributions are shown in Fig.~\ref{fig:posterior-nn}.

Similar to the previous cases, the data do not provide a meaningful constraint on $A$, confirming its secondary role in the dynamics of the scalar fields.
With respect to $n$, it's posterior distribution extends to the edge of the prior range, confirming that the current observations are unable to distinguish between the different polynomial exponents considered in this work.  In contrast, the remaining scalar-field parameters are constrained in a manner consistent with the fixed-$n$ analyses, yielding $\lambda>8.5$ and $\xi>10^{2.49}$ at the 95\% confidence level for DESI \ac{BAO}+\ac{BP}+\ac{PLK} dataset. 

\begin{table}[htbp]
\centering
\renewcommand{\arraystretch}{1.25}
\begin{adjustbox}{width=\textwidth}
\begin{tabular}{lcccc}
\toprule
& \multicolumn{2}{c}{$\Lambda \rm CDM$}& \multicolumn{2}{c}{$n=$free} \\
\cmidrule(lr){2-3} \cmidrule(lr){4-5}
\textbf{Parameter}
& \textbf{DESI \ac{BAO}}
& \textbf{DESI \ac{BAO}  + \ac{BP} + \ac{PLK}} & \textbf{DESI \ac{BAO}}
& \textbf{DESI \ac{BAO}  + \ac{BP} + \ac{PLK}}\\
\midrule
$\Omega_{b}h^2$ & $0.0220\pm 0.0004$ &$0.0226\pm 0.0001$& $0.0220\pm 0.0003$ &$0.0223\pm 0.0001$ \\
$\Omega_m$ &$0.298^{+0.007}_{-0.008}$ & $0.300\pm 0.004$ &$0.294^{+0.010}_{-0.007}$& $0.306^{+0.005}_{-0.010}$\\
$h$ & $0.68\pm{0.005}$& $0.69\pm 0.003$ & $0.67^{+0.02}_{-0.01}$& $0.68^{+0.02}_{-0.03}$ \\
$\lambda$ & & &$>5.13$&$>8.50$\\
$\log_{10}\xi$ & & &$> 1.41$ &$>2.49$ \\
$\log_{10}A$ & & &$---$&$---$ \\
$n$ & & & $---$ &$---$\\
\midrule
$\chi^2_{\min}$ &$10.28$ & $51.6$ &$5.6$ & $50.6$\\
\bottomrule
\end{tabular}
\end{adjustbox}
\caption{Marginalised constraints on the cosmological and model parameters with the polynomial exponent $n$ left free. Error bars denote 68\% confidence limits, while lower and upper bounds are reported at the 95\% confidence level for both the DESI \ac{BAO}  alone and the combined DESI \ac{BAO} + \ac{BP} + \ac{PLK} data set.}
\label{tab:nsummary}

\end{table}

\bibliographystyle{JHEP}
\bibliography{biblio}

\end{document}